\documentclass[fleqn,usenatbib]{mnras}

\usepackage[T1]{fontenc}

\DeclareRobustCommand{\VAN}[3]{#2}
\let\VANthebibliography\thebibliography
\def\thebibliography{\DeclareRobustCommand{\VAN}[3]{##3}\VANthebibliography}

\usepackage{graphicx}	
\usepackage{amsmath}	
\usepackage{amssymb}	
\usepackage{multirow}
\usepackage{cuted}
\usepackage{hyperref}
\usepackage{url}
\usepackage[dvipsnames]{xcolor}
\usepackage{soul}
\usepackage{todonotes}
\usepackage{longtable}
\usepackage{array}
\usepackage{pdflscape}
\usepackage[hypcap=false]{caption}
\usepackage{changepage} 
\usepackage{booktabs}

\usepackage[export]{adjustbox}
\usepackage{gensymb}

\newcommand{\Msun}{\mbox{$\mathrm{M}_{\sun}$}}

\newcommand{\allsdss}{SDSS~I~to~IV}
\newcommand{\DESI}{DESI}

\newcommand {\NtrainingCVs}{304}
\newcommand {\NnotdodgyCVs}{303}
\newcommand {\Nspectra}{98\,966\,000}
\newcommand {\Nspectraobjtype} {4\,018\,868}
\newcommand {\NspectraobjtypeTGT} {80\,767\,382}
\newcommand {\NspectraobjtypeSKY} {14\,123\,556}
\newcommand {\NspectraobjtypeBAD} {55\,315}
\newcommand {\NspectraobjtypeNON} {879}
\newcommand {\Nnonscience}{18\,198\,618}
\newcommand {\Nscience}{80\,767\,382}

\newcommand {\CMATN}{1024}
\newcommand {\CMAFN}{0}
\newcommand {\CMAFP}{0}
\newcommand {\CMATP}{59}
\newcommand {\CMATOT}{1083}

\newcommand {\NCVs}{1029}

\newcommand {\NCVswithperiods}{454}
\newcommand {\NnewCVs}{221}
\newcommand {\NnotnewCVs}{808}

\newcommand {\Newspec}{441}
\newcommand {\Newperiod}{84}

\newcommand {\Npeculiar}{12}
\newcommand {\NchrisCVs}{221}
\newcommand {\Nchristrainingspectra}{304}

\newcommand {\Nchrisfoundspectra}{254}
\newcommand {\Nrefcatspectra}{29\,154}

\newcommand {\Nmlspectraloz}{18\,511}

\newcommand {\Noutbursts}{43}

\newcommand {\Nnooutbursts}{128}

\newcommand {\Nuntyped}{156}
\newcommand {\Nmlselected}{293\,670}

\newcommand {\NknownCVs}{16\,277}
\newcommand {\Ngoodcvspectramissed}{48}

\newcommand {\Reliabilitypercent}{97.2}
\newcommand {\Nnotfound}{234}
\newcommand {\Nbadcvspectramissed}{186}
\newcommand {\NinWDsurvey}{580}
\newcommand {\Nknownspectra}{1531}

\newcommand {\NFUV}{410}
\newcommand {\Ncut}{406}
\newcommand {\Ncuttwo}{231}
\newcommand {\Ndedupespectra}{1803}
\newcommand {\NdedupeCVs}{1023}

\newcommand {\Nspike}{3859}

\newcommand {\Nnotinspected}{238\,279}

\newcommand {\NSPWD}{214}
\newcommand {\NSPSPCV}{65}
\newcommand {\NSPWDB}{67}

\newcommand {\Nnotthreesurveys}{709}

\newcommand{\Porb}{\mbox{$P_\mathrm{orb}$}}

\usepackage{newtxtext,newtxmath}

\title[1000 CVs from DESI]{1000 cataclysmic variables identified from DESI spectroscopy}

\author[K. Inight et al.]{
K.~Inight,$^{1}$\thanks{E-mail: keith.inight@gmail.com }
B.~T.~G\"ansicke,$^{1}$
A.~Aungwerojwit,$^{2}$
P.~Izquierdo,$^{1}$
C.~J.~Manser,$^{1}$
A.~D.~Myers,$^{3}$
\newauthor
A.~Swan,$^{1}$
J.~R.~Thorstensen,$^{4}$
J.~Aguilar,$^{5}$
S.~Ahlen,$^{6}$
D.~Bianchi,$^{7,8}$
D.~Brooks,$^{9}$
\newauthor
T.~Claybaugh,$^{5}$
A.~de la Macorra,$^{10}$
Biprateep~Dey,$^{11,12}$
P.~Doel,$^{9}$
A.~Font-Ribera,$^{13,14}$
\newauthor
J.~E.~Forero-Romero,$^{15,16}$
Satya~{Gontcho A Gontcho},$^{17}$
G.~Gutierrez,$^{18}$
J.~Guy,$^{5}$
R.~Joyce,$^{19}$
\newauthor
S.~Juneau,$^{19}$
S.~E.~Koposov,$^{20,21}$
A.~Kremin,$^{5}$
M.~Landriau,$^{5}$
L.~Le~Guillou,$^{22}$
\newauthor
A.~Meisner,$^{19}$
R.~Miquel,$^{13,14}$
J.~Moustakas,$^{23}$
J.~Najita,$^{19}$
I.~P\'erez-R\`afols,$^{24}$
\newauthor
W.~J.~Percival,$^{25,26,27}$
F.~Prada,$^{28}$
G.~Rossi,$^{29}$
E.~Sanchez,$^{30}$
D.~Schlegel,$^{5}$
M.~Schubnell,$^{31,32}$
\newauthor
H.~Seo,$^{33}$
J.~Silber,$^{5}$
G.~Tarl\'{e},$^{32}$
B.~A.~Weaver,$^{19}$
R.~Zhou,$^{5}$
H.~Zou$^{34}$
\\
\\
\parbox{\textwidth}{ \small
\textit{The authors' affiliations are shown at the end of this paper.}}
}

\date{Accepted XXX. Received YYY; in original form ZZZ}

\pubyear{2020}

\begin{document}
\label{firstpage}
\pagerange{\pageref{firstpage}--\pageref{lastpage}}
\maketitle

\begin{abstract}
Most cataclysmic variables (CVs) are discovered when they have an outburst generating an inherent selection bias against CVs that rarely, or never, outburst. CVs discovered by virtue of their spectroscopic characteristics are particularly valuable to offset this bias and we have used an established machine-learning technique to assist in searching \Nspectra\ spectra obtained by the Dark Energy Spectroscopic Survey (DESI) to find such CVs. DESI observations are much deeper than previous spectroscopic surveys and we have identified \NCVs\ CVs, \NnewCVs\ of which are new including ten of the AM\,CVn subtype. We have spectroscopically confirmed \Newspec\ CV candidates and obtained \Newperiod\,new or improved orbital periods. We present revised space density estimates based upon this new data. We have also added ten more to the eight known examples of an intriguing class of CVs which exhibit peculiar changes in accretion. 
\end{abstract}

known


\section{Introduction}\label{sec:introduction}

Cataclysmic variables (CVs) are close binaries consisting of a white dwarf accreting from a donor star that is filling its Roche lobe (see \citealt{2003cvs..book.....W} and \citealt{2025MNRAS.536.1057I} for an exhaustive and a short but recent overview on CVs). CVs form through a common envelope event \citep{1976IAUS...73...75P, 1979A&A....78..167M} and subsequently evolve towards shorter orbital periods via angular momentum loss driven by magnetic wind braking \citep{1981ApJ...248L..27P} and gravitational wave radiation \citep{1983ApJ...275..713R}. A dearth of CVs with periods between 2$-$3\,h, known as the period gap, has been noted since the early days of CV population studies \citep{1980MNRAS.190..801W}. The strength of this feature in the orbital period distribution varies among different observed samples \citep{2006MNRAS.373..484K, 2020MNRAS.494.3799P, 2023MNRAS.524.4867I} and its physical origin remains debated \citep[e.g.][]{2012ApJ...751...98V, 2022MNRAS.517.4916E}, with a reduction of the efficiency in magnetic braking being one of the favoured hypotheses \citep{1983ApJ...275..713R, 2024A&A...690L...1O, 2024A&A...682L...7S}. The orbital period keeps decreasing to $\Porb\simeq80$\,min \citep{2009MNRAS.397.2170G, 2011ApJS..194...28K}, at which point the donor mass falls below the limit required to  sustain nuclear burning ($\simeq0.075$\,\Msun, \citealt{2023A&A...671A.119C}). Further evolution results in an \textit{increase} in the period. These highly evolved systems, referred to as ``period bouncers'' are a dominant component in theoretical population models  \citep{1993A&A...271..149K, 1999MNRAS.309.1034K, 2001ApJ...550..897H}. However period bouncers remain scarce in observed CV samples \citep{2018MNRAS.473.3241H, 2020MNRAS.494.3799P, 2023MNRAS.525.3597I}, which suggests that the population models may require modifications \citep{1998PASP..110.1132P, 2025A&A...698L..22S}. Other gaps in the understanding of CV evolution include the fraction of CVs with nuclear evolved donors, and how they affect the properties of the overall CV population (e.g. \citealt{2002ApJ...567L..49T, 2021MNRAS.508.4106E, 2023MNRAS.523..305T, 2025A&A...703A.119T}); the influence of classical nova eruptions on the secular evolution of CVs \citep{2016MNRAS.455L..16S, 2022ApJ...938...31S} as well as the origin of the magnetic fields detected in many CV white dwarfs \citep{2021NatAs...5..648S, 2025A&A...698L..22S}. In addition to their importance in the context of binary evolution, CVs are natural laboratories for the general study of accretion physics \citep{2014apa..book.....G} and the shortest-period systems ($<65$\,min) are sources of gravitational wave radiation  which will be used to verify the performance of the \textit{LISA} space mission \citep{2018MNRAS.480..302K,2023MNRAS.525L..50S}.

Progress in all these areas of research critically depends on the availability of well-defined samples of known CVs with well-established physical properties. Historically, CVs were primarily detected because of their large-amplitude photometric variability caused by their thermally unstable accretion discs \citep{1981A&A...104L..10M}, with colours, X-ray emission, and spectroscopy representing alternative methods of identification \citep{2005ASPC..330....3G}. Ironically, a large fraction of CVs do not outburst at all because the strong magnetic field of their white dwarfs inhibits the formation of accretion discs \citep{1990SSRv...54..195C, 2000SSRv...93..611W, 2023MNRAS.524.4867I},  have such low mass transfer rates that their outburst recurrence times are decades \citep{2015PASJ...67..108K, 2023MNRAS.524.4867I} or have hot and stable accretion discs that again do not exhibit much photometric variability \citep{1998AJ....115.2527H, 2022MNRAS.510.3605I}.

Spectroscopic surveys and in particular the Sloan Digital Sky Survey (SDSS~--~\citealt{2000AJ....120.1579Y,2006AJ....131.2332G}) have been demonstrated to be largely immune to selection effects in their efficiency of identifying CVs, based on the simple fact that practically all known CVs exhibit Balmer and/or He emission lines  \citep{2002AJ....123..430S,2003AJ....126.1499S,2004AJ....128.1882S,2005AJ....129.2386S,2006AJ....131..973S,2007AJ....134..185S,2009AJ....137.4011S,2011AJ....142..181S,2023MNRAS.524.4867I,2023MNRAS.525.3597I,2025MNRAS.536.1057I}. The SDSS CV sample has been instrumental in confirming the accumulation (``spike'') of CVs near the orbital period minimum  \citep{2009MNRAS.397.2170G} that was predicted by population models but not present in CV samples dominated by outbursting systems \citep{1999MNRAS.309.1034K}. SDSS currently operates 2-m class telescopes at Apache Point and Las Cumbres observatories to obtain up to $500$ optical spectra in each observation. The five generations of the SDSS surveys have so far observed a total of 934 CVs including 349 new discoveries  \citep{2011AJ....142..181S, 2023MNRAS.524.4867I, 2023MNRAS.525.3597I,2025MNRAS.536.1057I}. 


The Dark Energy Spectroscopic Instrument (DESI~--~\citealt{2016arXiv161100036D, 2016arXiv161100037D}) on the Mayall 4-m telescope at Kitt Peak National Observatory has the potential to expand on the results from SDSS by identifying much larger numbers of intrinsically fainter CVs. DESI  is a multi-object spectrograph using 5000 fibres,  which are positioned by robotic actuators and are grouped into ten petals which feed ten identical three-arm spectrographs, each spanning 3600$-$9824\,\AA\ at a FWHM resolution of $\simeq$\,1.8\,\AA\ \citep{2022AJ....164..207D, 2023AJ....165....9S, 2024AJ....168..245P}. The inter-exposure sequence, which includes telescope slewing, spectrograph readout and focal plane reconfiguration, can be completed in as little as $\simeq2$\,min \citep{2022AJ....164..207D}, facilitating the acquisition of several 10\,000 spectra per night. DESI started main survey operations on 2021 May 14 with the primary objective of achieving spectroscopy of approximately 40\,million galaxies and quasars over five years in order to explore the nature of dark energy \citep{2024AJ....167...62D}. During sub-optimal observing conditions (e.g. poor seeing or high lunar illumination), observations switch focus to nearby bright galaxies \citep{2020RNAAS...4..187R,2023AJ....165..253H} and stars \citep{2020RNAAS...4..188A, 2023ApJ...947...37C}.

Here we analyse the entire set of \Nspectra\ spectra to be contained in DESI Data Release (DR) ~2 spanning the period 2020\,December\,14 to 2024\,April\,9 using machine learning, to identify a total of \NCVs\ CVs, and use this sample to estimate the space densities for different subtypes of CVs.

\section {CV sub-types} \label{sec:cv_subtypes}
CVs are categorised into a number of sub-types\footnote{AM\,CVn, CV, DN (dwarf nova), IP (intermediate polar), MCV (magnetic CV), NL (novalike), polar, SU\,UMa, U\,Gem, WZ\,Sge, Z Cam, ER\,UMa, classical and recurrent novae} based on their observed characteristics, with the sub-types often being named after a prototype CV that exemplifies particular characteristics. In this paper we have used the classifications from the literature where available, making additions and corrections where appropriate. 

We provide here a brief summary of the observational hallmarks of the different CV sub-types, for a more detailed description, see \citet{2023MNRAS.524.4867I}.

Non-(or weakly)-magnetic CVs ($B\lesssim1$\,MG) that exhibit disc outbursts are dwarf novae whilst those with steady hot discs are novalike variables (that sometimes exhibit ``negative superhumps'' in their light curves). Dwarf novae are further sub-classified into SU\,UMa CVs (typically short-period, $\Porb\lesssim3$\,h, that have relatively frequent outbursts, interspersed by longer and brighter superoutbursts during which they exhibit superhumps, photometric variability a few per cent longer than their orbital period \citealt{1988MNRAS.232...35W, 2009PASJ...61S.395K}), ER\,UMa CVs (systems with very short superoutburst recurrence times, \citealt{2015PASJ...67..108K}), WZ\,Sge CVs (which only have rare superoutbursts, \citealt{2015PASJ...67..108K}), U\,Gem CVs (typically long-period, $\Porb\gtrsim3$\,h, dwarf novae that do not show superoutbursts), and Z\,Cam CVs (that switch between outburst states and ``stand-still'' periods of constant brightness, \citealt{2001A&A...369..925B}). CVs with highly magnetic white dwarfs (MCVs) consist of polars (which typically have $B\gtrsim10$\,MG and the white dwarf spin period synchronised with the orbital period, $P_\mathrm{spin}=\Porb$, \citealt{1990SSRv...54..195C, 2025A&A...698A.106S}) and intermediate polars (IPs, with $10\gtrsim B\gtrsim1$\,MG, and $P_\mathrm{spin}<\Porb$, \citealt{1994PASP..106..209P}). Polar CVs whose orbital period is slightly different from the spin period are known as asynchronous polars. Finally, AM\,CVn CVs are ultra-short period ($\Porb\lesssim60$\,min) CVs with hydrogen-deficient donor stars \citep{2010PASP..122.1133S, 2025A&A...700A.107G}. 

\section{Observations}
The DESI galaxy and quasar samples result from observations of four main types of targets \citep{2023AJ....165...50M}: bright galaxies \citep{2023AJ....165..253H} from the bright-time BGS survey and  Luminous Red Galaxies \citep[LRGs;][]{2023AJ....165...58Z}, Emission Line Galaxies \citep[ELGs;][]{2023AJ....165..126R} and Quasars \citep[QSOs;][]{2023ApJ...944..107C} from the dark-time survey. These surveys focus on extragalactic objects and would not be expected to specifically target CVs. DESI also conducts a Milky Way survey \citep[MWS;][]{2023ApJ...947...37C}  that uses bright time to survey stars; mostly to an extinction-corrected limit of $r=19$ in the DESI Legacy Surveys imaging \citep{2019AJ....157..168D}, but extending as faint as $g < 20$, $r < 20$ or {\em Gaia} $G < 20$ for some bespoke samples. The MWS included a target class that is particularly relevant to CVs: $\simeq 70\,000$  (primarily) single  white dwarf candidates \citep[Section 4.4.1 in][]{2023ApJ...947...37C}, selected using similar criteria to \citet{2019MNRAS.482.4570G} in order to achieve a high degree of completeness. Given that CVs are binaries containing a white dwarf, this target class is expected to include a number of CVs with faint donors and low accretion rates. 

In addition, two secondary programs which can make use of spare fibres, ``SPCV'' and ``WD\_BINARIES\_BRIGHT / DARK'' (see Table C1 of \citealt{2026AJ....171..285D}) directly targeted CVs \citep{2023AJ....165...50M, 2024AJ....168...58D, 2025arXiv250314745D}.  The SPCV program targeted short-period CV candidates using a selection based entirely on \textit{Gaia} DR2 \citep{2018A&A...616A...1G} and identifying objects with unusually high photometric variability and the additional cuts on \textit{Gaia} parameters listed in Table \ref{tab:shortperiodsurvey}. Variability in the time-resolved \textit{Gaia} DR2 photometric observations is parameterised using the interquartile range (IQR) and median absolute deviation (MAD, \citealt{2019A&A...625A..97R}). SPCV  used a parallax significance cut (parallax divided by the standard deviation of the parallax) lower than the one provided by the \textit{Gaia} Collaboration because stricter cuts can remove stochastically variable objects like CVs. This yielded $\simeq 2300$ targets. Our own secondary program, WD\_BINARIES\_BRIGHT / DARK (henceforth ``close white dwarf binary'' , CWDB) deliberately targeted CVs and related objects based on the detection of an ultraviolet (UV) excess  and is described in detail in Appendix\,\ref{sec:CWDBC}. 

\begin{table}
\caption{Selection criteria for the short-period CV secondary target program.}\label{tab:shortperiodsurvey}
\begin{tabular}{ll}
\hline
Selection Criteria                 & Cut           \\ \hline
Faint limit                        &  $m_G<20$    \\
Bright limit                       &  $m_G>16$    \\
Parallax significance              &  $\varpi / \sigma_\varpi >3$  \\
Proper motion lower limit          & $\sqrt\mathrm{pmra^2+pmdec^2}>2$  \\
High galactic latitudes            & $\lvert b \rvert>20^\circ$ \\
Interquartile range                & $>0.23$ \\
Median absolute deviation          & $>0.14$ \\ \hline
\end{tabular}
\end{table}

Prior to DESI, one of the authors (JRT) obtained time-resolved spectroscopy using the MDM observatory at Kitt Peak (see \citealt{2020AJ....160....6T} for more details on the instrument and data reduction techniques) of a number of our CVs. We include here the so far unpublished periods resulting from these observations. 

\section{Methodology}

\subsection{DESI data analysed}

\begin{table}
\caption{\label{Table:numbers_of_spectra} Summary of the DESI spectra that were processed. 
}
\begin{tabular}{lrr}
\hline 
 & \multicolumn{2}{c}{Spectra}  \\ \hline
Total spectra from DESI             &                           &  \Nspectra          \\
Non-science spectra &&\Nnonscience\\
 \quad Sky spectra                             &\NspectraobjtypeSKY \\
 \quad Spectra from stuck fibers                &\Nspectraobjtype \\
 \quad Untargeted spectra&\NspectraobjtypeNON\\
 \quad Unusable (''BAD'') spectra&\NspectraobjtypeBAD \\

Science spectra processed by CNN       &                   &    \Nscience  \\
\quad Spectra selected by CNN &\Nmlselected   \\
\hline           
\end{tabular}
\end{table}
\begin{table*}
\caption{\label{Table:numbers_of_spectra2} Summary of the manual inspection performed on subsets of the \Nmlselected\ spectra selected by the CNN.  It should be noted that some objects had multiple DESI exposures, resulting in multiple spectra (Fig.\,\ref{fig:Number_of_spectra}) which need to be deduplicated.  The remaining \Nnotinspected\ spectra were not manually inspected.  *four of these spectra were of CVs identified as part of the main analysis and are therefore duplicates.
}
\begin{tabular}{lrrrrr}

\hline 
 & \multicolumn{3}{c}{Spectra}                   & \multicolumn{2}{c}{Objects} \\
 \cmidrule(lr){2-4} \cmidrule(lr){5-6}
 & Total & Non-CV spectra & CV spectra           & Non-CV         & CV         \\ \hline
Analysis  \\ 
\quad Spectra of known and candidate CVs &         1531 &         253 &         1278 &         19 &         676 \\ 
\quad Spectra with REF\_CAT catalogue value &          29154 &          28062 &          1092 &          9985 &          604 \\ 
\vspace{1mm} \quad Spectra with $Z<0.01$ & 18511 &          17040 &          1471 &          14208 &          861 \\ 
\vspace{1mm} \quad Deduplicated & 41578 & 39775 & 1803 & 6783 & 1023\\ 
Redshift validation of completeness  \\ 
\quad Random set & 9954 & 9954 & 0 & 9954 & 0\\ 
\quad $1.6<Z<1.7$ spike & 3859 & 3851 & 8 & 3851 & 8*\\ \hline

\end{tabular}
\end{table*}

\begin{figure} 
 \centering 
  \includegraphics[width=1\columnwidth]{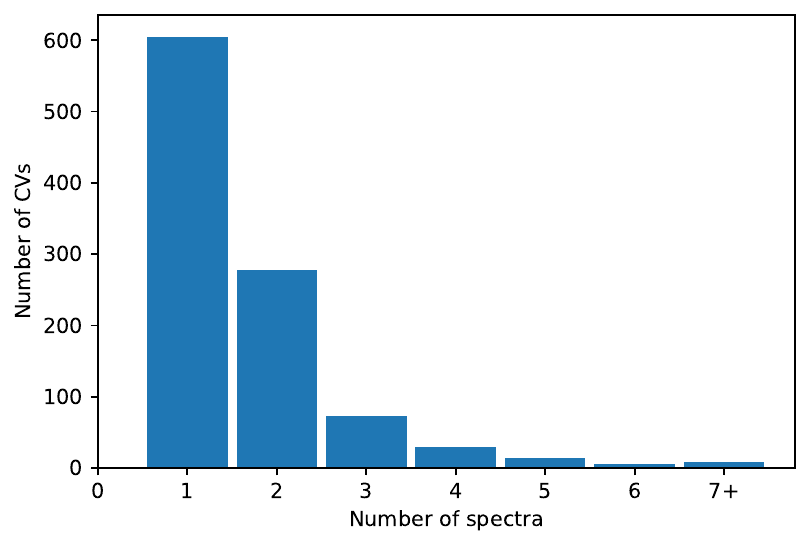} 
\caption{\label{fig:Number_of_spectra}
Histogram of individual DESI DR2 exposures for the \NCVs\ identified cataclysmic variables within this analysis.  Whilst the majority of CVs only have a single spectrum some have been observed multiple times. In general the CNN consistently selects all the spectra of a CV although exceptions can occur typically if an individual spectrum has a poor signal-to-noise ratio (S/N) or if the CV is in outburst. We do not coadd spectra as we wish to retain visibility of short-term changes. In addition coadded spectra could not be used for training the CNN.}
 \end{figure}

We analysed the entire set of \Nspectra\,spectra\footnote{Any reference to a DESI spectrum in this paper refers to the composite spectrum obtained by merging the three arms of the spectrograph in a single exposure.} in DESI DR2 (internally codenamed ``Loa''), spanning the period 2020 December 14 to 2024 April 9. The \Nspectra\ spectra include \NspectraobjtypeTGT\ science exposures (with \texttt{OBJTYPE="TGT"})\footnote{\texttt{OBJTYPE} and other names in this typeface refer to entries in the DESI data model}  with the remaining non-science exposures consisting of \NspectraobjtypeSKY\ ``sky'' spectra   (with \texttt{OBJTYPE="SKY"} although they may occasionally alight on a galaxy or star), \Nspectraobjtype\,spectra from ``stuck fibers'' (with \texttt{OBJTYPE=""} and \texttt{TARGETID}<0) and the remainder failed spectra (Table\,\ref{Table:numbers_of_spectra}).  We only considered the subset of science exposures.

\subsection{Techniques for shortlisting\label{sec:shortlisting}}
Whereas the identification of CVs among the SDSS spectroscopy almost exclusively relied on visual classification the volume of data produced by DESI prohibits human inspection of all the science spectra. We therefore used three techniques to shortlist candidate CV spectra for visual inspection (Table\,\ref{Table:numbers_of_spectra2}). The results of the application of each technique informed the design of subsequent techniques.

The first technique was to cross-match the coordinates of the \NspectraobjtypeTGT\ science spectra with a list of \NknownCVs\ known and candidate CVs that we have accumulated from the literature. We used a deliberately large search radius of three arcsec in the cross-match with the DESI observations to allow for high proper motion CVs and ones with inaccurate coordinates to be found. Multiple spectra that were found within the search radius were all taken forward for visual inspection.  This yielded \Nknownspectra\ spectra.

The second technique used machine learning (ML, see Section\,\ref{sec:mahine_learning} for details) adopting the well-established use of Convolutional Neural Networks (CNNs) and following very closely the approach described in section\,2 of \citet{2025MNRAS.536.1057I}  to filter the normalised \NspectraobjtypeTGT\,spectra. This yielded \Nmlselected\,spectra; still too large a number to manually inspect. We then assumed that many CVs would be close and bright enough to have been identified in all--sky surveys and therefore used the DESI attribute  \texttt{REF\_CAT}\footnote{Spectra with a \textit{Gaia}\,DR2  or, in a very small number of cases, a Siena Galaxy Atlas 2020 \citep{2023ApJS..269....3M} or Tycho--2 counterpart} together with a coordinate cross match with \textit{Gaia}\,DR3 to filter those spectra which had a catalogue entry yielding \Nrefcatspectra\,spectra. 

A third technique was used to identify the remaining CV candidates by analysing the DESI redshift ($Z$) that is calculated for each spectrum\footnote{https:/data.desi.lbl.gov/desi/spectro/redux/loa/zcatalog/v1/zall-tilecumulative-loa.fits}.  We therefore used \texttt{$\lvert Z \lvert<0.01$}  to filter the \Nmlselected\ spectra yielding a further 
\Nmlspectraloz\ candidates.

The \Nknownspectra, \Nrefcatspectra\ and  \Nmlspectraloz\ spectra were then manually inspected\footnote{These samples overlap and were deduplicated prior to manual inspection.}  (Table\,\ref{Table:numbers_of_spectra2}) to identify which ones were CV spectra (Section\,\ref{sec:IdentifyingCVs}). Where possible sub-types (see Section\,\ref{sec:cv_subtypes}),  were allocated to the CV spectra.

\subsection{Machine Learning} \label{sec:mahine_learning}

\subsubsection{Machine learning using the SDSS V model} \label{sec:ml_v1}

\citet{2025MNRAS.536.1057I} used a convolutional neural network (CNN) model to successfully identify CV candidates from SDSS spectra. The model was trained using the SDSS spectra of 623 known CVs, drawn from \citet{2023MNRAS.525.3597I,2023MNRAS.524.4867I} together with a randomly chosen subset of spectra representing a range of samples of non-CVs (galaxies, quasars, stars, white dwarfs and detached white dwarf plus main sequence binaries).  We replicate the approach here as closely as possible identifying, by cross matching coordinates, those SDSS CVs and non-CVs which also have a DESI spectrum.  We then randomly select our new training set from these samples (see Table\,\ref{Table:training}). During training it was discovered that the quality of 73 
of the CV spectra in Table\,\ref{Table:training} rendered them unusable due to the number of invalid data points (NaN) and these spectra were rejected.

\subsubsection{Normalisation of the DESI spectra} \label{sec:normalisation}

All DESI spectra were flux-normalised prior to the CNN analysis. The DESI spectrograph has three arms or cameras sensitive to blue, red and near-infrared light \citep{2024AJ....168..245P}. These ranges overlap and we create a single spectrum by merging each set of three spectra using $5780.1$\,\AA\ and $7570.1$\,\AA\ as transition wavelengths~--~i.e. the three spectra are clipped at the transition wavelengths and then concatenated into one merged spectrum.   These merged spectra are then filtered to remove defective data (where $\texttt{MASK} \neq 0$\  or $\texttt{IVAR}= 0$). If the filtered spectrum has less than 1000 data points or an arm is missing it is rejected. The filtered data points are replaced by interpolated values of flux using the \textsc{python} \textsc{specutils fluxcon} function. The spectra are then smoothed using a five-point \textsc{specutils boxsmooth} filter and finally flux-normalized by dividing by their mean flux.   

\subsubsection{Training the CNN with DESI DR2 data}
\label{sec:evaluationof CNN}
\begin{table*}
\caption{\label{Table:training} The spectroscopic samples used to train and test the CNN (the train/test dataset). The SDSS samples were constructed as detailed in \citet{2025MNRAS.536.1057I}. We cross-matched all the DESI spectra with the SDSS samples to obtain the matched DESI samples. Note that because some SDSS CVs have multiple spectra cross-matching each of them using coordinates will generate duplicate DESI spectra. The final samples are the subsets of the matched DESI samples used for training and testing. These final samples consist of random selections from the matched DESI samples with the exception of the CVs where all deduplicated spectra are retained.   }
\begin{tabular}{lrrr}
\hline
 & \multicolumn{3}{c}{Spectra}                      \\ \cline{2-4} 
                                         & SDSS sample & matched DESI sample &  final sample   \\ \hline

CVs  & 623  & 522 & 304 \\
Galaxies  & 50\,000  & 40\,661 & 1\,246 \\
Quasars  & 16\,713  & 14\,268 & 1\,200 \\
Stars  & 50\,000  & 26\,620 & 828 \\
Detached white dwarf plus main sequence binary  & 1\,602  & 718 & 706 \\
White Dwarfs  & 20\,341  & 15\,052 & 1\,200 \\
Total  & 139\,279  & 97\,841 & 5\,484 \\ \hline

\end{tabular}
\end{table*}

Following \citet{2025MNRAS.536.1057I} the CNN is trained by splitting the train/test dataset (see the final sample in Table\,\ref{Table:training}) into two randomly chosen samples with 80~per cent being used for training and the remaining 20~per cent being tested to evaluate the result. To explore the effect of the random selection of the test sample we repeated the random choice of samples 200 times and trained the CNN each of those times with the chosen training/test samples. We then selected the most reliable model (hereafter the trained CNN) based upon the confusion matrix. Following \citet{2025MNRAS.536.1057I} we analysed the incidence of CV spectra in the train/test database that were not found reliably. The quality of CV spectra that are not found reliably needs to be reviewed and such spectra removed from the train/test database if that improves the effectiveness of the CNN.  \NnotdodgyCVs\ of the \NtrainingCVs\ training CV spectra were found every time whilst  
one was not found in one or more of the 200 experiments. As there was only one we chose not to remove it from the train/test set.

The final confusion matrix (Fig.\,\ref{fig:conf1}) for the trained CNN  shows that 
no false positives were predicted and all true positives were found. In other words, on the test dataset the CNN achieved a perfect result.

\begin{figure}
\begin{tabular}{|ll|cc|}
\hline
\multicolumn{2}{|l|}{\multirow{2}{*}{Final model}}                                                                              & \multicolumn{2}{c|}{\begin{tabular}[c]{@{}c@{}}Predicted spectra \\ out of the test set\end{tabular}}                                                                                      \\ \cline{3-4} 
\multicolumn{2}{|l|}{}                                                                                                          & \multicolumn{1}{l|}{CV}                                                                                & \multicolumn{1}{l|}{Not a CV}                                                     \\ \hline
\multicolumn{1}{|l|}{\multirow{2}{*}{\begin{tabular}[c]{@{}l@{}}Actual spectra \\ out of the test set\end{tabular}}} & CV       & \multicolumn{1}{c|}{\begin{tabular}[c]{@{}c@{}}\CMATP\\ (True positive)\end{tabular}}   & \begin{tabular}[c]{@{}c@{}}\CMAFN \\ (False negative)\end{tabular} \\ \cline{2-4} 
\multicolumn{1}{|l|}{}                                                                                               & Not a CV & \multicolumn{1}{c|}{\begin{tabular}[c]{@{}c@{}}\CMAFP \\ (False positive)\end{tabular}} & \begin{tabular}[c]{@{}c@{}}\CMATN\\ (True negative)\end{tabular}   \\ \hline
\end{tabular}
 
 \caption{\label{fig:conf1} The confusion matrix resulting from training where the final CNN is applied to the \CMATOT\ test spectra (20 per cent) from the train/test dataset.  }
\end{figure}

\subsubsection{Validation}\label{sec:testing}

We validated the performance of the trained CNN using a sample of \NinWDsurvey\ DESI CV spectra (\NchrisCVs\, CVs). This sample was generated as a by-product of Swan et al. (submitted) where the DESI spectra of $\simeq64\,000$ white dwarf candidates were manually inspected and classified based on their spectroscopic features. We removed the \Nchristrainingspectra\ spectra that were present in the train/test dataset.  The trained CNN identified \Nchrisfoundspectra\ including at least one spectrum from all but two of the remaining CVs. The two non-detections are J222900.37+263706.86, whose spectrum got rejected in the normalisation step due to the lack of one arm (Section\,\ref{sec:normalisation}) and J142701.59$-$012310.25 whose individual spectra show very indistinct emission lines and was only spotted manually by Swan et al.  (submitted) in the higher-quality co-added spectrum based on four exposures.

\begin{table*}

\centering
\caption{\label{tab:cvlist}The complete table of CVs from DESI is contained in the supplementary data; an example is shown here. We provide the geometric distances from \citet{2021AJ....161..147B} and CV sub-types according to the definitions in Section\,3 in \citet{2023MNRAS.524.4867I}.   The references for the initial discovery (ID), spectrum (Sp)  and orbital and superhump periods ($P$~--~where known) are also shown.  Sub-type identifications in black are from VSX \citep{2017yCat....102027W}. : Tentative values, * Superhump periods  }

\begin{tabular}[c]{lllllrllll}

\hline
\multirow{2}{*}{DESI} &  \multirow{2}{*}{\begin{tabular}[c]{@{}l@{}}\textit{Gaia} EDR3\\ source\_id\end{tabular}} & \multirow{2}{*}{\begin{tabular}[c]{@{}l@{}}Period \\ (h)\end{tabular}} & \multirow{2}{*}{\begin{tabular}[c]{@{}l@{}}\textit{Gaia} EDR3\\$G$ (mag)\end{tabular}} & \multirow{2}{*}{\begin{tabular}[c]{@{}l@{}}Distance\\ (pc)\end{tabular}} & \multirow{2}{*}{\begin{tabular}[c]{@{}l@{}}Variable\\ Sub-type\end{tabular}}  & \multicolumn{3}{c}{References}                         \\ \cline{7-9} 

 &    &     &       &      &      &          \multicolumn{1}{l}{ID} & \multicolumn{1}{l}{Sp} & P \\ \hline

J000130.45+050623.5 & 2744979298990130944 &  & 20.41 & $1495$ & U Gem & 5 & 11 &  \\
J000322.38+140458.7 & 2768879344590703872 & 0.925(1) & 20.0 & $263$ & \textcolor{blue}{AM CVn} & 14 & 14 & 14 \\
\textcolor{blue}{J000339.34+064303.1} &  &  &  &  & \textcolor{blue}{DN} & 1 & 1 &  \\
J000442.44+091801.7 & 2752953781948080512 & \textcolor{blue}{32.43(1)} & 16.4 & $1354$ & \textcolor{blue}{CV} & 4 & 4 & 1 \\
J000558.72+294103.8 & 2861032167186378624 & \textcolor{blue}{1.7557(3)} & 19.37 & $569$ & \textcolor{blue}{Polar} & 16 & 8 & 1 \\
J000600.15+012129.8 & 2738755406045571968 & 1.52(10) & 19.57 & $352$ & WZ Sge: & 19 & 19 & 19 \\
\textcolor{blue}{J000622.84--002725.8} &  &  &  &  & \textcolor{blue}{DN} & 1 & 1 &  \\
J000659.65+192817.6 & 2798092891796120448 &  & 20.39 & $814$ & \textcolor{blue}{SU UMa:} & 5 & 1 &  \\
J000720.76+200721.6 &  &  &  &  & WZ Sge: & 18 & 2 &  \\
J000910.93+300308.2 & 2861111980562475136 &  & 19.55 & $1096$ & U Gem & 18 & 8 &  \\
J000938.25--121017.0 & 2421428928566171776 & 2.16(2)* & 20.46 & $801$ & SU UMa & 3 & 8 & 9 \\
J000949.31+295144.2 & 2861094289593259136 &  & 18.29 & $525$ & \textcolor{blue}{U Gem} & 15 & 1 &  \\
J001107.26+303235.9 & 2861506395998660096 & 1.5240(1)* & 20.06 & $793$ & SU UMa & 7 & 2 & 9 \\
J001133.73+045122.4 & 2742129944670172928 & 1.3485(2)* & 20.61 & $758$ & WZ Sge & 13 & 6 & 10 \\
J001158.28+315543.7 & 2861949606559861120 & 1.330(2) & 20.79 &  & SU UMa & 18 & 2 & 2 \\
J001204.51+020129.8 & 2546760133008036096 & \textcolor{blue}{3.2816(1)} & 16.45 & $1667$ & NL & 17 & 17 & 1 \\
J001231.54+280011.1 & 2859847683924640384 &  & 20.98 &  & WZ Sge: & 2 & 2 &  \\
J001310.61+212107.3 & 2798895359190442624 &  & 19.15 &  & U Gem & 18 & 1 &  \\
J001538.27+263656.6 & 2856493142666940544 & 2.436(1) & 17.66 & $557$ & SU UMa & 3 & 6 & 12 \\
J001636.89+185614.3 &  &  &  &  & \textcolor{blue}{DN} & 5 & 1 & \\ \hline
\end{tabular}
\\
\raggedright
References:\,1\,\textcolor{blue}{This work}, 2\,\citet{2023MNRAS.524.4867I}, 3\,\citet{2014MNRAS.443.3174B}, 4\,\citet{2007PhDT.......221A}, 5\,\citet{2014MNRAS.441.1186D}, 6\,\citet{2014AJ....148...63S}, 7\,\citet{1998IBVS.4578....1A}, 8\,\citet{2026ApJS..282...26H}, 9\,\citet{2012PASJ...64...21K}, 10\,\citet{2014PASJ...66...90K}, 11\,\citet{2020ApJS..249....3A}, 12\,\citet{2016AJ....152..226T}, 13\,\citet{2013ATel.5339....1H}, 14\,\citet{2022MNRAS.512.5440V}, 15\,\citet{2021AJ....161..242F}, 16\,\textcolor{blue}{MGAB catalog}, 17\,\citet{2020AJ....159...43H}, 18\,\textcolor{blue}{CRTS}, 19\,\citet{2025MNRAS.536.1057I} 
\end{table*}

\noindent
\subsection{Identifying CVs}\label{sec:IdentifyingCVs}
The \Nknownspectra, \Nrefcatspectra\ and  \Nmlspectraloz\ spectra were manually inspected\footnote{These samples overlap and were deduplicated prior to manual inspection.}  (Table\,\ref{Table:numbers_of_spectra2}) to identify which ones corresponded to genuine CVs. Additional available data sources were used in this process, including \textit{Gaia} (where a counterpart exists\footnote{Many DESI targets are too faint to have a \textit{Gaia} counterpart.}), \textit{GALEX} \citep{2017ApJS..230...24B},  ZTF light curves \citep{2014htu..conf...27B} and the Vizier \citep{2024ApJS..271...55K}, Simbad \citep{2000A&AS..143....9W} and VSX \citep{2006SASS...25...47W} databases, to assist the classification. We also (a) establish the position of each object in the Hertzsprung-Russell (HR) diagram \footnote{Where \textit{Gaia} photometry and parallex exist.} to aid classification \citep{2020MNRAS.492L..40A}, (b) compute the S/N of the spectrum from the uncertainties in the data using \textsc{specutils} \citep{nicholas_earl_2023_10016569}, (c) compute the synthetic  \textit{Gaia} $G$-band  magnitude using \textsc{pyhot} \citep{Fouesneau2026} and (d) obtain the Spectral Energy Distribution (hereafter SED) from the Vizier database. 

Where possible sub-types (see Section \ref{sec:cv_subtypes}),  were allocated to the CV spectra. 

\subsection{Completeness}
Machine learning is a very efficient tool in identifying samples of objects of interest among vast data sets, however, it is inherently difficult to determine how complete these samples are, which reduces their value for statistical analyses. To assess the completeness of the DESI CV sample that we produced, we need to investigate two aspects: (1) how many DESI spectra of CVs did the ML not correctly select for inspection and (2) how many CV spectra may be among the spectra flagged by the ML that we \textit{did not} manually inspect.

For the first point, we use the unique value assigned to each of the spectroscopic targets, the \texttt{TARGETID}. We crossmatched the coordinates of the \NdedupeCVs\ CVs from the three techniques with the complete set of \texttt{TARGETID}s and then searched for all spectra among the \Nscience\ science spectra, finding \Nnotfound\ spectra of the \NdedupeCVs\ CVs that were not selected by the CNN. Upon inspection of these spectra, \Ngoodcvspectramissed\ are clearly recognisable as CVs indicating that the overall reliability of the CNN in flagging CV spectra is \Reliabilitypercent\ per cent. The remaining \Nbadcvspectramissed\ spectra did not show any features that would allow them to be classified as CVs, reasons for that failure include poor quality spectra or a technical failure such as missing one or more of the three constituent spectra.

Having confirmed that the CNN reliably flags DESI spectra as CV candidates, we turn to the second issue: the large number of false positives. In Section\,\ref{sec:testing}, we inspected less than $\simeq15$~per cent of the total \Nmlselected\ flagged by the CNN (techniques two and three). We therefore needed to determine whether any CVs selected by the CNN were missed by both the second and third techniques. We selected a random sample of 9954 spectra from the full sample of the \Nmlselected\ that satisfy the following criteria:  $Z>0.01$, no \texttt{REF\_CAT} entry and not being among the \NdedupeCVs\ CVs identified from the three techniques. We checked that the distribution of redshift and apparent magnitude of the random sample reflected that of the full sample. Manual inspection of this sample yielded no CV spectra which confirms that filtering on \texttt{REF\_CAT}  and $Z<0.01$ does not systematically exclude candidate CVs and that the sample is $\simeq 99$ per cent complete.

\begin{figure} 
 \centering 
  \includegraphics[width=1\columnwidth]{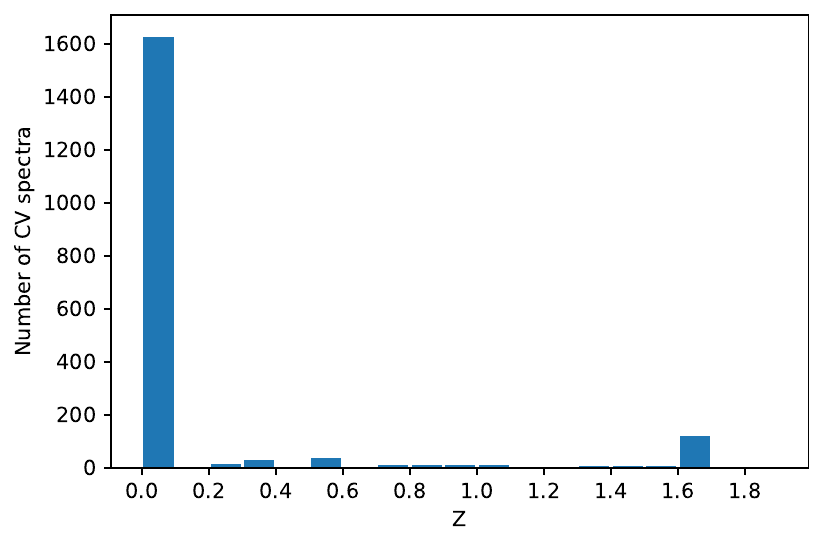} 
\caption{\label{fig:Number_of_spectra_redshifts}
The redshift distribution of the \Ndedupespectra\ DESI CV spectra, as measured by the DESI Redrock pipeline, shows a spike at $1.6<Z<1.7$. This spike is not seen in the full sample of \Nmlselected\ candidate CV spectra selected by the CNN.}
 \end{figure} 

\begin{figure} 
 \centering 
  \includegraphics[width=1\columnwidth]{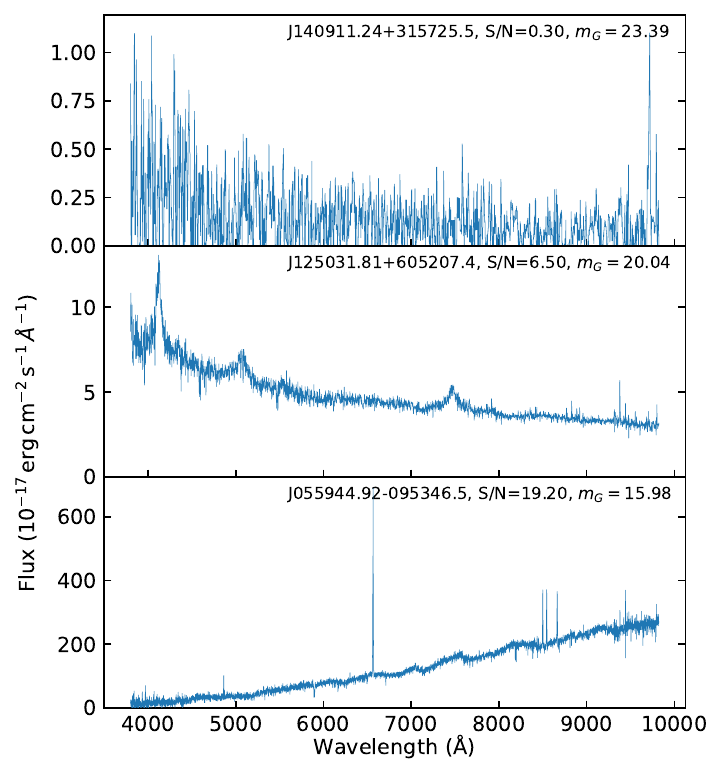} 
\caption{\label{fig:Spike_plots}
Three typical examples of non-CV spectra selected by the CNN with $1.6<Z<1.7$. Top panel: A very common example having a low S/N. The second panel shows a quasar spectrum whilst the bottom panel shows narrow lines with a negligible redshift which could be an active star or a low-redshift galaxy.  }
 \end{figure}

Inspecting the redshift distribution of the DESI spectra of the  CVs that we identified, we noted that $\simeq15$\,per cent of spectra had spuriously large redshifts measured by the DESI pipeline (Fig.\,\ref{fig:Number_of_spectra_redshifts}). In particular we noted a spike between $1.6<Z<1.7$ where the pipeline was confused by the CV spectra. We therefore manually inspected the \Nspike\ spectra selected by the CNN with $1.6<Z<1.7$ resulting in eight CV spectra of which four were additional spectra  of CVs already found by the three techniques and four were of CVs missed by the three techniques. The remainder were typically low S/N spectra with no CV-related features, genuine extragalactic objects, and a small number of objects with narrow Balmer emission lines that may be low-redshift galaxies (Fig.\,\ref{fig:Spike_plots}).  

The manual inspection of the deduplicated CV candidates picked by the three techniques yielded 1014 individual CVs. After adding the two CVs that were not selected by the machine learning program CNN model (see section\,\ref{sec:testing}) and the four CVs identified from the spike the overall number of CVs observed was \NCVs.

\section{Results}
\subsection{Cataclysmic variables found in \texorpdfstring{\DESI}{xx} DR2} \label{subsec:CVtable}

\begin{figure} 
 \centering 
  \includegraphics[width=1\columnwidth]{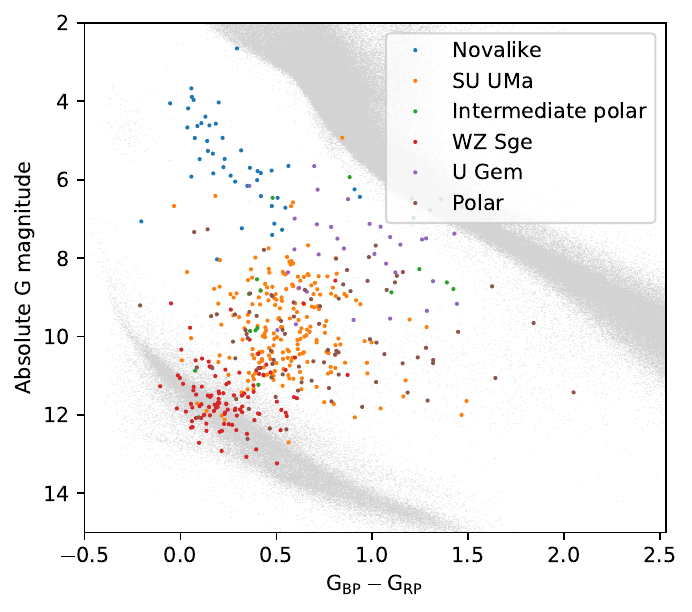} 
\caption{\label{fig:HR_diagram}
HR diagram of the 667 DESI CVs that  have \textit{Gaia} photometry  and parallaxes.   }
 \end{figure}

We found \NCVs\ CVs among the DESI DR1 spectra, of which \NnotnewCVs\ were previously known, leaving \NnewCVs\ new, unpublished systems. The full details on the \NCVs\ DESI CVs are included in the supplementary data, a sample of which is shown in Table\,\ref{tab:cvlist}. The location of CVs with  \textit{Gaia} parallaxes are shown in the HR diagram in Fig.\,\ref{fig:HR_diagram}.

Where possible we have obtained photometric orbital periods of any CVs that do not already have published periods. We used Analysis of Variance techniques \citep{1996ApJ...460L.107S} to compute a periodogram and then inspected the light curves phase-folded on candidate periods to ensure that we identify the orbital period rather than a higher harmonic of it. In most cases, we processed the entire available  light curve data, however, for a small number of CVs we selected specific sections of the light curve, e.g. superoutbursts of SU\,UMa CVs showing evidence for superhumps, or magnetic CVs in a high state. Where we suspect that the light curve is the sum of two or more periods we will remove (``pre-whiten'') the signature of the first period before a further analysis. 

When we detected multiple cyclotron humps in the spectrum of a polar  we estimated the magnetic field strength ($B$) by matching the wavelengths of the humps to the harmonics of the cyclotron resonant frequency given by $f=eB/2\pi m_e$ where $e$ and $m_e$  are the charge and mass of an electron respectively.

\subsection{Individual DESI CVs}
Whilst analysing the \NCVs\ CVs in DESI some systems with unusual or noteworthy characteristics were identified and these are discussed below.

\begin{figure} 
 \centering 
 \includegraphics[width=1.0\columnwidth]{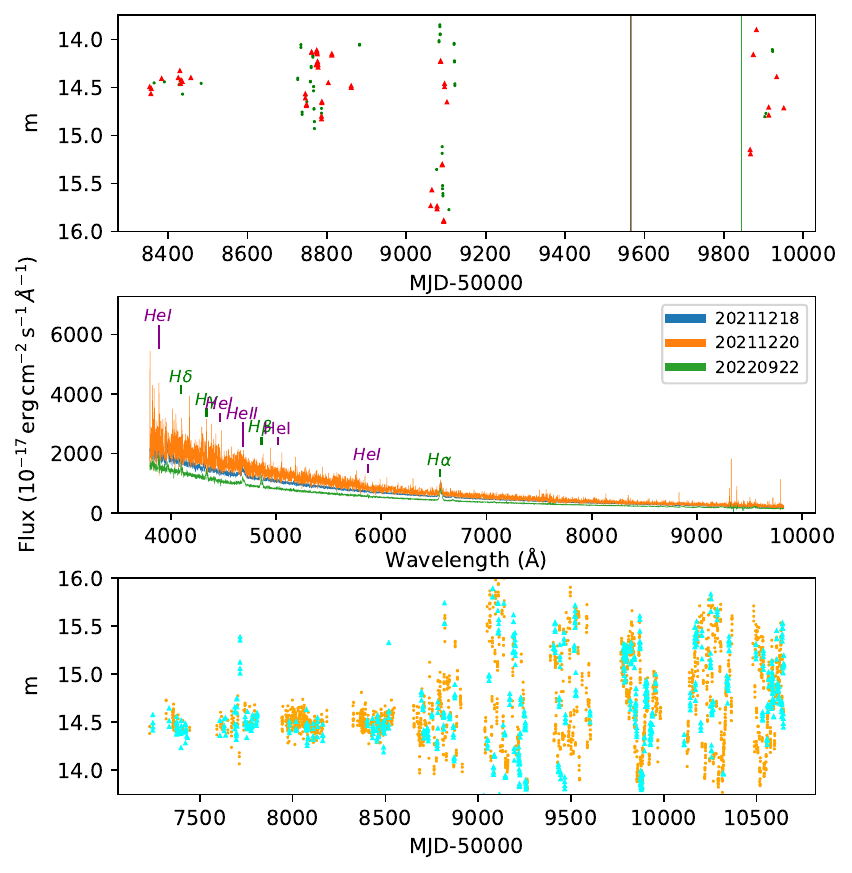} 
\caption{\label{fig:J020639.27-020342.8}
Top panel: The ZTF light curve of the novalike J020639.27$-$020342.8 with the $r$-band filter shown as red triangles and $g$-band data as green dots.  Center panel: Two spectra of J020639.27$-$020342.8 obtained on the same night. Bottom panel: The ATLAS light curve of J020639.27$-$020342.8 with the $o$-band filter shown as orange dots and $c$-band data as cyan triangles. The Z\,Cam standstill is clearly visible.}
 \end{figure}

\begin{figure} 
 \centering 
 \includegraphics[width=1.0\columnwidth]{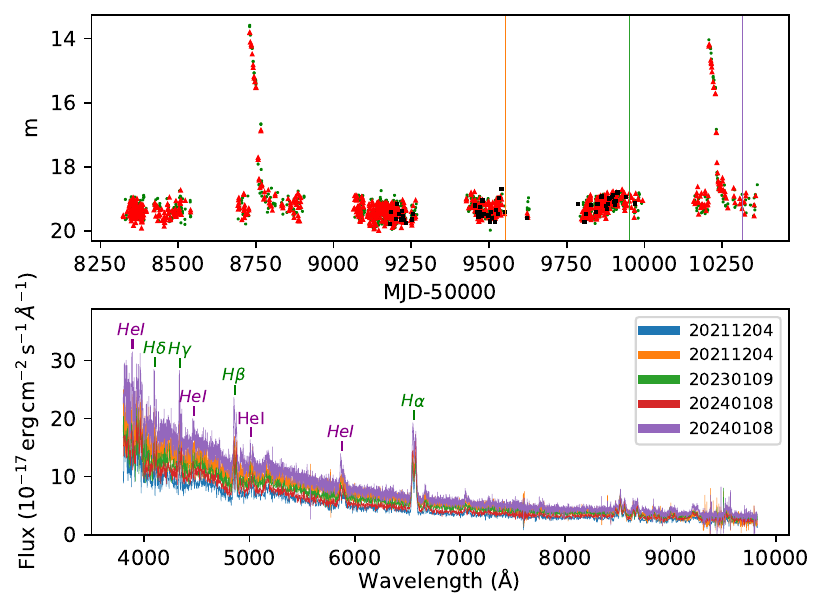} 
\caption{\label{fig:J025612.87-103359.4}
Top panel: The ZTF light curve of the WZ\,Sge J025612.87$-$103359.4 with the $r$-band filter shown as red triangles, $g$-band data as green dots and $i$-band data as black squares. Bottom panel: The spectra of J025612.87$-$103359.4.}
 \end{figure} 
 
\begin{figure} 
 \centering 
 \includegraphics[width=1.0\columnwidth]{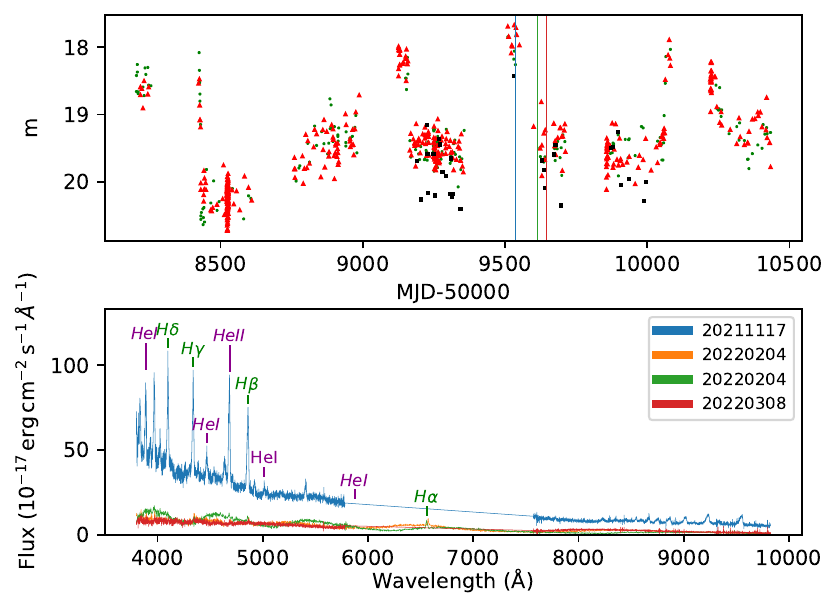} 
\caption{\label{fig:J093322.31-030114.2}
Top panel: The ZTF light curve of the polar  J093322.31$-$030114.2 with the $r$-band filter shown as red triangles, $g$-band data as green dots and $i$-band data as black squares. Bottom panel: The spectra of J093322.31$-$030114.2.}
 \end{figure} 
 
\begin{figure} 
 \centering 
 \includegraphics[width=1.0\columnwidth]{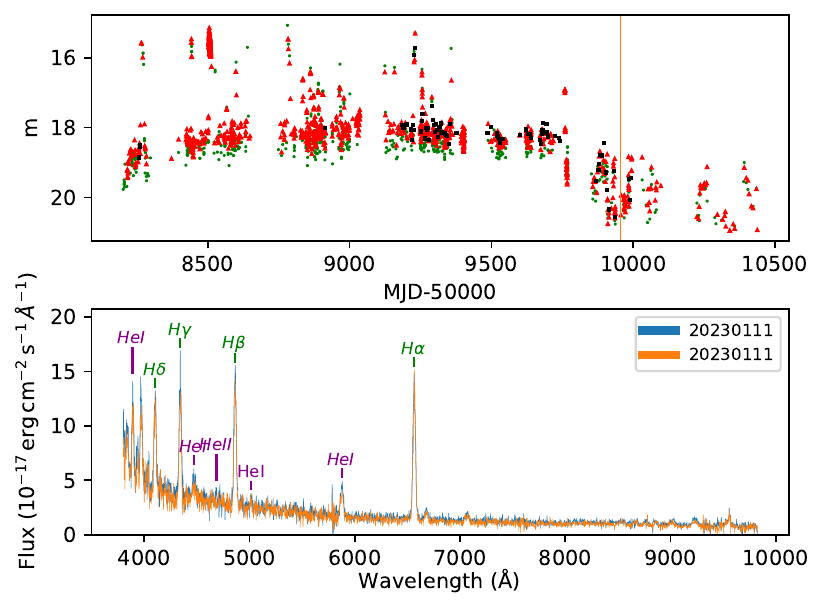} 
\caption{\label{fig:J094325.88+520128.8}
Top panel: The ZTF light curve of the SU\,UMa J094325.88+520128.8 with the $r$-band filter shown as red triangles, $g$-band data as green dots and $i$-band data as black squares. Bottom panel: The spectra of J094325.88+520128.8.}
 \end{figure} 
 
\begin{figure} 
 \centering 
 \includegraphics[width=1.0\columnwidth]{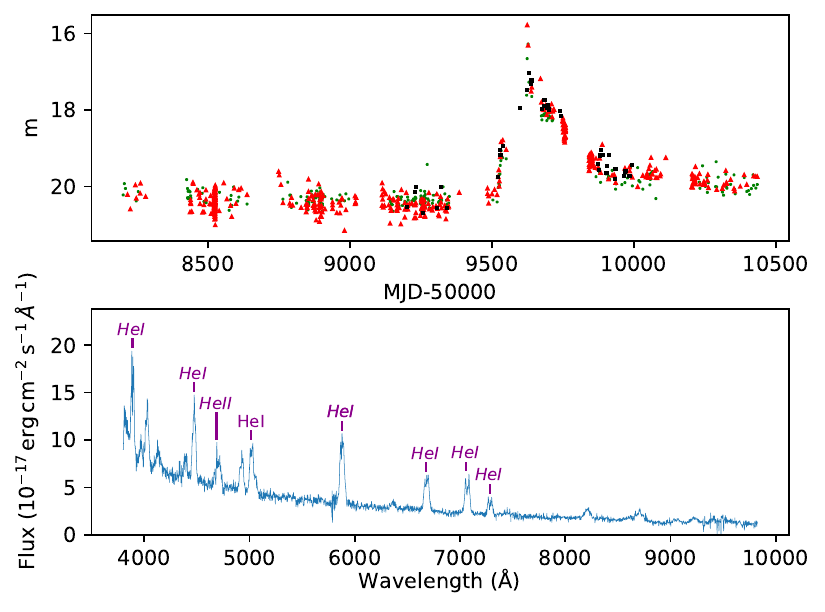} 
\caption{\label{fig:J094708.84+233151.8}
Top panel: The ZTF light curve of the AM\,CVn J094708.84+233151.8 with the $r$-band filter shown as red triangles, $g$-band data as green dots and $i$-band data as black squares. Bottom panel: The spectrum of J094708.84+233151.8.}
 \end{figure} 
 
\begin{figure} 
 \centering 
 \includegraphics[width=1.0\columnwidth]{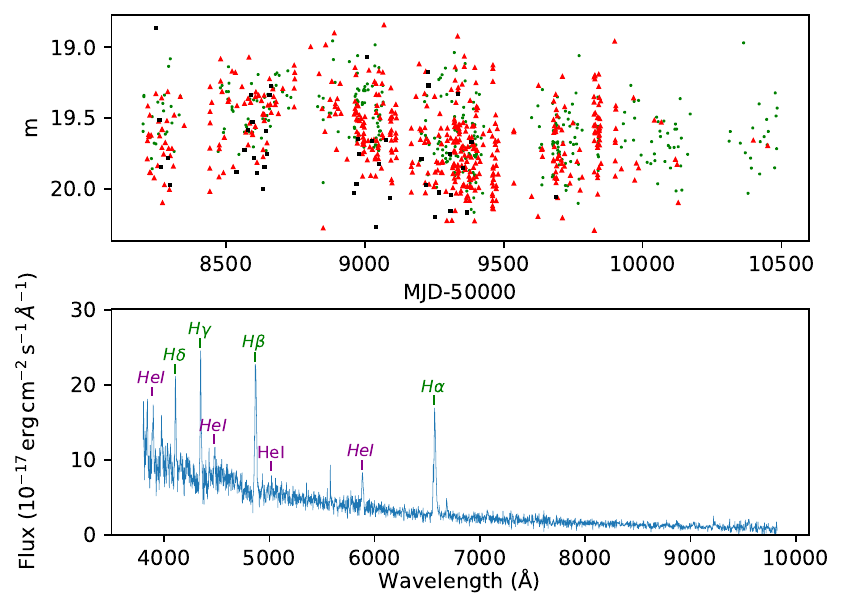} 
\caption{\label{fig:J140508.17+250147.0}
Top panel: The ZTF light curve of J140508.17+250147.0 with the $r$-band filter shown as red triangles, $g$-band data as green dots and $i$-band data as black squares. Bottom panel: The spectrum of J140508.17+250147.0.}
 \end{figure} 
 
\begin{figure} 
 \centering 
 \includegraphics[width=1.0\columnwidth]{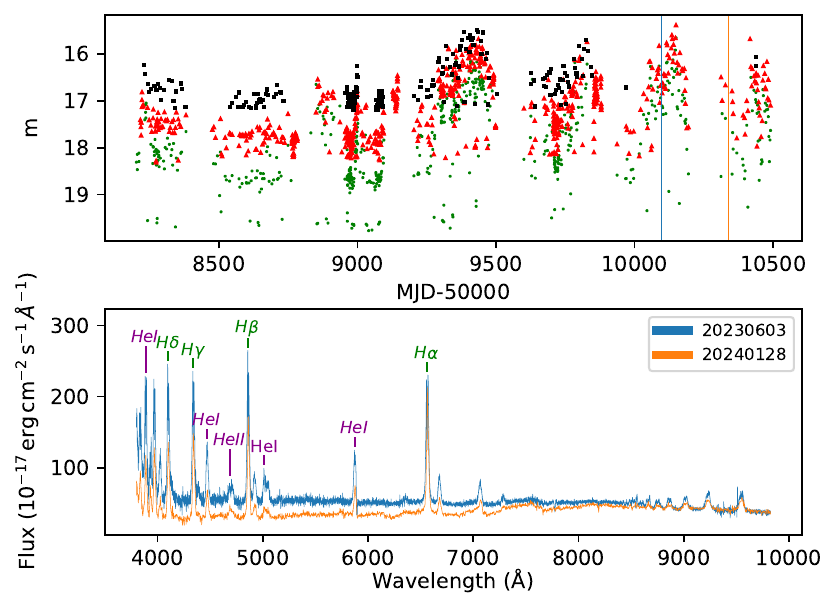} 
\caption{\label{fig:J154453.62+255348.9}
Top panel: The ZTF light curve of the intermediate polar J154453.62+255348.9 with the $r$-band filter shown as red triangles, $g$-band data as green dots and $i$-band data as black squares. Bottom panel: The spectra of J154453.62+255348.9.}
 \end{figure} 
 
\begin{figure} 
 \centering 
 \includegraphics[width=1.0\columnwidth]{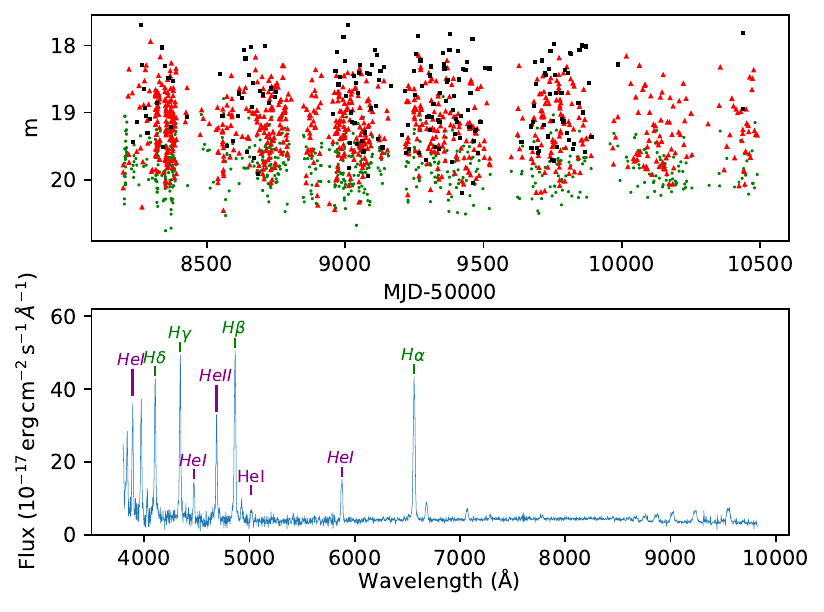} 
\caption{\label{fig:J163125.82+735411.6}
Top panel: The ZTF light curve of the novalike J163125.82+735411.6 with the $r$-band filter shown as red triangles, $g$-band data as green dots and $i$-band data as black squares. Bottom panel: The spectrum of J163125.82+735411.6.}
 \end{figure} 

\begin{figure} 
 \centering 
 \includegraphics[width=1.0\columnwidth]{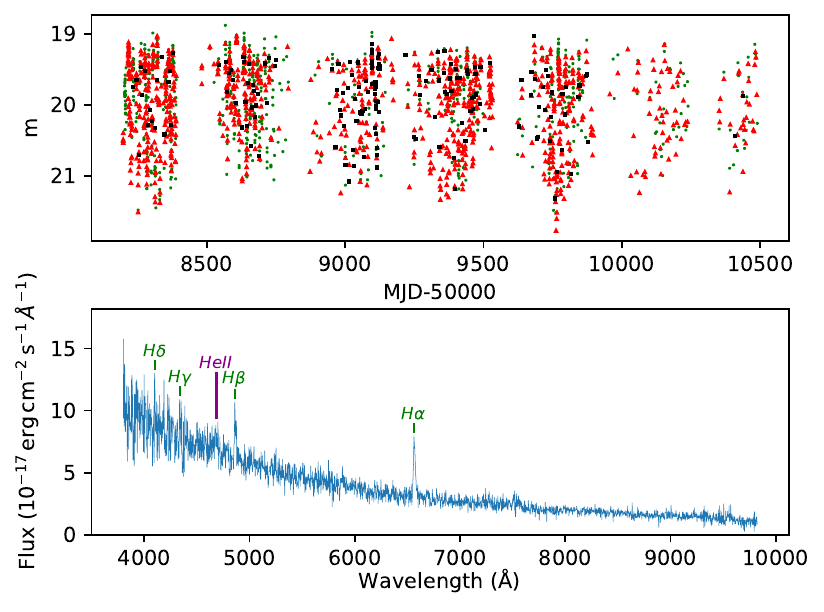} 
\caption{\label{fig:J170550.16+254611.8}
Top panel: The ZTF light curve of the dwarf nova J170550.16+254611.8 with the $r$-band filter shown as red triangles, $g$-band data as green dots and $i$-band data as black squares. Bottom panel: The spectrum of J170550.16+254611.8.}
 \end{figure} 

\begin{figure} 
 \centering 
 \includegraphics[width=1.0\columnwidth]{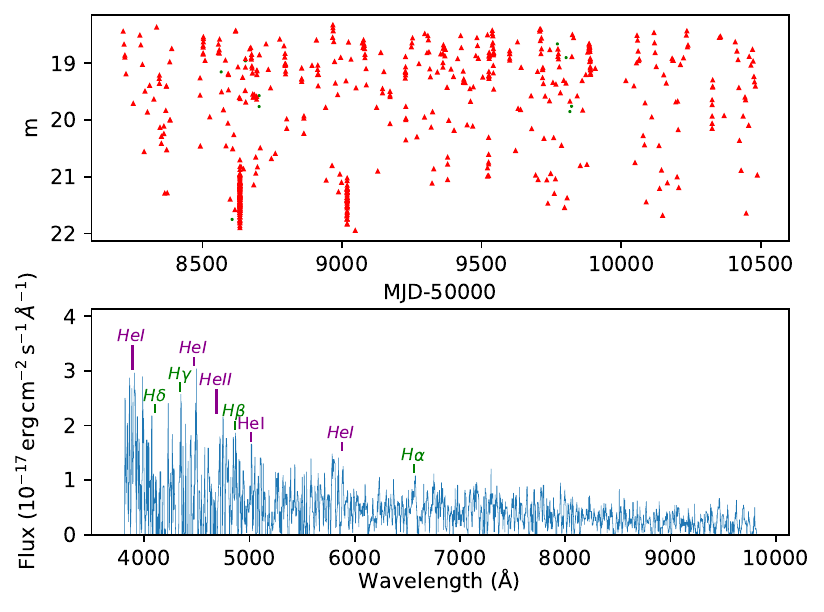} 
\caption{\label{fig:J173104.69+124428.3}
Top panel: The ZTF light curve of the dwarf nova J173104.69+124428.3 with the $r$-band filter shown as red triangles, $g$-band data as green dots and $i$-band data as black squares. Bottom panel: The spectrum of J173104.69+124428.3.}
 \end{figure} 
 
\begin{figure} 
 \centering 
 \includegraphics[width=1.0\columnwidth]{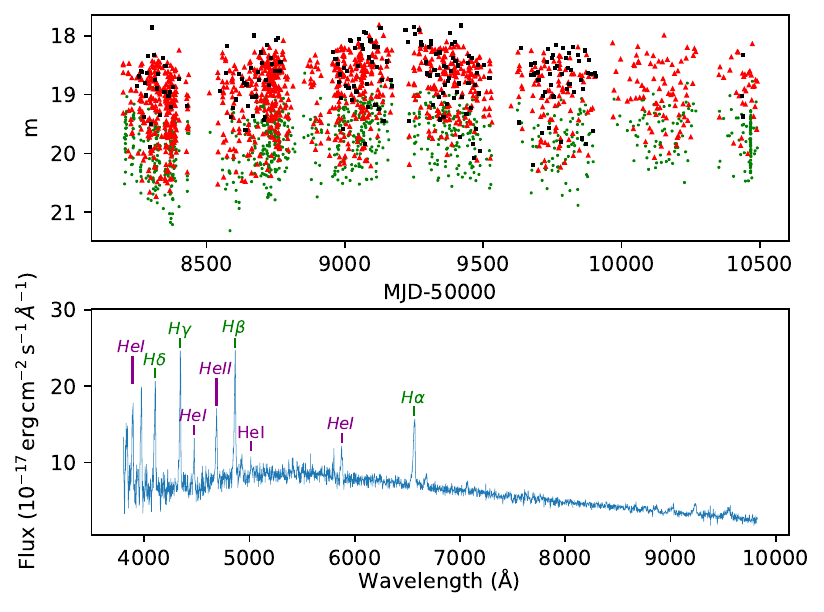} 
\caption{\label{fig:J173118.62+680524.6}
Top panel: The ZTF light curve of the polar J173118.62+680524.6 with the $r$-band filter shown as red triangles, $g$-band data as green dots and $i$-band data as black squares. Bottom panel: The spectrum of J173118.62+680524.6.}
 \end{figure} 

\begin{figure} 
 \centering 
 \includegraphics[width=1.0\columnwidth]{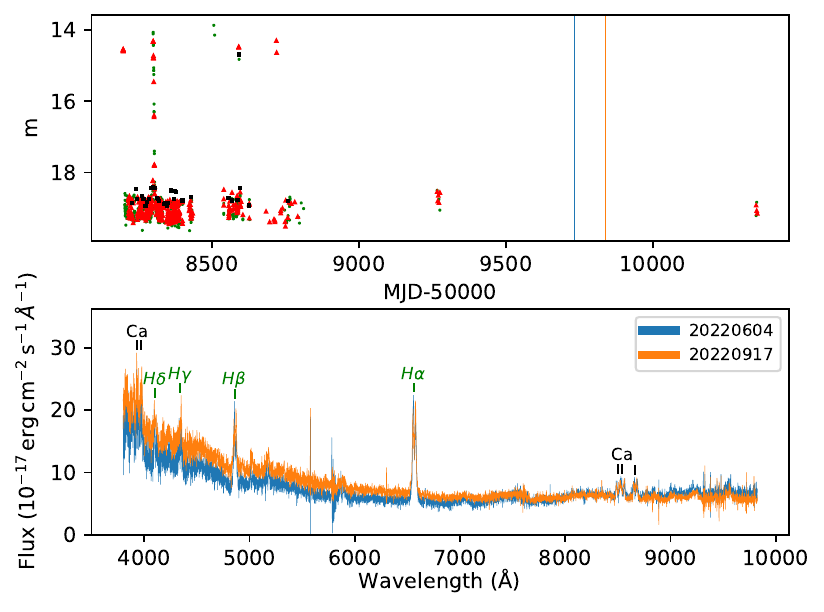} 
\caption{\label{fig:J180546.34+314017.7}
Top panel: The ZTF light curve of the polar J180546.34+314017.7 with the $r$-band filter shown as red triangles, $g$-band data as green dots and $i$-band data as black squares. Bottom panel: The spectra of J180546.34+314017.7.}
 \end{figure} 
 
\begin{figure} 
 \centering 
 \includegraphics[width=1.0\columnwidth]{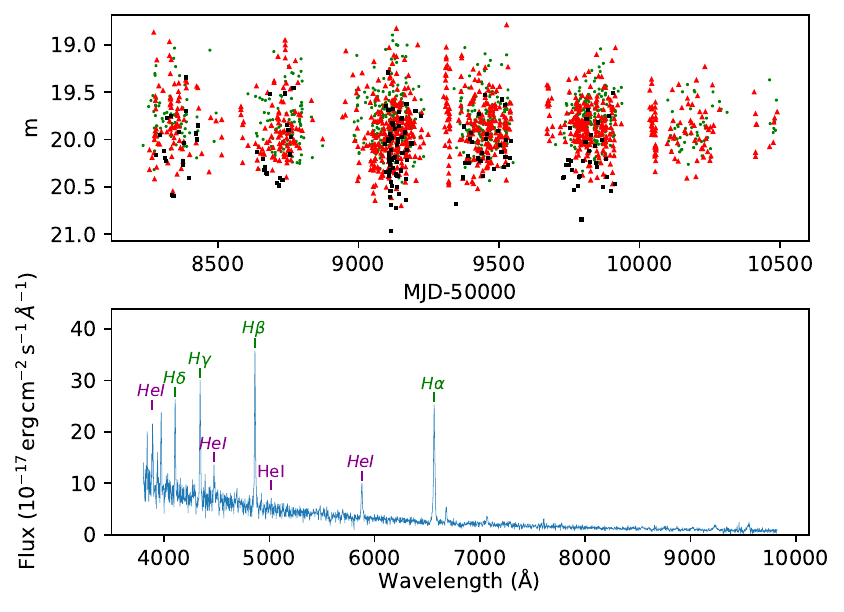} 
\caption{\label{fig:J223501.45+264507.3}
Top panel: The ZTF light curve of the intermediate polar J223501.45+264507.3 with the $r$-band filter shown as red triangles, $g$-band data as green dots and $i$-band data as black squares. Bottom panel: The spectrum of J223501.45+264507.3.}
 \end{figure}

\subsubsection{J020639.27--020342.8}

There are three DESI spectra (Fig.\,\ref{fig:J020639.27-020342.8}). The first and last (2021 December 18 and 2022 August 22) show double-peaked Balmer and \ion{He}{II} emission lines above a blue continuum and with a synthetic magnitude of $m=14.4$ consistent with the quiescent \textit{Gaia} $G$-band magnitude. The third spectrum (2021 December 20) shows a stronger continuum and the emission lines have virtually disappeared; the synthetic magnitude is $m=13.9$. The object is in the novalike area of the HR diagram. The ATLAS light curve shows clear Z\,Cam behaviour with a three-year standstill consistent with \citet{2001PASP..113..764D}.

\subsubsection{J025612.87--103359.4}

The DESI spectra (Fig.\,\ref{fig:J025612.87-103359.4}) show double-peaked Balmer and \ion{He}{I} emission lines and broad absorption lines from the white dwarf. The object is located closer to the white dwarf cooling sequence than the main sequence. The ZTF light curve shows two superoutbursts ($\Delta m \simeq6$)  and no normal outbursts. We obtained a period of $1.59384(3)$\,h from the ZTF light curve. \citet{2011ApJS..194...28K} predict a donor type of M6.0 for this period which is not seen in the observed spectroscopy indicating that this is probably a period bouncer.

\subsubsection{J093322.31--030114.2}

There are four DESI spectra (Fig.\,\ref{fig:J093322.31-030114.2}). One shows Balmer, \ion{He}{I} and \ion{He}{II} emission lines where the  \ion{He}{II} line is stronger than H$\beta$. The other three spectra are dominated by cyclotron humps consistent with a polar with a magnetic field of $B\simeq39$\,MG. The object is nearer the white dwarf cooling sequence than the main sequence in the HR diagram. The ZTF light curve shows strong short term variability ($\Delta m \simeq1$) due to cyclotron beaming. The ZTF light curve also shows longer term state changes of $\Delta m\simeq1.5$ and also five short-lived high states of $\Delta m \simeq 2$ reminiscent of AR\,UMa \citep{2024ApJ...965...96M}. We obtained a period of $2.4525(2)$\,h from the ZTF light curve.

\subsubsection{J094325.88+520128.8}

This is a previously-studied SU\,UMa \citep{2004AJ....128.1882S}. We obtained a superhump period of $1.746(1)$\,h from the ZTF light curve (Fig.\,\ref{fig:J094325.88+520128.8}). The ZTF light curve shows an unusual dimming of $\Delta m \simeq 2$ over two years implying a large reduction in the average accretion rate.

\subsubsection{J094708.84+233151.8}

The DESI spectrum (Fig.\,\ref{fig:J094708.84+233151.8}) shows multiple-peaked \ion{He}{I} and \ion{He}{II} emission lines and no hydrogen lines, identifying this system as an  AM\,CVn. It is on the white dwarf cooling sequence. The ZTF light curve shows a peculiar prolonged outburst of $\Delta m \simeq 5$ with flux reverting to one magnitude brighter than the preceding level. This object was identified as a CV by \textit{Gaia} Alerts (Gaia21fgn) due to the unusual transient. Similar long-lasting brightenings have been detected in a few other AM\,CVn systems (e.g. \citealt{2020ApJ...900L..37R}), sometimes super-imposed by short outbursts (e.g. \citealt{2025MNRAS.537.3078A}).

\subsubsection{J140508.17+250147.0}

The DESI spectrum (Fig.\,\ref{fig:J140508.17+250147.0}) shows asymmetric Balmer and \ion{He}{I} emission lines above the signature of the white dwarf. The object is next to the white dwarf cooling sequence. The SED shows a blue, hot, object consistent with the DESI spectrum.  The ZTF light curve shows variability but no outbursts. We obtained a period of $4.1902(1)$\,h from the ZTF light curve which is inconsistent with the other information and may not be the orbital period. Further spectroscopic radial velocity observations may yield a shorter period reminiscent of V455\,And \citep{2005A&A...430..629A}, FS\,Aur \citep{2003PASP..115..725T} and GW\,Lib \citep{2021MNRAS.502..581C}.

\subsubsection{J154453.62+255348.9}

One of the DESI spectra (Fig.\,\ref{fig:J154453.62+255348.9}) shows asymmetric Balmer, \ion{He}{I} and \ion{He}{II} emission lines together with the signature of the donor. It is close to the main sequence in the HR diagram. The ZTF light curve shows state changes but no outbursts.  The \textit{TESS} light curve not only confirms the orbital period from \citet{2017RNAAS...1...29T} but also reveals a second period of either $14.15$\,min or $13.84$\,min. The second period is likely to be the spin period of the white dwarf indicating that this is an intermediate polar.

\subsubsection{J163125.82+735411.6}

The DESI spectra (Fig.\,\ref{fig:J163125.82+735411.6}) show asymmetric Balmer, \ion{He}{I} and \ion{He}{II} emission lines. The \ion{He}{II} line is over half the strength of the H$\beta$ line suggesting that this might be a magnetic CV \citep{1992PhDT.......119S}. The object is nearer the main sequence than the white dwarf cooling sequence in the HR diagram. The \textit{TESS} light curve shows a period at $34.2(5)$\,h, a very strong period at $2.925(3)$\,h and a weak period at $2.694(3)$\,h. The difference between the latter two periods is too large for the object to be an asynchronous polar. We found a period of $5.85018(2)$\,h (or possibly half that) from the  ZTF $r$-band light curve. The amplitude and shape of the phase-folded light curve, using the $5.85$\,h period is suggestive of cyclotron beaming. This is unlikely to be a novalike as they rarely show clear orbital modulation. We conclude that this system is most likely a magnetic CV, and the detection of two distinct periods, 34\,h and 5.85\,h, in two independent data sets (TESS, ZTF) suggests that it might be an extreme example of an IP with both long orbital and spin periods, and  $P_\mathrm{spin} /P_\mathrm{orb}  \simeq 0.2$.

\subsubsection{J170550.16+254611.8}

The DESI spectrum (Fig.\,\ref{fig:J170550.16+254611.8}) shows double-peaked Balmer and \ion{He}{II} emission lines above a blue continuum. The \ion{He}{II} line is over half the strength of the H$\beta$ line implying that this is a magnetic CV.   It is closer to the white dwarf cooling sequence than the main sequence in the HR diagram. The ZTF light curve shows variability of $\Delta m \simeq 2$ and what appear to be frequent outbursts and which may be ER\,UMa supercycles. We obtained a period of $28.74(2)$\,d from the ZTF light curve which is likely to be the supercycle period.

\subsubsection{J173104.69+124428.3}

The DESI spectrum (Fig.\,\ref{fig:J173104.69+124428.3}) shows double-peaked Balmer, \ion{He}{I} and \ion{He}{II}  emission lines. The object is bright and closer to the main sequence than the white dwarf cooling sequence. The ZTF $r$-band light curve shows frequent large-amplitude variability of $\Delta m \simeq 4$. We cannot recover a period from the light curve and therefore interpret the variability as being due to outbursts.

\subsubsection{J173118.62+680524.6}

The DESI spectrum (Fig.\,\ref{fig:J173118.62+680524.6}) shows asymmetric Balmer, \ion{He}{I} and \ion{He}{II} emission lines above a curved feature that is likely to be due to cyclotron radiation. The object is close to the main sequence in the HR diagram however this is misleading as the cyclotron radiation dominates that of the donor. The ratio of the equivalent width of the \ion{He}{II} line to the H$\beta$ line is $0.87$ implying that this is a magnetic CV. This system is probably the optical counterpart to the X-ray source XMMSL2\,J173118.4+680524 \citep{2008A&A...480..611S} whose coordinates coincide within $1.4$\,arcsec. The ZTF light curve shows variability of $\Delta m \simeq 2.5$, which we believe to be caused by cyclotron beaming, and no outbursts. We obtained a period of $1.54633(3)$\,h from the ZTF light curve. The spectrum is reminiscent of BL\,Hyi \citep{1995A&A...301..447S} and may have a similarly small magnetic field with $B\simeq10-20$\,MG.

\subsubsection{J180546.34+314017.7}

The DESI spectrum (Fig.\,\ref{fig:J180546.34+314017.7}) shows double-peaked Balmer emission lines and also {\ion{Ca}{} emission lines. The signatures of both the white dwarf and the donor are visible. The object is located midway between the main sequence and the white dwarf cooling sequence.  The ZTF light curve shows five outbursts, which may be superoutbursts,  all with $\Delta m \simeq 5$. The ATLAS light curve shows 14 outbursts also with $\Delta m \simeq 5$;  several of these are definitely superoutbursts whilst the remainder are  indeterminate. This object resembles RZ\,Leo \citep{2001A&A...372..563M} which also has infrequent WZ\,Sge-type outbursts.

\subsubsection{J223501.45+264507.3}

The DESI spectrum (Fig.\,\ref{fig:J223501.45+264507.3}) shows narrow Balmer and \ion{He}{I} emission lines. The object is close to the white dwarf cooling sequence in the HR diagram. The ZTF light curve shows variability but no outbursts. We obtained a period of $4.72341(1)$\,h  ($f_{1}=5.08109\,\mathrm{d^{-1}}$) from the ZTF photometry and the folded light curve shows two unequal humps which we take as the orbital period. Prewhitening with this period and the first six harmonics reveals three other signals, $f_{2}=15.57619\,\mathrm{d^{-1}}$,  $f_{3}=20.65732\,\mathrm{d^{-1}}$ and $f_{4}= 25.73840\,\mathrm{d^{-1}}$. Following \citet{1986MNRAS.219..347W} if we take $\Omega=f_{1}$ as the orbital period and $\omega=f_{4}$ as the white dwarf spin period then $f_{3}=\omega-\Omega$ and $f_2=\omega-2\Omega$.   This implies that this is an intermediate polar with an unusually long spin period of $55.9$\,min comparable with V1062\,Tau  with a spin period of $63$\,min \citep{2010PASP..122.1285T}.

\subsection{Unusual state changes}\label{sec:Unusual_state_changes}

Section 6.6 in \citet{2023MNRAS.524.4867I} describes eight CVs identified in SDSS that exhibited peculiar state changes. These CVs display erratic light curve morphologies which are neither disc outbursts nor clearly identifiable state changes. Furthermore these spectra are inconsistent with being either magnetic CVs or novalike variables. During the manual inspections for this work we identified a further \Npeculiar\ CVs showing similar erratic light curve morphologies (Fig.\ref{fig:Unusual_state_changes}).  Noting the eight out of 507 CVs from \citet{2023MNRAS.524.4867I} and the \Npeculiar\ out of \NCVs\,from this work that are of this type suggests that $\simeq1$\,per cent of CVs fall into this class.  IR\,Com is a well-studied \citep{2014MNRAS.442L..23M} example showing a period when accretion effectively stopped and can be regarded as a prototype for this class of CVs. 

\begin{figure*} 
 \centering 
  \includegraphics[width=2.1\columnwidth]{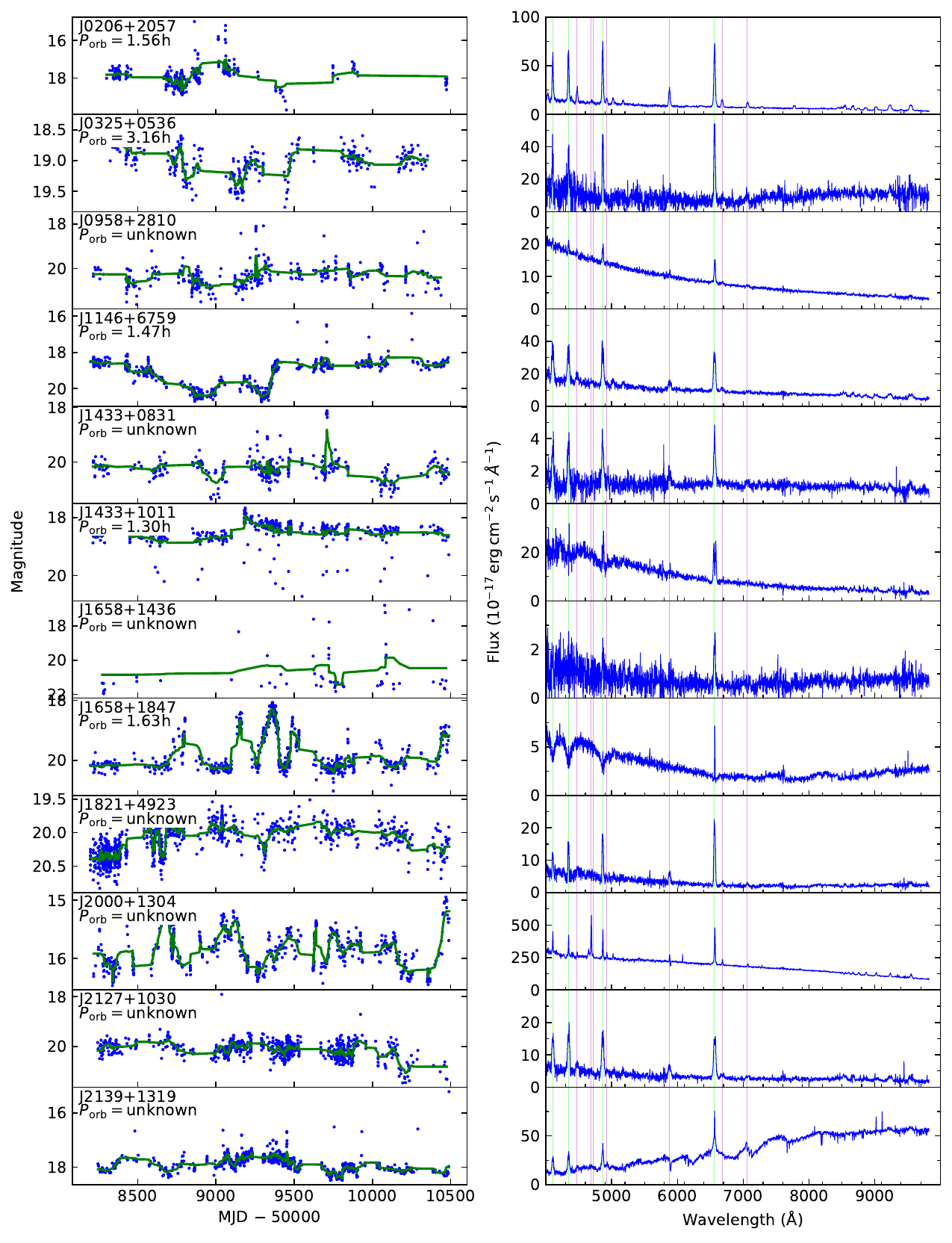} 
\caption{\label{fig:Unusual_state_changes}
ZTF light curves and spectra of CVs which show peculiar changes in flux. We merged the data points from the three ZTF filters by finding the mean magnitude for each filter and then adjusting individual values accordingly. The vertical green and magenta lines mark the hydrogen and helium transitions, respectively.}
 \end{figure*}

The origin of this peculiar behaviour is not clear. Our sample consists of widely varying orbital periods and positions on the HR diagram (Fig.\,\ref{fig:Unusual_state_changes_HRand periods}) illustrating that the peculiar behaviour is not associated with any particular stage of CV evolution.  The cause of the high/low state transitions is still debated, but a likely contender is star spots near the inner Lagrangian point \citep{1994ApJ...427..956L,1998ApJ...499..348K,2000A&A...361..952H}. 

\begin{figure} 
 \centering 
   \includegraphics[width=1.0\columnwidth]{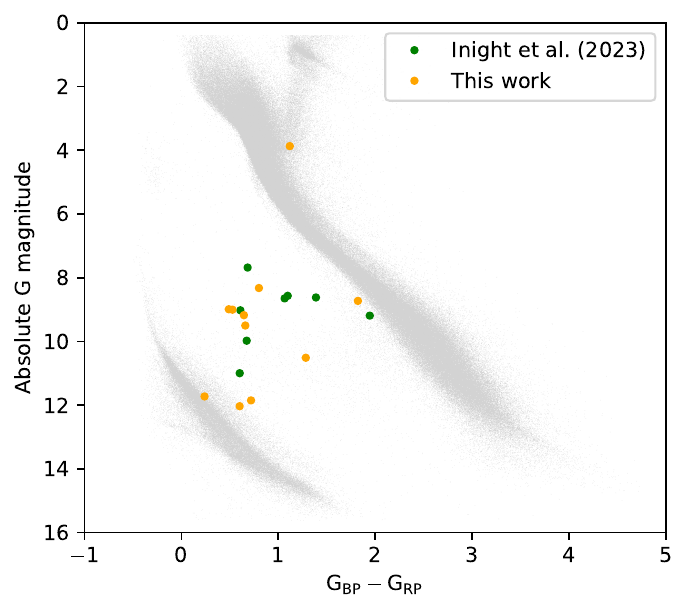} 
 \includegraphics[width=1.0\columnwidth]{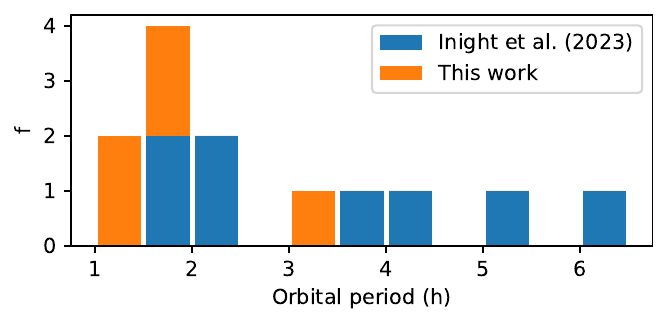} 
\caption{\label{fig:Unusual_state_changes_HRand periods}
CVs with peculiar state changes from \citet{2023MNRAS.524.4867I} (8,\,orange) and from this work (\Npeculiar, blue). Top panel: HR diagram computed from the  \textit{Gaia} photometry and Bailer-Jones distances \citep{2021yCat.1352....0B}.  J165803.75+143634.2 is not shown as it has no \textit{Gaia} counterpart.
Bottom panel: Histogram of the orbital periods where known.}
 \end{figure}

\subsection{CVs that defy classification}
The method of classification is largely settled with only minor subclassifications being debated (see for example \citealt{2023MNRAS.524.4867I}). However there are a small number of CVs that are not easily placed into one of the defined categories. We found the following in DESI.

\subsubsection{J084108.10+102536.3}

The DESI spectrum shows Balmer and \ion{He}{II} emission lines. It is midway between the main sequence and the white dwarf sequence in the HR diagram. The ZTF light curve  (Fig.\,\ref{fig:J084108.10+102536.3}) shows four outbursts. The ATLAS light curve shows at least one superoutburst with superhumps. The classification is ambiguous; the presence of superhumps indicates an SU\,UMa but the presence of the donor spectrum and the position in the HR diagram suggests an earlier CV.

\subsubsection{J094606.92--005601.8}

The DESI spectra (Fig.\,\ref{fig:J094606.92-005601.8}) show double-peaked Balmer and \ion{He}{I} emission lines. Signatures of both a white dwarf and the donor are also visible. The object is nearer the main sequence than the white dwarf cooling sequence. The ZTF light curve shows a single superoutburst of $\Delta m \simeq 6$ and no other outbursts. We have tentatively classified this as an SU \,UMa by virtue of its superoutburst which fulfils the definition of an SU\,UMa. However this light curve is normally associated with an old, low accretion rate WZ\,Sge. The spectroscopy suggests that, far from being an old WZ\,Sge this is a much younger object with a significantly sized donor. This object is reminiscent of RZ\,Leo \citep{2001PASJ...53..905I} whose spectra shows both the white dwarf and donor, and which also has rare outbursts.

\subsubsection{J100243.11--024636.0}

The DESI spectrum shows double-peaked Balmer emission lines together with the signature of a white dwarf. There is probably also the signature of the donor which would be consistent with the SED.   The system has no \textit{Gaia} counterpart. The ZTF light curve (Fig.\,\ref{fig:J100243.11-024636.0}) shows a  superoutburst of $\Delta m \simeq 5$ at $\mathrm{MJD}=58423$ together with four further outbursts resulting in a classification of SU\,UMa. This object resembles J115215.79+491441.7 (see below) with a light curve associated with an older CV contrasting with the spectrum of an earlier one. This object is also reminiscent of RZ\,Leo \citep{2001PASJ...53..905I}.  

\subsubsection{J115215.79+491441.7}

This is BC\,UMa. The DESI spectrum shows double-peaked Balmer emission lines together with the signatures of the white dwarf and donor (Fig.\,\ref{fig:J115215.79+491441.7}).  It is closer to the white dwarf cooling sequence than the main sequence. The ZTF light curve shows a superoutburst together with a second probable superoutburst but no ordinary outbursts.   \citet{2003PASP..115.1308P} confirms that this object contains an M-dwarf donor. Again we have an ambiguity as the light curve and period suggest an old low accretion rate system whilst the spectrum is that of a much earlier system. Again this object is reminiscent of RZ\,Leo \citep{2001PASJ...53..905I}.

\subsubsection{J150441.75+084752.4}

This is a previously known CV with a period of 1.941\,h \citep{2015AJ....149..128T}. The DESI spectrum shows double-peaked Balmer, \ion{Ca}{} and \ion{He}{I} emission lines together with the signatures of the white dwarf and the donor. It is midway between the main sequence and the white dwarf cooling sequence.  The ZTF light curve  (Fig.\,\ref{fig:J150441.75+084752.4}) shows no outbursts, however the ATLAS light curve  shows an outburst of $\Delta m \simeq 2.5$ on $\mathrm{MJD}=58277$. On the strength of a single outburst this object can now be regarded as a dwarf nova but it is ambiguous as it combines a very low level of accretion whilst still being relatively early with a relatively high mass donor having just dropped below the period gap.

\begin{figure} 
 \centering 
 \includegraphics[width=1.0\columnwidth]{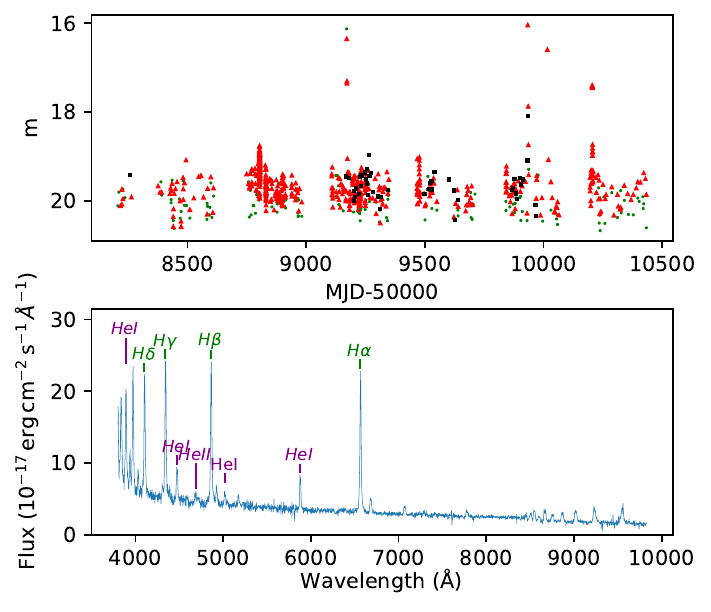} 
\caption{\label{fig:J084108.10+102536.3}
Top panel: The ZTF light curve of J084108.10+102536.3 with the $r$-band filter shown as red triangles, $g$-band data as green dots and $i$-band data as black squares. Bottom panel: The spectrum of J084108.10+102536.3.}
 \end{figure} 
 
\begin{figure} 
 \centering 
 \includegraphics[width=1.0\columnwidth]{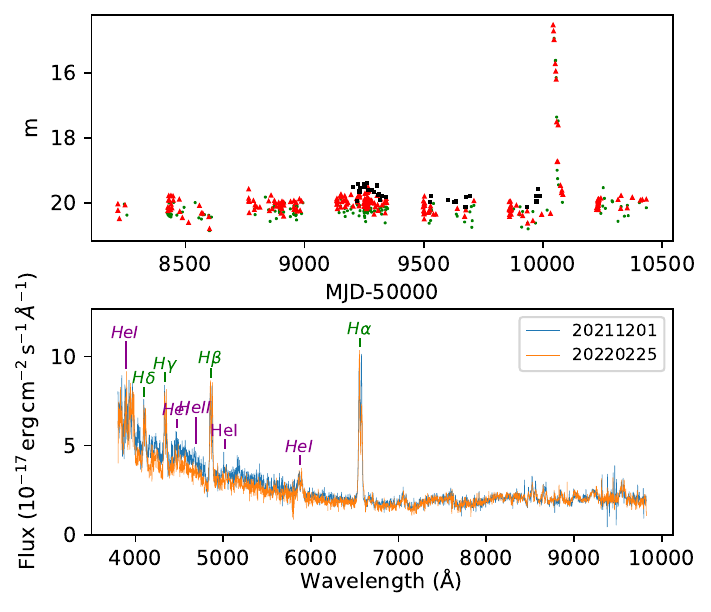} 
\caption{\label{fig:J094606.92-005601.8}
Top panel: The ZTF light curve of J094606.92$-$005601.8 with the $r$-band filter shown as red triangles, $g$-band data as green dots and $i$-band data as black squares. Bottom panel: The spectra of J094606.92$-$005601.8.}
 \end{figure} 
 
\begin{figure} 
 \centering 
 \includegraphics[width=1.0\columnwidth]{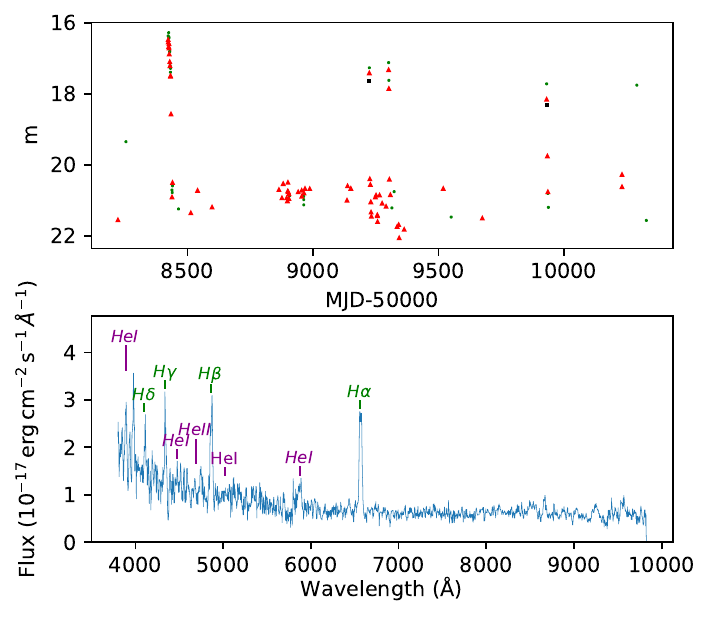} 
\caption{\label{fig:J100243.11-024636.0}
Top panel: The ZTF light curve of J100243.11$-$024636.0 with the $r$-band filter shown as red triangles, $g$-band data as green dots and $i$-band data as black squares. Bottom panel: The spectrum of J100243.11$-$024636.0.}
 \end{figure} 
 
\begin{figure} 
 \centering 
 \includegraphics[width=1.0\columnwidth]{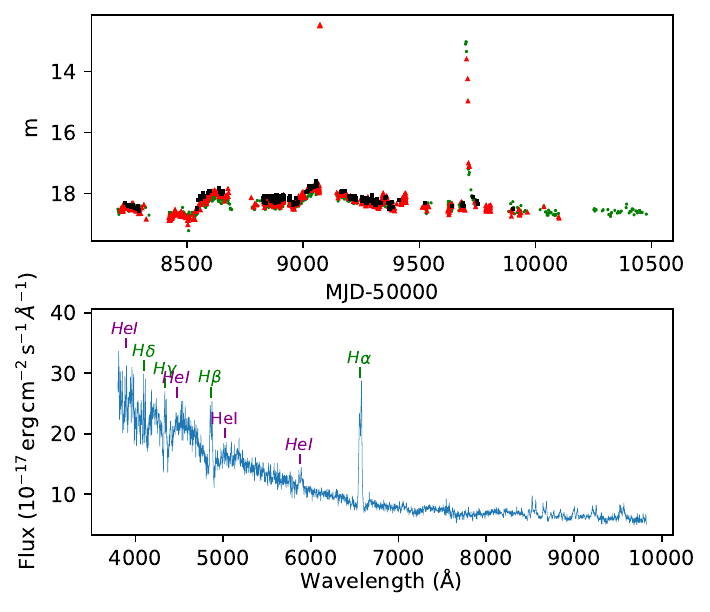} 
\caption{\label{fig:J115215.79+491441.7}
Top panel: The ZTF light curve of J115215.79+491441.7 with the $r$-band filter shown as red triangles, $g$-band data as green dots and $i$-band data as black squares. Bottom panel: The spectrum of J115215.79+491441.7.}
 \end{figure} 
 
\begin{figure} 
 \centering 
 \includegraphics[width=1.0\columnwidth]{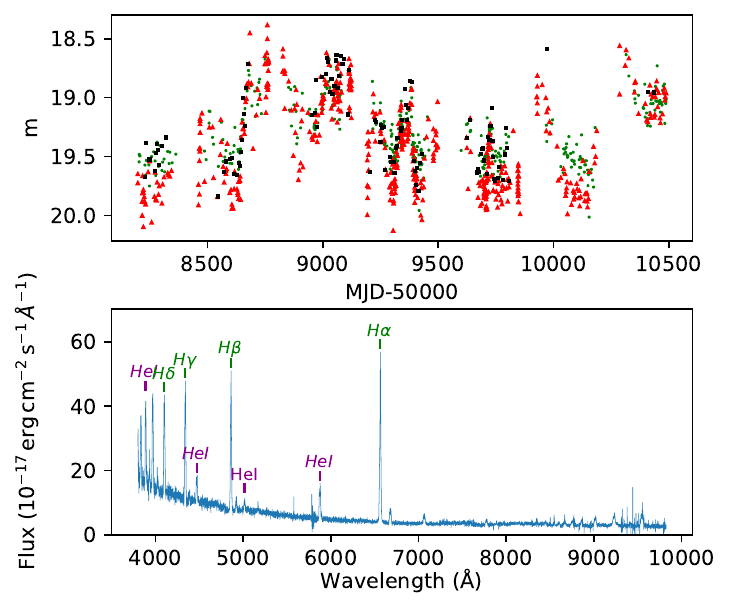} 
\caption{\label{fig:J150441.75+084752.4}
Top panel: The ZTF light curve of J150441.75+084752.4 with the $r$-band filter shown as red triangles, $g$-band data as green dots and $i$-band data as black squares. Bottom panel: The spectrum of J150441.75+084752.4.}
 \end{figure}

\subsection{Highly evolved donors}

We identified five unusual CVs whose spectra resemble an F--type star with superimposed Balmer emission lines (Fig.\,\ref{fig:Generate_F_type_plots}). The absolute magnitude of these five CVs is too low for the donor to be an F-type star and hence there is likely to have been a period of unstable mass transfer onto the white dwarf. The donor is therefore highly evolved and likely to consist of a helium core with a thin hydrogen envelope. Similar CVs have been identified previously based on a candidate selection that involves a number of cuts on position in the \textit{Gaia} HR diagram and periodic variability in the ZTF light curves \citep{2021MNRAS.508.4106E,2021MNRAS.505.2051E}. Spectroscopic follow-up of the resulting stars was then used to confirm that they were CVs. In contrast, we have identified these CVs from spectroscopy and then used light curves for confirmation. 

We note that J045138.25$-$074243.1 has an unusually short period  of $1.7$\,h given its position on the HR diagram and is reminiscent of the 55\,min period CV ZTF\,J1813+4251, which also has an extremely evolved and hot donor star  \citep{2022Natur.610..467B}.

\begin{figure} 
 \centering 
   \includegraphics[width=1.0\columnwidth]{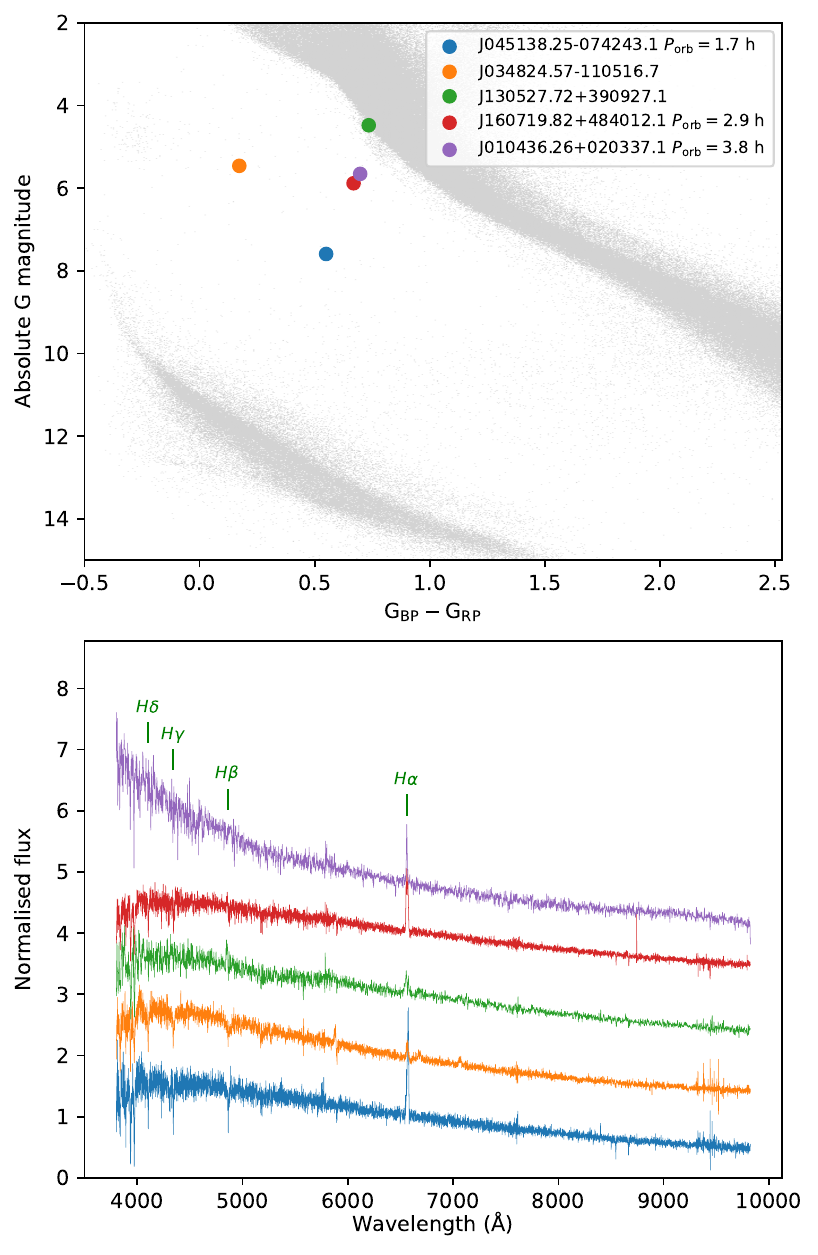} 

\caption{\label{fig:Generate_F_type_plots}
Five CVs whose spectra resemble F--type stars with superimposed Balmer emission lines but which have absolute magnitudes fainter than those F--type stars. Top panel: HR diagram with periods where known, showing that the objects are relatively bright CVs close to the main sequence. These are very likely CVs with highly nuclear evolved donor stars, see \citet{2021MNRAS.505.2051E, 2021MNRAS.508.4106E}. Bottom panel: Normalised spectral flux of the CVs.}
 \end{figure}

\section{Comparison with previous work}\label{sec:Houetal}

\citet{2026ApJS..282...26H} searched six million spectra obtained as part of DESI DR1 \citep{2026AJ....171..285D} from 2021~May to 2022~June and found 412 CVs. They used a Bagging-TopPush Algorithm to analyse the subset of DESI spectra targeted by the Milky Way survey. Our CNN identified 409 of the 412 CVs in \citet{2026ApJS..282...26H}, and failed to identify three of them. We show the DESI spectra of these three systems in Fig. \ref{fig:Houetallfails}. One of them is AM\,CVn, the prototype of this class of ultra-short period CVs, and its spectrum is dominated by absorption rather than emission features. Such a type of spectrum was not available in our training set. The other two spectra do not exhibit any clear CV features. We inspected the DESI spectra of the 412 CVs from \citet{2026ApJS..282...26H}, and we argue that thirteen of them are in fact not CVs (Table\,\ref{tab:Houetalfails}).

\begin{figure} 
 \centering 
  \includegraphics[width=1\columnwidth]{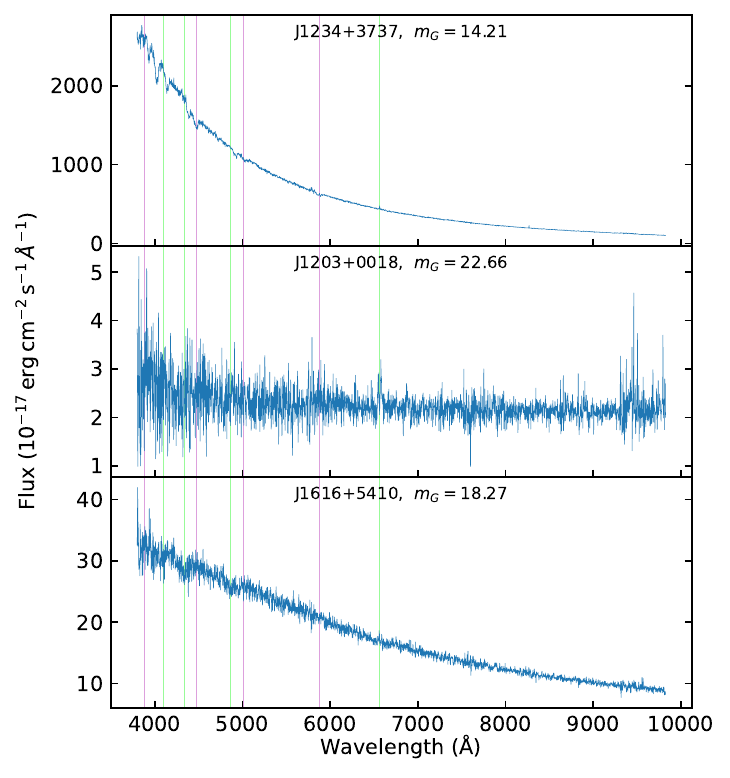} 
\caption{\label{fig:Houetallfails}
The three CVs from \citet{2026ApJS..282...26H} that were not identified by our CNN include AM\,CVn which, unlike most members of that subclass, has helium absorption lines (top), a low S/N spectrum (middle) and a spectrum that does not exhibit any CV features (bottom panel).}
 \end{figure}

\begin{table*}
\caption{\label{tab:Houetalfails} Objects identified as CVs by \citet{2026ApJS..282...26H} that we have not chosen to include in our list of CVs.}
\begin{tabular}{llp{20mm}p{80mm}}
\hline 
Name & DESI \texttt{TARGETID} & Our classification & Notes  \\ \hline
J0235+0343 & 2305843027865769037 & Detached binary & This is FS\,Cet (Feige 24) and it is classified as a detached binary in VSX with a period of $4.23$\,d \citep{1978ApJ...223..260T}.  The narrow emission lines sitting in the absorption lines from the white dwarf are characteristic of a detached binary \\
J0820+8342 & 39633565725493461 & Unclear & The DESI spectrum shows single-peaked Balmer and \ion{He}{I} emission lines. We have no \textit{Gaia} or light curve data. The spectrum, albeit noisy,  is not  typical of CVs.   \\
J0836+0446 & 39627901582185773 & Detached binary & The DESI spectrum shows narrow Balmer emission lines sitting in broad absorption lines  and also the signature of the donor. This indicates that this is a detached binary. \\
J0908$-$0125 & 39627750788565263 & Detached binary & The DESI spectrum shows narrow Balmer emission lines sitting in broad absorption troughs. The donor signature is not visible in the spectrum, but is visible in the SED.  \\
J1023+0038 & 39627805461320147 & Millisecond pulsar & This is AY\,Sex, a known millisecond pulsar \citep{2015MNRAS.453.3461S}. \\
J1139+6630 & 2305843017086402997 & Detached binary & The DESI spectrum shows narrow Balmer emission lines sitting in broad absorption troughs. The donor signature is not visible in the spectrum, but is visible in the SED.  \\
J1203+0018 & 39627793801152964 & Unclear & The DESI spectrum does not show any CV features such as emission lines. This spectrum was not detected by our CNN. \\
J1249+0357 & 39627884561698984 & Detached binary & The DESI spectrum shows narrow Balmer emission lines and also the signature of the donor. There is no evidence of a disc or accretion and we concur with \citet{2010MNRAS.402..620R} that this is a detached binary. \\
J1439$-$0106 & 39627764256478032 & Detached binary & \citet{2025MNRAS.536.1057I} tentatively classified this as a novalike.  The three spectra show  narrow Balmer, \ion{He}{I} emission lines sitting in the absorption lines from the white dwarf. Unlike the SDSS spectra presented in \citet{2025MNRAS.536.1057I} the calcium infrared triplet clearly consists of single-peaked emission lines that we assume originate in the donor rather than the accretion disc. This suggests that this is a detached binary. The position in the HR diagram is on the edge of the novalike region consistent with a detached binary.   \\
J1515+1916 & 39628245351533090 & Detached binary or subdwarf or chromospherically active M\,dwarf. & The DESI spectrum shows narrow Balmer emission lines sitting in broad absorption troughs.  \citet{2011AJ....142..181S} found that this object had absorption with very narrow emission lines and could be a pre-CV with the emission caused by irradiation from the secondary by a hot white dwarf. We concur with \citet{2012AJ....144...81T} and \citet{2023MNRAS.524.4867I} that this is not a CV.  \\
J1559+0356 & 39627885354422152 & Detached binary & The DESI spectrum shows narrow Balmer emission lines sitting in broad absorption troughs. This is a detached system \citep{2011AandA...536A..43N,2013MNRAS.429..256P,2016MNRAS.458.3808R}. \\
J1616+5410 & 39633311974295183 & Unclear & The DESI spectrum does not show any CV features such as emission lines. This spectrum was not detected by our CNN. \\
J1638+0102 & 39627813073985818 & Detached binary & This is MGAB-V1230 and is classified as a detached eclipsing binary in VSX. Gabriel Murawski concluded that the narrow emission lines sitting in the absorption lines are due to reflection from the donor of radiation from the white dwarf. \\ \hline

\end{tabular}
\end{table*}

\section{Spectroscopy as an unbiased selection method}\label{sec:selection_bias}
A large fraction of CVs have been identified as a result of their outbursts \citep{2005ASPC..330....3G}. The growing number of  wide-field sky photometric surveys such as CRTS and ZTF produce CV discoveries on an industrial scale \citep{2014MNRAS.441.1186D,2021AJ....162...94S}, introducing a bias towards CVs with clear signatures of variability among the known population of these systems. Statistical analyses of such biased samples are at risk of drawing wrong conclusions on the overall properties of CVs. 

We used the ZTF photometry \citep{2019PASP..131a8002B} to investigate the variability of the \NnewCVs\ new CVs that we spectroscopically identified. ZTF light curves were available for 171 of them, of which only \Noutbursts\ displayed outbursts, the remaining \Nnooutbursts\ did not show large-amplitude variability. This large fraction of non-outbursting CVs among the new DESI discoveries highlights the importance of spectroscopy as a largely unbiased method to identify CVs. Such unbiased CV samples can be used for quantitative tests of CV population models \cite[e.g.][]{2009MNRAS.397.2170G}, and result in the discovery of new sub-classes of CVs (e.g. Section\,\ref{sec:Unusual_state_changes}) and individual rare systems that may be missing links in the evolution of CVs \citep[e.g.][]{2023MNRAS.524.4867I, 2025MNRAS.543.2116C}.

\begin{figure*} 
 \centering 
  \includegraphics[width=1\textwidth]{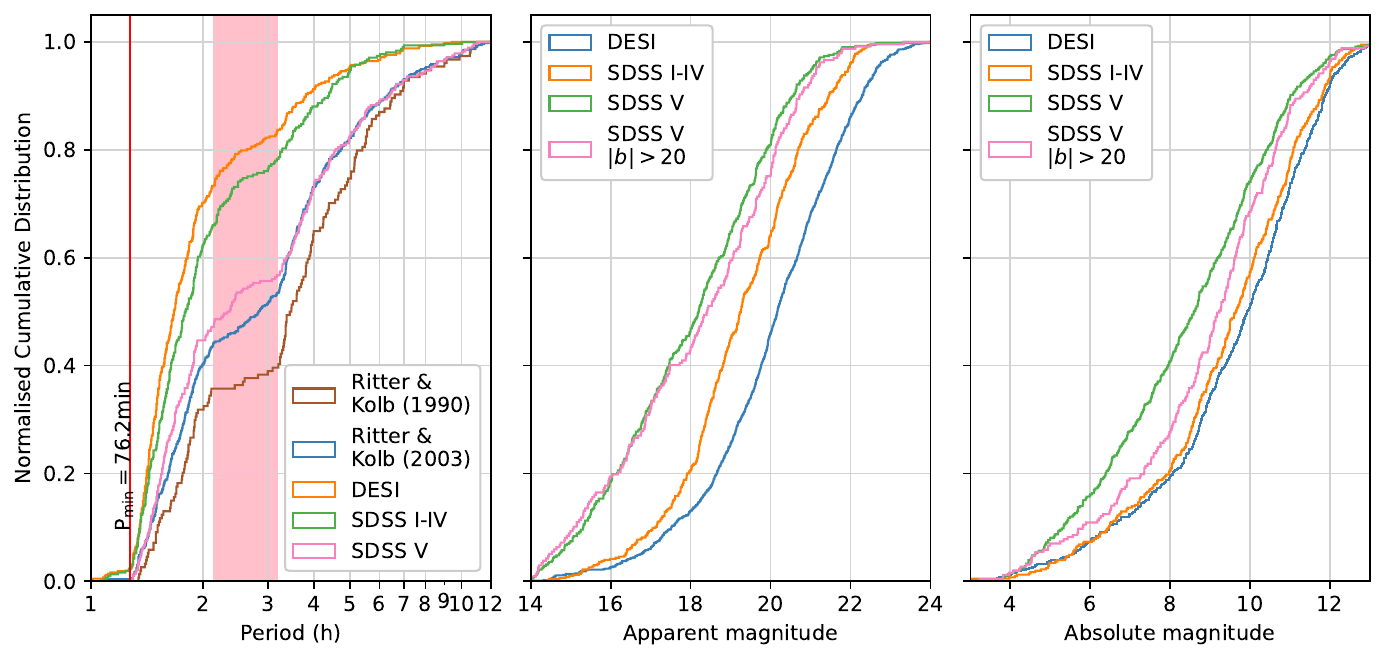} 
\caption{\label{fig:Period_distribution}
Left panel: The normalised cumulative orbital period distributions of the 448 DESI CVs (orange), 507 and 291 CVs observed by SDSS\,I-IV (green, \citealt{2023MNRAS.524.4867I}) and SDSS\,V (magenta, \citealt{2025MNRAS.536.1057I}), 168 CVs from version 5 of the Ritter and Kolb catalogue (brown, \citealt{1990A&AS...85.1179R}) and 572 CVs from version 7 of the Ritter and Kolb catalogue (blue, \citealt{2003A&A...404..301R}). The red vertical line shows the period minimum \citep{1983ApJ...268..825P,2006MNRAS.373..484K} and the red shaded area is the period gap \citep{1980MNRAS.190..801W,2006MNRAS.373..484K}. The largely untargeted DESI and SDSS\,I-IV surveys contain much larger proportions of short-periods systems, in particular DESI which is currently the deepest CV survey. In contrast, SDSS\,V, which was a targeted CV survey, including follow-up of many photometrically identified systems, has the lowest proportion of short-period systems, highlighting the bias introduced by selection methods based on variability. Centre and right panels: The distribution of the absolute and apparent magnitude of CVs respectively from this work is shown in blue. The apparent magnitude is a synthetic $G$-band magnitude calculated from the spectra whilst the absolute magnitude uses \textit{Gaia} parallaxes and magnitudes where they exist. The apparent magnitude plot demonstrates that DESI has consistently observed CVs to a greater depth (around two magnitudes) than SDSS, resulting in a larger proportion of intrinsically faint (Figure\,\ref{fig:Faint_CVs}) compared to any previous survey.     
}
 \end{figure*} 

\begin{figure} 
 \centering 
  \includegraphics[width=1\columnwidth]{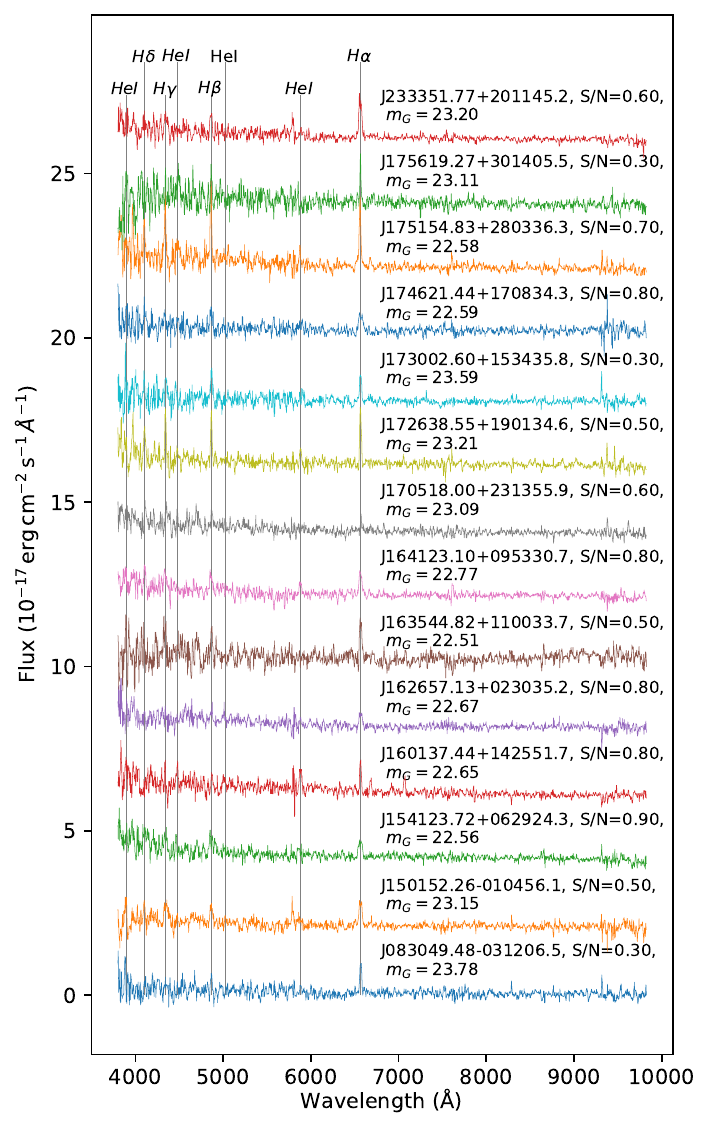} 
\caption{\label{fig:Faint_CVs}
The spectra of a selection of the faintest DESI CVs, having $\mathrm{S/N}<1$ across the full spectral range of the DESI spectrum, no \textit{Gaia} counterpart and no ZTF light curve data. For each CV we provide the S/N ratio and the synthetic magnitude (based on a \textit{Gaia} $G$-band filter). The spectra are offset vertically by suitable amounts. The magnitude limit of DESI at which it can spectroscopically identify CVs is $1-2$\,mag deeper than any other previous survey.}
\end{figure}

\section{Comparison with previous CV samples}

The orbital period distribution is a key diagnostic of CV evolution, which has been discussed for the past 50 years \citep{1956ApJ...123...44C, 1980MNRAS.190..801W, 1997MNRAS.290L..55V, 1999MNRAS.309.1034K, 2006MNRAS.373..484K, 2009MNRAS.397.2170G, 2024A&A...682L...7S}. \NCVswithperiods\ of the CVs in the DESI sample have a known orbital period (Table\,\ref{tab:cvlist}), obtained either from the literature, or measured from archival photometry (Section\,\ref{subsec:CVtable}). We compare in Fig.\,\ref{fig:Period_distribution} their orbital period distribution with those of the CVs observed by SDSS\,I-IV \cite{2023MNRAS.524.4867I}, SDSS\,V \citep{2025MNRAS.536.1057I}, and the \citep{1990A&AS...85.1179R} and \citep{2003A&A...404..301R} catalogues. A distinct feature of the (largely, see Section\,\ref{sec:targetting}) unbiased DESI and SDSS\,I-IV surveys are the large fractions of short-period CVs, with that fraction being even higher in the DESI sample compared to the SDSS\,I-IV sample. 

Inspecting the distributions in apparent and absolute magnitude (middle and right panels of Fig.\,\ref{fig:Period_distribution}, respectively) illustrates that DESI is the deepest CV survey  (by up to two magnitudes), resulting in a larger fraction of intrinsically faint CVs. These fainter systems are, overall, characterised by cooler white dwarfs, lower-mass and cooler donor stars, and lower accretion rates, all features of CVs that have evolved towards shorter periods~--~explaining why the DESI sample contains the highest proportion of short-period CVs. It is interesting to note that DESI, despite its depth, does not uncover CVs with absolute magnitudes fainter than the known bottom end of the CV luminosity function.

In contrast, the SDSS\,V sample contains a much lower number of short-period CVs, which is the result of it being a \textit{targeted} survey, following up known CV candidates which were in large part identified by photometric variability. The impact of the variability-selection is particularly high at lower galactic latitudes, where the target list of SDSS\,V contains large numbers of distant but absolutely bright CVs. 

The ``period gap'', i.e. the paucity of CVs with orbital periods in the range $\simeq2-3$\,h, is of particular importance in CV evolution models, usually associated with a change in the rate of angular momentum loss at $\simeq3$\,h resulting in temporary detached configuration of the systems \citep{1983ApJ...275..713R}. The exact mechanisms at work, as well as the location of the period gap boundaries have been subject to much discussion \citep[e.g.][]{1993A&A...271..149K, 1993A&A...279L...5K, 2011ApJS..194...28K, 2015ApJ...809...80G, 2024A&A...682L...7S}. Fig.\,\ref{fig:Period_distribution} suggests that the dearth of CVs with periods within the gap is a consistent feature, however, it also illustrates that care has to be taken in the selection of samples of observed CVs to underpin theoretical models. It is apparent that the well-defined lower and upper edges of the period gap seen in the early samples, dominated by intrinsically bright CVs, is gradually washed out in homogeneous samples drawn from deep spectroscopic surveys (with the exception of SDSS-V for the reasons outlined above).

\section{The effectiveness of the \texorpdfstring{\DESI}{xx} targeting strategies in finding CVs  }\label{sec:targetting}

In contrast to the study by \citet{2026ApJS..282...26H}, who only analysed DESI DR1 spectra of MWS targets, our study covers the entire volume of DESI DR2 spectra (\Nscience) and hence includes all target categories of DESI. The experience from SDSS is that the spectra of many CVs were serendipitously obtained as part of other programs, in particular quasar searches. 

Here we review how effective each of the targeting methods described in Section\,\ref{sec:introduction} were in identifying CVs. We first examine the targeting classes that specifically search for CVs namely the white dwarf class, the short-period CV class and our close white dwarf binary (CWDB, see Appendix\,\ref{sec:CWDBC}) class. Of the \NCVs\ CVs found in DESI these three targeting classes yielded \NSPWD, \NSPSPCV\ and \NSPWDB\ CVs respectively (with significant overlaps~--~see Fig.\,\ref{fig:Analyse_carton_frequencies_2}). The bulk of the \NCVs\ were therefore found because they fulfilled the selection criteria of one or more of the DESI ``main'' surveys (i.e. the bright-time MWS, BGS or dark-time LRG, ELG and QSO surveys). This is reminiscent of the experience of SDSS where many CVs were found because they had similar colours to quasars \citep{2023MNRAS.524.4867I}. It is even more remarkable that almost all the CVs were serendipitously targeted by at least one of the DESI main surveys (Fig.\,\ref{fig:Analyse_carton_frequencies_3}).

We investigated the apparently poor performance of the CWDB target class which only yielded \NSPWDB\ CVs. We first inspected the distribution of the DESI CVs within the \textit{Gaia} HR diagram (Fig.\,\ref{fig:desi_ourCVs_hrd}, left), which demonstrates that most systems are located between the main sequence and the white dwarf cooling sequence and a small number are mixed into the main sequence. However, moving to the UV-optical colour-magnitude diagram, (Fig.\,\ref{fig:desi_ourCVs_hrd}, right), it became apparent that only \NFUV\ out of \NCVs\ had a \textit{GALEX} far-UV detection. The cut in Eq.\,\ref{eq:fuv-cut} left \Ncut\ CVs, and the astrometric quality cut further reduced this to \Ncuttwo\ CVs. Noting that we removed the single white dwarf targets from the CWDB target class to avoid target duplication, it transpires that the CWDB only adds a relatively small number of genuine cataclysmic variables. A brief inspection of the DESI spectra obtained via the CWDB target class that were not flagged as CV candidates reveals a large fraction of detached white dwarf binaries, motivating an extension of the work presented here to this class of systems.

In summary the lack of \textit{GALEX} observations is a key limitation of the UV-based target selection, which is in part due to the incomplete sky coverage scope of the \textit{GALEX} survey which was curtailed in 2009 when the far-UV detector failed. Another limitation is that many CVs in our sample are simply fainter than the magnitude limit of \textit{GALEX}. A UV-based target selection for white dwarf binaries will be worthwhile revisiting once the \textit{UVEX} mission \cite{2021arXiv211115608K} completes a sufficiently large area of its imaging survey.

\begin{figure} 
 \centering 
   \includegraphics[width=1.0\columnwidth]{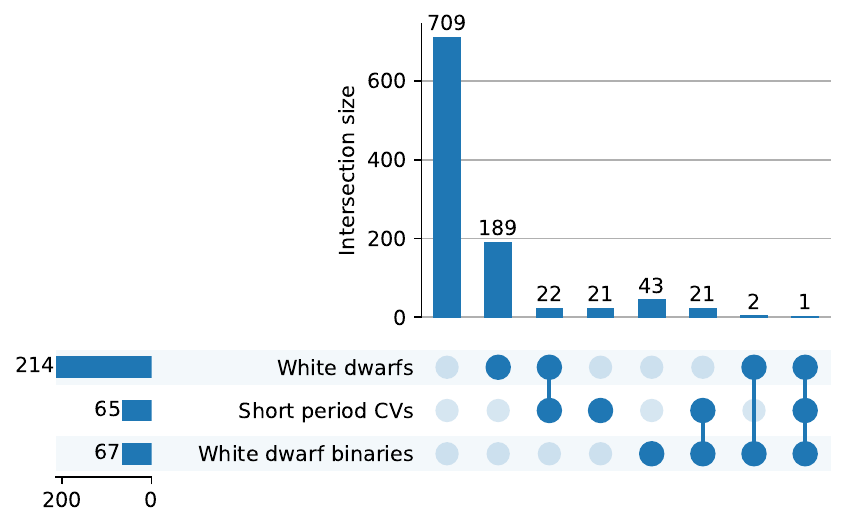} 

\caption{\label{fig:Analyse_carton_frequencies_2}
Each of the \NCVs\,CVs was targeted by one or more target classes.  Here we analyse the three classes that specifically targeted CVs using an UpSet diagram \citep{6876017}.  The UpSet diagram shows the number of CVs targeted by each target class and the size of the intersections of classes. It is noteworthy that the bulk, \Nnotthreesurveys, of CVs were found serendipitously by surveys targeting other objects rather than by the three classes specifically targeting CVs.    }
 \end{figure} 
\begin{figure*} 
 \centering 
   \includegraphics[width=0.75\textwidth]{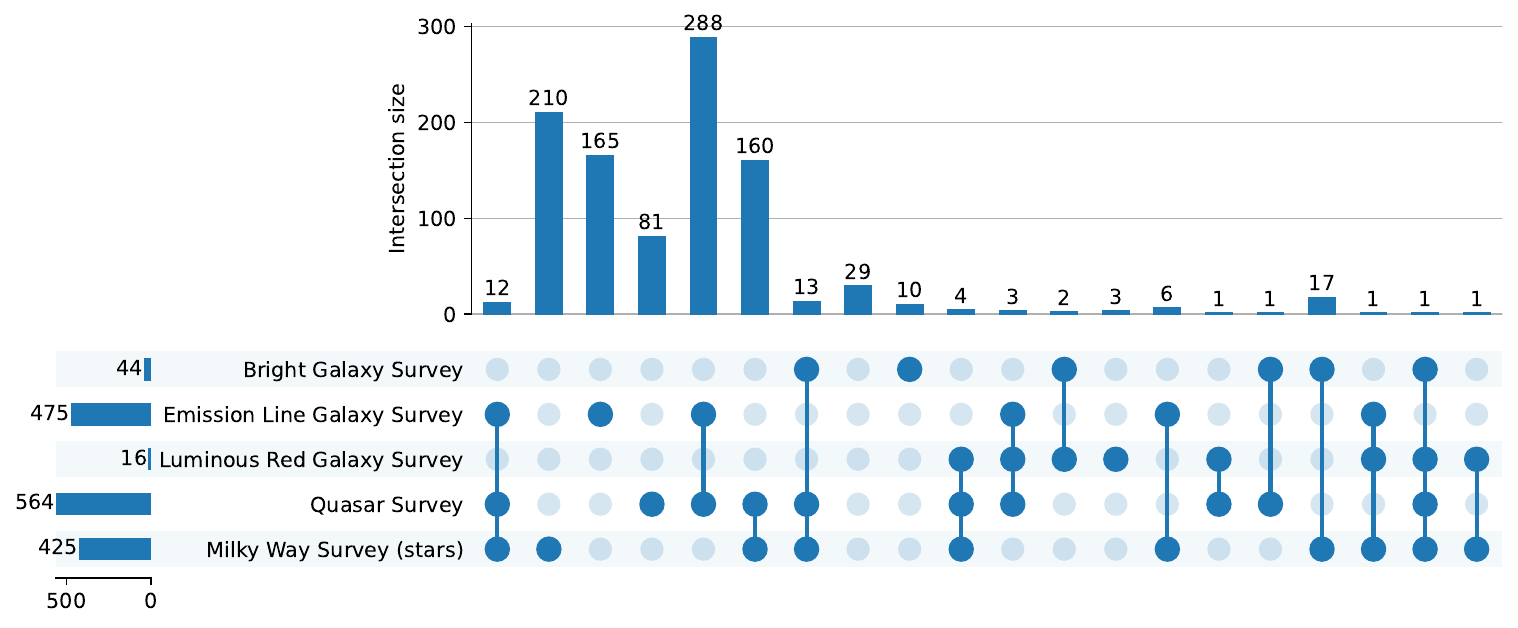} 

\caption{\label{fig:Analyse_carton_frequencies_3}
Each of the \NCVs\,CVs was targeted by one or more surveys.  Here we analyse the five main  surveys using an UpSet diagram \citep{6876017}.  The UpSet diagram shows the number of CVs targeted by each category and the size of the intersections of categories. It is noteworthy that all but 29 of the CVs were targeted by at least one of the main surveys.   }
 \end{figure*}

\section{Space density}\label{sec:space_density}

We have followed the approach described in detail in section\,6.4 of \citet{2023MNRAS.524.4867I} to estimate the space density of the CV sub-types and give a brief summary here.

The space density ($\rho$) of a type of object is the expected number of systems to be found in a cubic parsec. A simplified  assumption is that the Galaxy is axially symmetric and the density drops exponentially with the height ($|z|$) above the galactic plane according to a scale height ($h$) which in turn will depend upon the age of the object type:
\begin{equation} \label{eq:scale1}
\rho\left(z \right)=\rho _{0}\times\exp\left( -\frac{\left | z \right |}{h} \right) 
\end{equation}

To estimate the galactic mid-plane space density, $\rho _{0}$, for an object type observed $N_{\mathrm{obs}}$ times by DESI we need to firstly determine the size of the volume that DESI surveyed ($V_{\mathrm{eff}}$) which is a function of the maximum distance at which DESI can observe that object type ($R_{\mathrm{lim}}$) which in turn depends upon the average absolute magnitude of the object type. $V_{\mathrm{eff}}$ also depends upon the size of the DESI footprint. Secondly we need to estimate the proportion of CVs in the DESI footprint that were actually observed by DESI (the completeness); the inverse of the completeness is then applied as a correction factor ($C_{\mathrm{orr}}$). Bringing this together the estimate of the space density is calculated from
\begin{equation}\label{equ:spaced}
\rho _{0} =\frac{N_{\mathrm{obs}}\,\times C_{\mathrm{orr}}}{V_{\mathrm{eff}}(h,R_{\mathrm{lim}})}
\end{equation} 

With two exceptions we have followed the approach described in \citet{2023MNRAS.524.4867I} to estimate the space density of the CV sub-types. \citet{2023MNRAS.524.4867I} firstly calculated the limiting distance for each sub-type based on the assumption that 
\allsdss\ could reliably detect CVs with $m_\mathrm{G}<21$. DESI uses a larger aperture telescope than SDSS and we calculated limiting magnitudes based on an assumption of $m_\mathrm{G}<22.5$.

\citet{2023MNRAS.524.4867I} assumed scale heights for CVs of each sub-type and we use the same assumptions here.  The next step was to estimate the completeness of our sample. DESI, in common with SDSS, uses a large number of selection functions to determine which objects are targeted and in practice it is not possible to analyse these to determine completeness. \citet{2023MNRAS.524.4867I} estimated completeness from the proportion of a known sample of CVs that were rediscovered by \allsdss. We used a similar approach but used the set of SDSS CVs from \citet{2023MNRAS.524.4867I,2025MNRAS.536.1057I} as our reference. We noted that the completeness varied as a function of orbital period (see Fig.\,\ref{fig:completeness}) with longer-period CVs ($P_\mathrm{orb}>2$\,h) being far less complete. In the space density estimates we therefore used values for completeness of 0.6 for systems that tend to have shorter periods (SU\,UMa, WZ\,Sge and polar) and $0.2$ for the remainder (U\,Gem, novalike and intermediate polar).

\citet{2023MNRAS.524.4867I} then determined the footprint of \allsdss\ taking account of galactic latitude by modelling the celestial sphere using $2^8$ \textsc{astropy} HEALPix pixels \citep{2005ApJ...622..759G} and then analysing the HEALPixs which were encompassed by one or more plates. We use the same approach here using the DESI exposures instead of plates. We simplified the problem by firstly ignoring whether each exposure occurred in bright time or dark time and secondly also duplication caused by the number of passes. For each sub-type this results in the effective volume spanned by the survey footprint which is then used to estimate the space density.

\begin{figure} 
 \centering 
  \includegraphics[width=\columnwidth]{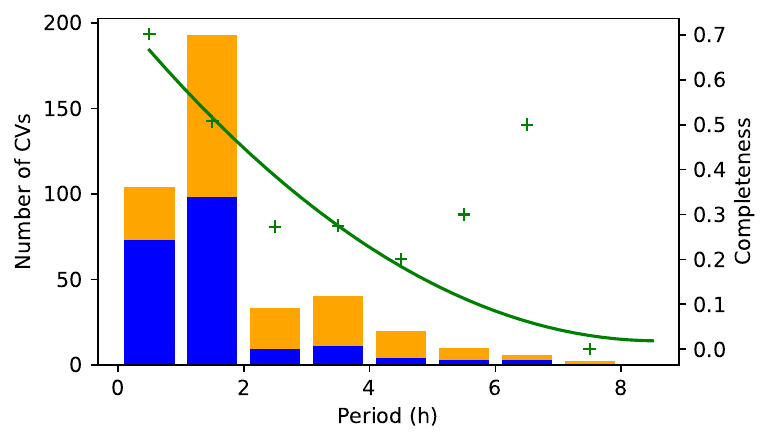} 
\caption{\label{fig:completeness}
Analysis of CVs by orbital period showing the number of CVs observed by DESI (blue) compared with the number reported in \citet{2023MNRAS.524.4867I,2025MNRAS.536.1057I}. The completeness is the ratio of these two numbers and is shown in green together with a weighted (by number of CVs in each bin) quadratic fit. A linear fit was tried but proved to be poor fit whilst higher order polynomials exhibited physically implausible peaks and troughs. It is evident that shorter period systems have a much great completeness than longer period systems  }
 \end{figure} 

\begin{figure*} 
 \centering 
  \includegraphics[width=0.9\textwidth]{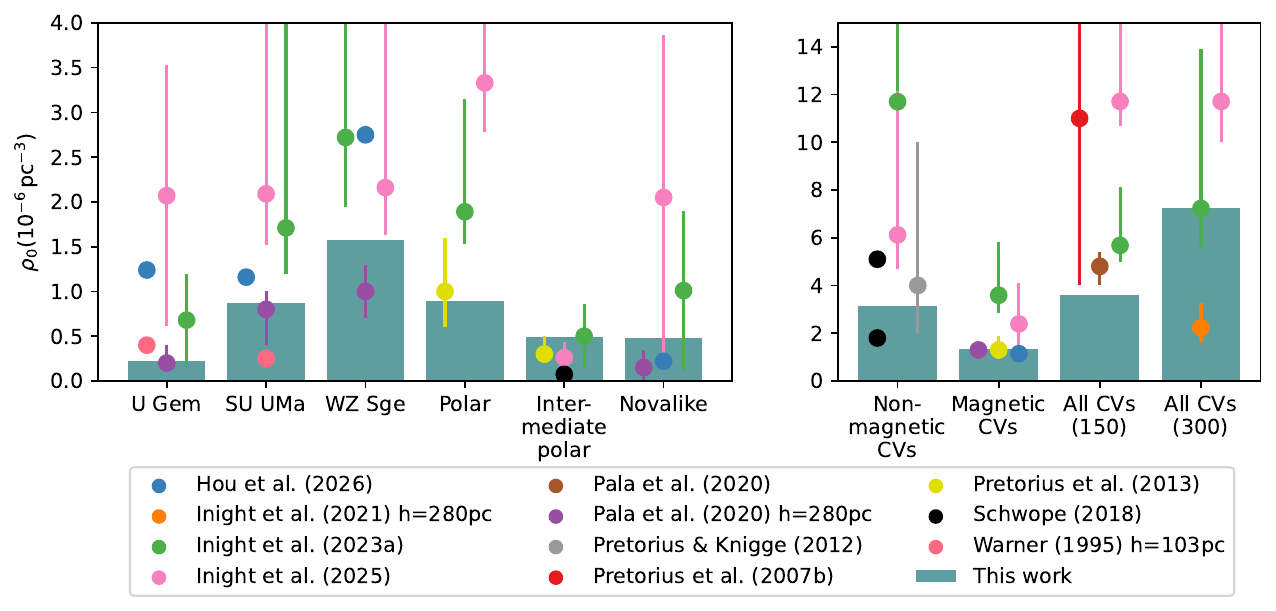} 
\caption{\label{fig:space_density}
Comparison of the space density estimates for CVs from this work (bar chart) with those from eleven previous papers. All estimates use the assumptions for the scale heights from \citet{2007MNRAS.374.1495P} unless otherwise stated. Left panel: Estimates for individual sub-types. Note that the values from \citet{2020MNRAS.494.3799P} are understated as they do not take account of completeness.  Right panel: Composite estimates. \citet{2021MNRAS.504.2420I} specifically excluded selection effects.  Note that our estimate for all CVs within $150$\,pc in the third bar suffers from a small sample of only seven CVs.}
 \end{figure*} 

\begin{table*}
\centering
\caption{Estimates of the space densities ($\rho_0$ ) of the CV sub-types in DESI for  three  assumptions of the scale height (120, 260 and 450\,pc) together with $\rho_0$ (P2007) which assumes the scale heights from \citet{2007MNRAS.374.1495P}. $N$ is the number of DESI objects that are closer than the limiting distance. Note that the values of $\rho_0$ for individual sub-types are slightly understated ($\simeq20$\,per cent)  as we have not taken account of the remaining unclassified CVs and dwarf novae. 
This does not apply to the ``All CVs'' values which have been calculated for 150\,pc and 300\,pc limiting distances to enable comparison with \citet{2020MNRAS.494.3799P} and \citet{2021MNRAS.504.2420I}, respectively. These are calculated values and their precision should not be used to infer their uncertainty which is dominated by the uncertainty in the scale height. }
\label{tab:spacedensity}
\begin{tabular}{lllllll} 
\hline
\multirow{2}{*}{CV subtype} & \multirow{2}{*}{\begin{tabular}[l]{@{}l@{}}Limiting distance\\ (pc)\end{tabular}} & \multirow{2}{*}{$N$} & \multicolumn{4}{c}{Space Density $\mathrm{(10^{-6} pc^{-3})}$}       \\ 
\cline{4-7}
                      &                                                                                     &                    & $\rho_0(120)$ & $\rho_0(260)$ & $\rho_0(450)$ & $\rho_0(P2007)$            \\
\hline
    U Gem & 963 & 6 & 0.26 & 0.09 & 0.05 & 0.24 \\
SU UMa & 483 & 47 & 1.85 & 0.88 & 0.64 & 0.88 \\
WZ Sge & 441 & 69 & 3.19 & 1.59 & 1.19 & 1.59 \\
Polar & 274 & 10 & 1.36 & 0.85 & 0.71 & 0.90 \\
Intermediate polar & 791 & 10 & 0.66 & 0.26 & 0.17 & 0.50 \\
Novalike & 2867 & 37 & 0.52 & 0.11 & 0.04 & 0.49 \\
[0.5ex]
All CVs & 150 & 8 & 4.87 & 3.65 & 3.27 & 3.65 \\
All CVs & 300 & 85 & 10.43 & 6.19 & 5.03 & 7.29  
    \\\hline
\end{tabular}
\end{table*}

The estimates for the space density of short period systems (SU\,UMa and WZ\,Sge) are consistent with previous estimates as shown in Fig.\,\ref{fig:space_density}. However these estimates are lower bounds as they do not take account of the  \Nuntyped\,CVs where we could not identify the subtype, many of which do not exhibit transients and are therefore likely to be short-period systems.


\section{Conclusions}\label{sec:conclusions}

We have successfully used a methodology developed for SDSS to search for CVs within DESI resulting in the discovery of \NnewCVs\ new CVs, spectroscopic confirmation of \Newspec\ CV candidates and \Newperiod\ new or improved orbital periods. The new CVs are particularly valuable as they are predominantly under-represented systems with low accretion rates. The new CVs include \Npeculiar\ additional members of a putative new class of CVs with a prototype of IR\,Com which show peculiar state changes. In passing we note five other CVs that do not naturally fall into any of the established classifications.   

Many of the new CVs are too faint to have a \textit{Gaia} counterpart or ZTF light curve. In these cases, with the spectrum the only evidence, it is difficult to determine the CV subtype. Future deeper photometric surveys such as LSST, are likely to provide new evidence to determine sub-types and orbital periods. 

Our targeting strategy was limited by the coverage and depth of \textit{GALEX} far-UV observations. We were fortunate that other surveys, which were targeting distant galaxies, serendipitously also surveyed our CVs. Future space-based far-UV surveys, such as \textit{UVEX}, will be extremely valuable for the efficient targeting of faint CVs.   

\section{Acknowledgements}
This project has received funding from the European Research Council (ERC) under the European Union’s Horizon 2020 research and innovation programme (Grant agreement No. 101020057). ADM was supported by the U.S.\ Department of Energy, Office of Science, Office of High Energy Physics, under Award Number DE-SC0019022. AA acknowledges support from Naresuan University (NU), and National Science Research and Innovation Fund (NSRF), grant no. R2569B032.

This research used data obtained with the Dark Energy Spectroscopic Instrument (DESI). DESI construction and operations is managed by the Lawrence Berkeley National Laboratory. This material is based upon work supported by the U.S. Department of Energy, Office of Science, Office of High-Energy Physics, under Contract No. DE–AC02–05CH11231, and by the National Energy Research Scientific Computing Center, a DOE Office of Science User Facility under the same contract. Additional support for DESI was provided by the U.S. National Science Foundation (NSF), Division of Astronomical Sciences under Contract No. AST-0950945 to the NSF’s National Optical-Infrared Astronomy Research Laboratory; the Science and Technology Facilities Council of the United Kingdom; the Gordon and Betty Moore Foundation; the Heising-Simons Foundation; the French Alternative Energies and Atomic Energy Commission (CEA); the National Council of Humanities, Science and Technology of Mexico (CONAHCYT); the Ministry of Science and Innovation of Spain (MICINN), and by the DESI Member Institutions: www.desi.lbl.gov/collaborating-institutions. The DESI collaboration is honored to be permitted to conduct scientific research on I’oligam Du’ag (Kitt Peak), a mountain with particular significance to the Tohono O’odham Nation. Any opinions, findings, and conclusions or recommendations expressed in this material are those of the author(s) and do not necessarily reflect the views of the U.S. National Science Foundation, the U.S. Department of Energy, or any of the listed funding agencies.

Based on observations obtained with the Samuel Oschin Telescope 48-inch and the 60-inch Telescope at the Palomar Observatory as part of the Zwicky Transient Facility project. ZTF is supported by the National Science Foundation under Grants No. AST-1440341 and AST-2034437 and a collaboration including current partners Caltech, IPAC, the Oskar Klein Center at Stockholm University, the University of Maryland, University of California, Berkeley , the University of Wisconsin at Milwaukee, University of Warwick, Ruhr University, Cornell University, Northwestern University and Drexel University. Operations are conducted by COO, IPAC, and UW.

CRTS is supported by the U.S.\ National Science Foundation under grants AST-0909182 and CNS-0540369. The work at Caltech was supported in part by the NASA Fermi grant 08-FERMI08-0025, and by the Ajax Foundation. The CSS survey is funded by the National Aeronautics and Space Administration under Grant No. NNG05GF22G issued through the Science Mission Directorate NEOs Observations Program. 

This research has made use of the International Variable Star Index (VSX) data base, operated at AAVSO, Cambridge, Massachusetts, USA. 

This paper includes data collected by the TESS mission. Funding for the TESS mission is provided by the NASA's Science Mission Directorate.

This work has made use of data from the European Space Agency (ESA) mission \textit {Gaia} (\url{https://www.cosmos.esa.int/gaia}), processed by the {\it Gaia} Data Processing and Analysis Consortium (DPAC, \url{https://www.cosmos.esa.int/web/gaia/dpac/consortium}). Funding for the DPAC has been provided by national institutions, in particular the institutions participating in the {\it Gaia} Multilateral Agreement.  This research has made use of NASA's Astrophysics Data System. This research has made use of the VizieR catalogue access tool, CDS, Strasbourg, France.

This work has made use of data from the Asteroid Terrestrial-impact Last Alert System (ATLAS) project. The Asteroid Terrestrial-impact Last Alert System (ATLAS) project is primarily funded to search for near earth asteroids through NASA grants NN12AR55G, 80NSSC18K0284, and 80NSSC18K1575; byproducts of the NEO search include images and catalogs from the survey area. This work was partially funded by Kepler/K2 grant J1944/80NSSC19K0112 and HST GO-15889, and STFC grants ST/T000198/1 and ST/S006109/1. The ATLAS science products have been made possible through the contributions of the University of Hawaii Institute for Astronomy, the Queen’s University Belfast, the Space Telescope Science Institute, the South African Astronomical Observatory, and The Millennium Institute of Astrophysics (MAS), Chile.

\section{Data Availability}
DESI data will be publicly available at the end of the proprietary period. Files in .csv format of the \Nmlselected\,spectra identified by the CNN and the \NCVs\,CVs accompany this publication. For each CV spectrum we also include a .png file showing the spectrum, HR diagram, SED, image and relevant data. We also provide a .csv file of the spectra used for training and testing. The other data used in this article are available from the sources referenced in the text.


\nocite{*}

\bibliographystyle{mnras}

\bibliography{refs,others,refs_supp}

@BOOK{2003cvs..book.....W,
       author = {{Warner}, Brian},
        title = "{Cataclysmic Variable Stars}",
     publisher= {Cambridge University Press},
         year = 2003,
          doi = {10.1017/CBO9780511586491},
       adsurl = {https://ui.adsabs.harvard.edu/abs/2003cvs..book.....W}
}

@ARTICLE{6876017,
  author={Lex, Alexander and Gehlenborg, Nils and Strobelt, Hendrik and Vuillemot, Romain and Pfister, Hanspeter},
  journal={IEEE Transactions on Visualization and Computer Graphics}, 
  title={UpSet: Visualization of Intersecting Sets}, 
  year={2014},
  volume={20},
  number={12},
  pages={1983-1992},
  doi={10.1109/TVCG.2014.2346248}}

@software{nicholas_earl_2023_10016569,
  author       = {Nicholas Earl and
                  Erik Tollerud and
                  Ricky O'Steen and
                  brechmos and
                  Wolfgang Kerzendorf and
                  Ivo Busko and
                  shaileshahuja and
                  Dan D'Avella and
                  Thomas Robitaille and
                  P. L. Lim and
                  Adam Ginsburg and
                  Derek Homeier and
                  Brigitta Sipőcz and
                  Jesse Averbukh and
                  James Tocknell and
                  Brian Cherinka and
                  Sara Ogaz and
                  Robel Geda and
                  James Davies and
                  Kyle Conroy and
                  Hans Moritz Günther and
                  Kyle Barbary and
                  Jonathan Foster and
                  Michael Droettboom and
                  Duy Tuong Nguyen and
                  E. M. Bray and
                  Andy Casey and
                  Peter Teuben and
                  Steve Crawford and
                  Henry Ferguson},
  title        = {astropy/specutils: v1.12.0},
  month        = oct,
  year         = 2023,
  publisher    = {Zenodo},
  version      = {v1.12.0},
  doi          = {10.5281/zenodo.10016569},
  url          = {https://doi.org/10.5281/zenodo.10016569}
}

@article{Fouesneau2026,
  author = {Fouesneau, Morgan},
  title = {Pyphot: A tool for computing photometry from spectra},
  journal = {Journal of Open Source Software},
  year = {2026},
  volume = {11},
  number = {119},
  pages = {8814},
  doi = {10.21105/joss.08814},
  url = {https://doi.org}
}

@ARTICLE{2025arXiv250314745D,
       author = {{DESI Collaboration} and {Abdul-Karim}, M. and {Adame}, A.~G. and {Aguado}, D. and {Aguilar}, J. and {Ahlen}, S. and {Alam}, S. and {Aldering}, G. and {Alexander}, D.~M. and {Alfarsy}, R. and {Allen}, L. and {Allende Prieto}, C. and {Alves}, O. and {Anand}, A. and {Andrade}, U. and {Armengaud}, E. and {Avila}, S. and {Aviles}, A. and {Awan}, H. and {Bailey}, S. and {Baleato Lizancos}, A. and {Ballester}, O. and {Bault}, A. and {Bautista}, J. and {BenZvi}, S. and {Beraldo e Silva}, L. and {Bermejo-Climent}, J.~R. and {Beutler}, F. and {Bianchi}, D. and {Blake}, C. and {Blum}, R. and {Bolton}, A.~S. and {Bonici}, M. and {Brieden}, S. and {Brodzeller}, A. and {Brooks}, D. and {Buckley-Geer}, E. and {Burtin}, E. and {Canning}, R. and {Carnero Rosell}, A. and {Carr}, A. and {Carrilho}, P. and {Casas}, L. and {Castander}, F.~J. and {Cereskaite}, R. and {Cervantes-Cota}, J.~L. and {Chaussidon}, E. and {Chaves-Montero}, J. and {Chen}, S. and {Chen}, X. and {Claybaugh}, T. and {Cole}, S. and {Cooper}, A.~P. and {Cousinou}, M.-C. and {Cuceu}, A. and {Davis}, T.~M. and {Dawson}, K.~S. and {de Belsunce}, R. and {de la Cruz}, R. and {de la Macorra}, A. and {de Mattia}, A. and {Deiosso}, N. and {Della Costa}, J. and {Demina}, R. and {Demirbozan}, U. and {DeRose}, J. and {Dey}, A. and {Dey}, B. and {Ding}, J. and {Ding}, Z. and {Doel}, P. and {Douglass}, K. and {Dowicz}, M. and {Ebina}, H. and {Edelstein}, J. and {Eisenstein}, D.~J. and {Elbers}, W. and {Emas}, N. and {Escoffier}, S. and {Fagrelius}, P. and {Fan}, X. and {Fanning}, K. and {Fawcett}, V.~A. and {Fern\textbackslash'andez-Garc\textbackslash'ia}, E. and {Ferraro}, S. and {Findlay}, N. and {Font-Ribera}, A. and {Forero-Romero}, J.~E. and {Forero-S\textbackslash'anchez}, D. and {Frenk}, C.~S. and {G\textbackslash''ansicke}, B.~T. and {Galbany}, L. and {Garc\textbackslash'ia-Bellido}, J. and {Garcia-Quintero}, C. and {Garrison}, L.~H. and {Gazta\textbackslash\raisebox{-0.5ex}\textasciitildenaga}, E. and {Gil-Mar\textbackslash'in}, H. and {Gnedin}, O.~Y. and {Gontcho}, S. Gontcho A and {Gonzalez-Morales}, A.~X. and {Gonzalez-Perez}, V. and {Gordon}, C. and {Graur}, O. and {Green}, D. and {Gruen}, D. and {Gsponer}, R. and {Guandalin}, C. and {Gutierrez}, G. and {Guy}, J. and {Hahn}, C. and {Han}, J.~J. and {Han}, J. and {He}, S. and {Herrera-Alcantar}, H.~K. and {Honscheid}, K. and {Hou}, J. and {Howlett}, C. and {Huterer}, D. and {Ir\textbackslashv\{s\}i\textbackslashv\{c\}}, V. and {Ishak}, M. and {Jacques}, A. and {Jimenez}, J. and {Jing}, Y.~P. and {Joachimi}, B. and {Joudaki}, S. and {Joyce}, R. and {Jullo}, E. and {Juneau}, S. and {Kara\textbackslashc\{c\}ayl\{\textbackslashi\}}, N.~G. and {Karim}, T. and {Kehoe}, R. and {Kent}, S. and {Khederlarian}, A. and {Kirkby}, D. and {Kisner}, T. and {Kitaura}, F.-S. and {Kizhuprakkat}, N. and {Kong}, H. and {Koposov}, S.~E. and {Kremin}, A. and {Krolewski}, A. and {Lahav}, O. and {Lai}, Y. and {Lamman}, C. and {Lan}, T.-W. and {Landriau}, M. and {Lang}, D. and {Lange}, J.~U. and {Lasker}, J. and {Le Goff}, J.~M. and {Le Guillou}, L. and {Leauthaud}, A. and {Levi}, M.~E. and {Li}, S. and {Li}, T.~S. and {Lodha}, K. and {Lokken}, M. and {Luo}, Y. and {Magneville}, C. and {Manera}, M. and {Manser}, C.~J. and {Margala}, D. and {Martini}, P. and {Maus}, M. and {McCullough}, J. and {McDonald}, P. and {Medina}, G.~E. and {Medina-Varela}, L. and {Meisner}, A. and {Mena-Fern\textbackslash'andez}, J. and {Menegas}, A. and {Mezcua}, M. and {Miquel}, R. and {Montero-Camacho}, P. and {Moon}, J. and {Moustakas}, J. and {Mu\textbackslash\raisebox{-0.5ex}\textasciitildenoz-Guti\textbackslash'errez}, A. and {Mu\textbackslash\raisebox{-0.5ex}\textasciitildenoz-Santos}, D. and {Myers}, A.~D. and {Myles}, J. and {Nadathur}, S. and {Najita}, J. and {Napolitano}, L. and {Newman}, J.~A. and {Nikakhtar}, F. and {Nikutta}, R. and {Niz}, G. and {Noriega}, H.~E. and {Padmanabhan}, N. and {Paillas}, E. and {Palanque-Delabrouille}, N. and {Palmese}, A. and {Pan}, J. and {Pan}, Z. and {Parkinson}, D. and {Peacock}, J. and {Percival}, W.~J. and {P\textbackslash'erez-Fern\textbackslash'andez}, A. and {P\textbackslash'erez-R\textbackslash`afols}, I. and {Peterson}, P.},
        title = "{Data Release 1 of the Dark Energy Spectroscopic Instrument}",
      journal = {arXiv e-prints},
         year = 2025,
        month = mar,
          eid = {arXiv:2503.14745},
        pages = {arXiv:2503.14745},
          doi = {10.48550/arXiv.2503.14745},
archivePrefix = {arXiv},
       eprint = {2503.14745},
 primaryClass = {astro-ph.CO},
       adsurl = {https://ui.adsabs.harvard.edu/abs/2025arXiv250314745D}
}

@ARTICLE{2011AandA...536A..43N,
       author = {{Nebot G{\'o}mez-Mor{\'a}n}, A. and {G{\"a}nsicke}, B.~T. and {Schreiber}, M.~R. and {Rebassa-Mansergas}, A. and {Schwope}, A.~D. and {Southworth}, J. and {Aungwerojwit}, A. and {Bothe}, M. and {Davis}, P.~J. and {Kolb}, U. and {M{\"u}ller}, M. and {Papadaki}, C. and {Pyrzas}, S. and {Rabitz}, A. and {Rodr{\'\i}guez-Gil}, P. and {Schmidtobreick}, L. and {Schwarz}, R. and {Tappert}, C. and {Toloza}, O. and {Vogel}, J. and {Zorotovic}, M.},
        title = "{Post common envelope binaries from SDSS. XII. The orbital period distribution}",
      journal = {\aap},
         year = 2011,
        month = dec,
       volume = {536},
          eid = {A43},
        pages = {A43},
          doi = {10.1051/0004-6361/201117514},
archivePrefix = {arXiv},
       eprint = {1109.6662},
 primaryClass = {astro-ph.SR},
       adsurl = {https://ui.adsabs.harvard.edu/abs/2011A&A...536A..43N}
}

@ARTICLE{2026ApJS..284...38Y,
       author = {{Yang}, Mingkuan and {Yuan}, Hailong and {Yang}, Xiaozhen and {Bai}, Zhongrui and {He}, Yuji and {Xiong}, Jianping and {Li}, Jiao and {Wang}, Mengxin and {Dong}, Yiqiao and {Jiang}, Ziyue and {Liu}, Qian and {Li}, Ganyu and {Zhou}, Ming and {Zhang}, Haotong and {Chen}, Xuefei},
        title = "{Discovery and Characterization of White Dwarf─FGK Main-sequence Binaries within the Optical Main-sequence Locus}",
      journal = {\apjs},
         year = 2026,
        month = jun,
       volume = {284},
       number = {2},
          eid = {38},
        pages = {38},
          doi = {10.3847/1538-4365/ae5bc2},
archivePrefix = {arXiv},
       eprint = {2605.13112},
 primaryClass = {astro-ph.SR},
       adsurl = {https://ui.adsabs.harvard.edu/abs/2026ApJS..284...38Y}
}

@ARTICLE{2026AJ....171..285D,
       author = {{DESI Collaboration} and {Abdul Karim}, M. and {Adame}, A.~G. and {Aguado}, D. and {Aguilar}, J. and {Ahlen}, S. and {Alam}, S. and {Aldering}, G. and {Alexander}, D.~M. and {Alfarsy}, R. and {Allen}, L. and {Allende Prieto}, C. and {Alves}, O. and {Anand}, A. and {Andrade}, U. and {Armengaud}, E. and {Avila}, S. and {Aviles}, A. and {Awan}, H. and {Bailey}, S. and {Baleato Lizancos}, A. and {Ballester}, O. and {Bault}, A. and {Bautista}, J. and {Bean}, R. and {Behera}, J. and {BenZvi}, S. and {Beraldo e Silva}, L. and {Bermejo-Climent}, J.~R. and {Beutler}, F. and {Bianchi}, D. and {Blake}, C. and {Blum}, R. and {Bolton}, A.~S. and {Bonici}, M. and {Brieden}, S. and {Brodzeller}, A. and {Brooks}, D. and {Buckley-Geer}, E. and {Burtin}, E. and {Bystr{\"o}m}, A. and {Canning}, R. and {Carnero Rosell}, A. and {Carr}, A. and {Carrilho}, P. and {Casas}, L. and {Castander}, F.~J. and {Cereskaite}, R. and {Cervantes-Cota}, J.~L. and {Chaussidon}, E. and {Chaves-Montero}, J. and {Chen}, S. and {Chen}, X. and {Circosta}, C. and {Claybaugh}, T. and {Cole}, S. and {Cooper}, A.~P. and {Cousinou}, M.-C. and {Cuceu}, A. and {Davis}, T.~M. and {Dawson}, K.~S. and {de Belsunce}, R. and {de la Cruz}, R. and {de la Macorra}, A. and {de Mattia}, A. and {Deiosso}, N. and {Della Costa}, J. and {Demina}, R. and {Demirbozan}, U. and {DeRose}, J. and {Dey}, A. and {Dey}, B. and {Ding}, J. and {Ding}, Z. and {Doel}, P. and {Douglass}, K. and {Dowicz}, M. and {Ebina}, H. and {Edelstein}, J. and {Eisenstein}, D.~J. and {Elbers}, W. and {Emas}, N. and {Escoffier}, S. and {Fagrelius}, P. and {Fan}, X. and {Fanning}, K. and {Favole}, G. and {Fawcett}, V.~A. and {Fern{\'a}ndez-Garc{\'\i}a}, E. and {Ferraro}, S. and {Findlay}, N. and {Font-Ribera}, A. and {Forero-Romero}, J.~E. and {Forero-S{\'a}nchez}, D. and {Frenk}, C.~S. and {G{\"a}nsicke}, B.~T. and {Galbany}, L. and {Garc{\'\i}a-Bellido}, J. and {Garcia-Quintero}, C. and {Garrison}, L.~H. and {Gazta{\~n}aga}, E. and {Gil-Mar{\'\i}n}, H. and {Gloudemans}, A. and {Gnedin}, O.~Y. and {Gontcho A Gontcho}, S. and {Gonzalez}, D. and {Gonzalez-Morales}, A.~X. and {Gonzalez-Perez}, V. and {Gordon}, C. and {Graur}, O. and {Green}, D. and {Gruen}, D. and {Gsponer}, R. and {Guandalin}, C. and {Gutierrez}, G. and {Guy}, J. and {Hahn}, C. and {Han}, J.~J. and {Han}, J. and {He}, S. and {Herrera-Alcantar}, H.~K. and {Heydenreich}, S. and {Honscheid}, K. and {Hou}, J. and {Howlett}, C. and {Huterer}, D. and {Ir{\v{s}}i{\v{c}}}, V. and {Ishak}, M. and {Jacques}, A. and {Jiang}, L. and {Jimenez}, J. and {Jing}, Y.~P. and {Joachimi}, B. and {Joudaki}, S. and {Joyce}, R. and {Jullo}, E. and {Juneau}, S. and {Kara{\c{c}}ayl{\i}}, N.~G. and {Karim}, T. and {Kehoe}, R. and {Kent}, S. and {Khederlarian}, A. and {Kirkby}, D. and {Kisner}, T. and {Kitaura}, F.-S. and {Kizhuprakkat}, N. and {Kong}, H. and {Koposov}, S.~E. and {Kremin}, A. and {Krolewski}, A. and {Lahav}, O. and {Lai}, Y. and {Lamman}, C. and {Lan}, T.-W. and {Landriau}, M. and {Lang}, D. and {Lange}, J.~U. and {Lasker}, J. and {Le Goff}, J.~M. and {Le Guillou}, L. and {Leauthaud}, A. and {Levi}, M.~E. and {Li}, S. and {Li}, T.~S. and {Liu}, W. and {Lodha}, K. and {Lokken}, M. and {Luo}, Y. and {Magneville}, C. and {Manera}, M. and {Manser}, C.~J. and {Margala}, D. and {Martini}, P. and {Maus}, M. and {McCullough}, J. and {McDonald}, P. and {Medina}, G.~E. and {Medina-Varela}, L. and {Meisner}, A. and {Mena-Fern{\'a}ndez}, J. and {Menegas}, A. and {Meneses-Rizo}, J. and {Mezcua}, M. and {Miquel}, R. and {Montero-Camacho}, P. and {Moon}, J. and {Moustakas}, J. and {Mu{\~n}oz-Guti{\'e}rrez}, A. and {Mu noz-Santos}, D. and {Myers}, A.~D. and {Myles}, J. and {Nadathur}, S. and {Najita}, J. and {Napolitano}, L. and {Newman}, J.~A. and {Nikakhtar}, F. and {Nikutta}, R. and {Niz}, G. and {Noriega}, H.~E. and {Nugent}, P.},
        title = "{Data Release 1 of the Dark Energy Spectroscopic Instrument}",
      journal = {\aj},
         year = 2026,
        month = may,
       volume = {171},
       number = {5},
          eid = {285},
        pages = {285},
          doi = {10.3847/1538-3881/ae4c43},
archivePrefix = {arXiv},
       eprint = {2503.14745},
 primaryClass = {astro-ph.CO},
       adsurl = {https://ui.adsabs.harvard.edu/abs/2026AJ....171..285D}
}

@ARTICLE{2026ApJS..282...26H,
       author = {{Hou}, Wen and {Lin}, Jia-Mao and {Luo}, A.-Li and {Dai}, Zhi-Bin and {Dong}, Yi-Qiao and {Ma}, Shu-Guo},
        title = "{Spectroscopic Identification and Population Properties of Cataclysmic Variables from DESI DR1}",
      journal = {\apjs},
         year = 2026,
        month = feb,
       volume = {282},
       number = {2},
          eid = {26},
        pages = {26},
          doi = {10.3847/1538-4365/ae2024},
       adsurl = {https://ui.adsabs.harvard.edu/abs/2026ApJS..282...26H}
}

@ARTICLE{2025MNRAS.543.2116C,
       author = {{Castro Segura}, N. and {Pelisoli}, I. and {G{\"a}nsicke}, B.~T. and {Coppejans}, D.~L. and {Steeghs}, D. and {Aungwerojwit}, A. and {Inight}, K. and {Romero}, A. and {Sahu}, A. and {Dhillon}, V.~S. and {Munday}, J. and {Parsons}, S.~G. and {Kennedy}, M.~R. and {Green}, M.~J. and {Brown}, A.~J. and {Dyer}, M.~J. and {Pike}, E. and {Garbutt}, J.~A. and {Jarvis}, D. and {Kerry}, P. and {Littlefair}, S.~P. and {McCormac}, J. and {Sahman}, D.~I. and {Buckley}, D.~A.~H.},
        title = "{A sibling of AR Scorpii: SDSS J230641.47+244055.8 and the observational blueprint of white dwarf pulsars}",
      journal = {\mnras},
         year = 2025,
        month = nov,
       volume = {543},
       number = {3},
        pages = {2116-2129},
          doi = {10.1093/mnras/staf1511},
archivePrefix = {arXiv},
       eprint = {2506.20455},
 primaryClass = {astro-ph.SR},
       adsurl = {https://ui.adsabs.harvard.edu/abs/2025MNRAS.543.2116C}
}

@ARTICLE{2025A&A...703A.119T,
       author = {{Tovmassian}, Gagik and {Belloni}, Diogo and {Pala}, Anna F. and {Kupfer}, Thomas and {Yu}, Weitian and {G{\"a}nsicke}, Boris T. and {Waagen}, Elizabeth O. and {Gonz{\'a}lez-Carballo}, Juan-Luis and {Szkody}, Paula and {de Martino}, Domitilla and {Schreiber}, Matthias R. and {Long}, Knox S. and {Bedard}, Alan and {Bednarz}, Slawomir and {Berenguer}, Jordi and {Bernacki}, Krzysztof and {Bolzoni}, Simone and {Botana-Alb{\'a}}, Carlos and {Cantrell}, Christopher and {Cooney}, Walt and {Cynamon}, Charles and {De la Fuente Fern{\'a}ndez}, Pablo and {Dufoer}, Sjoerd and {Ma{\~n}anes}, Esteban Fern{\'a}ndez and {Garc{\'\i}a-Cuesta}, Faustino and {Farf{\'a}n}, Rafael Gonzalez and {Fleurant}, Pierre A. and {G{\'o}mez}, Enrique A. and {Green}, Matthew J. and {Hambsch}, Franz-Josef and {Jordanov}, Penko and {Kardasis}, Emmanuel and {Lane}, David and {Lee}, Darrell and {Lima}, Isabel J. and {Mart{\'\i}nez}, Fernando Lim{\'o}n and {Locatelli}, Gianpiero and {Martin-Velasco}, Jose-Luis and {Mendicini}, Daniel J. and {Michaud}, Michel and {Ort{\'\i}z}, Mois{\'e}s Montero Reyes and {Aimar}, Mario Morales and {Myers}, Gordon and {Nogues}, Ramon Naves and {Pappa}, Giuseppe and {Pearce}, Andrew and {Pierce}, James and {Popowicz}, Adam and {Rodrigues}, Claudia V. and {Rodr{\'\i}guez}, Nieves C. and {Amat}, David Quiles and {Reina-Lorenz}, Esteban and {Salto-Gonz{\'a}lez}, Jos{\'e}-Luis and {Shears}, Jeremy and {Sikora}, John and {Steenkamp}, Andr{\'e} and {Stubbings}, Rod and {Young}, Brad and {Walton}, Ivan L.},
        title = "{Revisiting the extremely long-period cataclysmic variables V479 Andromedae and V1082 Sagittarii}",
      journal = {\aap},
         year = 2025,
        month = nov,
       volume = {703},
          eid = {A119},
        pages = {A119},
          doi = {10.1051/0004-6361/202556385},
archivePrefix = {arXiv},
       eprint = {2508.21358},
 primaryClass = {astro-ph.SR},
       adsurl = {https://ui.adsabs.harvard.edu/abs/2025A&A...703A.119T}
}

@ARTICLE{2025A&A...700A.107G,
       author = {{Green}, Matthew J. and {van Roestel}, Jan and {Wong}, Tin Long Sunny},
        title = "{A catalogue of ultracompact mass-transferring white dwarf binaries}",
      journal = {\aap},
         year = 2025,
        month = aug,
       volume = {700},
          eid = {A107},
        pages = {A107},
          doi = {10.1051/0004-6361/202554925},
archivePrefix = {arXiv},
       eprint = {2505.10535},
 primaryClass = {astro-ph.SR},
       adsurl = {https://ui.adsabs.harvard.edu/abs/2025A&A...700A.107G}
}

@ARTICLE{2025A&A...699A.153R,
       author = {{Rebassa-Mansergas}, Alberto and {Solano}, Enrique and {Brown}, Alex J. and {Parsons}, Steven G. and {Murillo-Ojeda}, Raquel and {Raddi}, Roberto and {Camisassa}, Maria and {Torres}, Santiago and {van Roestel}, Jan},
        title = "{Magnitude-limited catalogue of unresolved white dwarf-main sequence binaries from Gaia DR3}",
      journal = {\aap},
         year = 2025,
        month = jul,
       volume = {699},
          eid = {A153},
        pages = {A153},
          doi = {10.1051/0004-6361/202554700},
archivePrefix = {arXiv},
       eprint = {2505.15895},
 primaryClass = {astro-ph.SR},
       adsurl = {https://ui.adsabs.harvard.edu/abs/2025A&A...699A.153R}
}

@ARTICLE{2025A&A...698L..22S,
       author = {{Schreiber}, Matthias R. and {Belloni}, Diogo},
        title = "{The incidence of magnetic cataclysmic variables can be explained by the late appearance of white dwarf magnetic fields}",
      journal = {\aap},
         year = 2025,
        month = jun,
       volume = {698},
          eid = {L22},
        pages = {L22},
          doi = {10.1051/0004-6361/202554828},
archivePrefix = {arXiv},
       eprint = {2505.24153},
 primaryClass = {astro-ph.SR},
       adsurl = {https://ui.adsabs.harvard.edu/abs/2025A&A...698L..22S}
}

@ARTICLE{2025A&A...698A.106S,
       author = {{Schwope}, Axel D.},
        title = "{PolarCat: Catalog of polars, low-accretion rate polars, and candidate objects}",
      journal = {\aap},
         year = 2025,
        month = jun,
       volume = {698},
          eid = {A106},
        pages = {A106},
          doi = {10.1051/0004-6361/202554519},
archivePrefix = {arXiv},
       eprint = {2505.10337},
 primaryClass = {astro-ph.SR},
       adsurl = {https://ui.adsabs.harvard.edu/abs/2025A&A...698A.106S}
}

@ARTICLE{2025MNRAS.537.3078A,
       author = {{Aungwerojwit}, Amornrat and {G{\"a}nsicke}, Boris T. and {Breedt}, E. and {Arjyotha}, S. and {Hermes}, J.~J. and {Hambsch}, F.-J. and {Kumar}, A. and {Ram{\'\i}rez}, S.~H. and {Wilson}, T.~G. and {Dhillon}, V.~S. and {Marsh}, T.~R. and {Poshyachinda}, S. and {Scaringi}, S. and {Haislip}, J.~B. and {Reichart}, D.~E.},
        title = "{Follow-up on three poorly studied AM CVn stars}",
      journal = {\mnras},
         year = 2025,
        month = mar,
       volume = {537},
       number = {4},
        pages = {3078-3090},
          doi = {10.1093/mnras/staf173},
archivePrefix = {arXiv},
       eprint = {2501.16424},
 primaryClass = {astro-ph.SR},
       adsurl = {https://ui.adsabs.harvard.edu/abs/2025MNRAS.537.3078A}
}

@ARTICLE{2025MNRAS.536.1057I,
       author = {{Inight}, Keith and {G{\"a}nsicke}, Boris T. and {Schwope}, Axel and {Anderson}, Scott F. and {Breedt}, Elm{\'e} and {Brownstein}, Joel R. and {Demasi}, Sebastian and {Friedrich}, Susanne and {Hermes}, J.~J. and {Long}, Knox S. and {Mulvany}, Timothy and {Adamane Pallathadka}, Gautham and {Salvato}, Mara and {Scaringi}, Simone and {Schreiber}, Matthias R. and {Stringfellow}, Guy S. and {Thorstensen}, John R. and {Tovmassian}, Gagik and {Zakamska}, Nadia L.},
        title = "{Cataclysmic variables from Sloan Digital Sky Survey - V (2020-2023) identified using machine learning}",
      journal = {\mnras},
         year = 2025,
        month = jan,
       volume = {536},
       number = {2},
        pages = {1057-1076},
          doi = {10.1093/mnras/stae2524},
archivePrefix = {arXiv},
       eprint = {2406.19459},
 primaryClass = {astro-ph.SR},
       adsurl = {https://ui.adsabs.harvard.edu/abs/2025MNRAS.536.1057I}
}

@ARTICLE{2024AJ....168..245P,
       author = {{Poppett}, Claire and {Tyas}, Luke and {Aguilar}, J. and {Bebek}, Christopher and {Bramall}, D. and {Claybaugh}, T. and {Edelstein}, J. and {Fagrelius}, P. and {Heetderks}, H. and {Jelinsky}, P. and {Jelinsky}, S. and {Lafever}, Robin and {Lambert}, A. and {Lampton}, M. and {Levi}, Michael E. and {Martini}, P. and {Rockosi}, C. and {Schmoll}, J. and {Sharples}, Ray M. and {Sirk}, Martin and {Wishnow}, Edward and {Yu}, Jiaxi and {Ahlen}, S. and {Bault}, A. and {BenZvi}, S. and {Brooks}, D. and {Cole}, S. and {de la Macorra}, A. and {Dey}, Arjun and {Doel}, P. and {Fanning}, K. and {Font-Ribera}, A. and {Forero-Romero}, J.~E. and {Gazta{\~n}aga}, E. and {Gontcho A Gontcho}, S. and {Gonzalez-Morales}, A.~X. and {Hahn}, C. and {Honscheid}, K. and {Jimenez}, J. and {Juneau}, S. and {Kirkby}, D. and {Kremin}, A. and {Landriau}, M. and {Le Guillou}, L. and {Manera}, M. and {Meisner}, A. and {Miquel}, R. and {Moustakas}, J. and {Mueller}, E. and {Mu{\~n}oz-Guti{\'e}rrez}, A. and {Myers}, A.~D. and {Nie}, J. and {Niz}, G. and {Palanque-Delabrouille}, N. and {Percival}, W.~J. and {Prada}, F. and {Rabinowitz}, D. and {Rezaie}, M. and {Rossi}, G. and {Sanchez}, E. and {Schlafly}, Edward F. and {Schlegel}, D. and {Schubnell}, M. and {Seo}, H. and {Sprayberry}, D. and {Tarl{\'e}}, G. and {Vargas-Maga{\~n}a}, M. and {Weaver}, B.~A. and {Zhou}, R.},
        title = "{Overview of the Fiber System for the Dark Energy Spectroscopic Instrument}",
      journal = {\aj},
         year = 2024,
        month = dec,
       volume = {168},
       number = {6},
          eid = {245},
        pages = {245},
          doi = {10.3847/1538-3881/ad76a4},
       adsurl = {https://ui.adsabs.harvard.edu/abs/2024AJ....168..245P}
}

@ARTICLE{2024A&A...690L...1O,
       author = {{Ort{\'u}zar-Garz{\'o}n}, Valentina and {Schreiber}, Matthias R. and {Belloni}, Diogo},
        title = "{Suggested magnetic braking prescription derived from field complexity fails to reproduce the cataclysmic variable orbital period gap}",
      journal = {\aap},
         year = 2024,
        month = oct,
       volume = {690},
          eid = {L1},
        pages = {L1},
          doi = {10.1051/0004-6361/202451829},
archivePrefix = {arXiv},
       eprint = {2409.05673},
 primaryClass = {astro-ph.SR},
       adsurl = {https://ui.adsabs.harvard.edu/abs/2024A&A...690L...1O}
}

@ARTICLE{2024AJ....168...58D,
       author = {{DESI Collaboration} and {Adame}, A.~G. and {Aguilar}, J. and {Ahlen}, S. and {Alam}, S. and {Aldering}, G. and {Alexander}, D.~M. and {Alfarsy}, R. and {Allende Prieto}, C. and {Alvarez}, M. and {Alves}, O. and {Anand}, A. and {Andrade-Oliveira}, F. and {Armengaud}, E. and {Asorey}, J. and {Avila}, S. and {Aviles}, A. and {Bailey}, S. and {Balaguera-Antol{\'\i}nez}, A. and {Ballester}, O. and {Baltay}, C. and {Bault}, A. and {Bautista}, J. and {Behera}, J. and {Beltran}, S.~F. and {BenZvi}, S. and {Beraldo e Silva}, L. and {Bermejo-Climent}, J.~R. and {Berti}, A. and {Besuner}, R. and {Beutler}, F. and {Bianchi}, D. and {Blake}, C. and {Blum}, R. and {Bolton}, A.~S. and {Brieden}, S. and {Brodzeller}, A. and {Brooks}, D. and {Brown}, Z. and {Buckley-Geer}, E. and {Burtin}, E. and {Cabayol-Garcia}, L. and {Cai}, Z. and {Canning}, R. and {Cardiel-Sas}, L. and {Carnero Rosell}, A. and {Castander}, F.~J. and {Cervantes-Cota}, J.~L. and {Chabanier}, S. and {Chaussidon}, E. and {Chaves-Montero}, J. and {Chen}, S. and {Chen}, X. and {Chuang}, C. and {Claybaugh}, T. and {Cole}, S. and {Cooper}, A.~P. and {Cuceu}, A. and {Davis}, T.~M. and {Dawson}, K. and {de Belsunce}, R. and {de la Cruz}, R. and {de la Macorra}, A. and {Della Costa}, J. and {de Mattia}, A. and {Demina}, R. and {Demirbozan}, U. and {DeRose}, J. and {Dey}, A. and {Dey}, B. and {Dhungana}, G. and {Ding}, J. and {Ding}, Z. and {Doel}, P. and {Doshi}, R. and {Douglass}, K. and {Edge}, A. and {Eftekharzadeh}, S. and {Eisenstein}, D.~J. and {Elliott}, A. and {Ereza}, J. and {Escoffier}, S. and {Fagrelius}, P. and {Fan}, X. and {Fanning}, K. and {Fawcett}, V.~A. and {Ferraro}, S. and {Flaugher}, B. and {Font-Ribera}, A. and {Forero-Romero}, J.~E. and {Forero-S{\'a}nchez}, D. and {Frenk}, C.~S. and {G{\"a}nsicke}, B.~T. and {Garc{\'\i}a}, L. {\'A}. and {Garc{\'\i}a-Bellido}, J. and {Garcia-Quintero}, C. and {Garrison}, L.~H. and {Gil-Mar{\'\i}n}, H. and {Golden-Marx}, J. and {Gontcho A Gontcho}, S. and {Gonzalez-Morales}, A.~X. and {Gonzalez-Perez}, V. and {Gordon}, C. and {Graur}, O. and {Green}, D. and {Gruen}, D. and {Guy}, J. and {Hadzhiyska}, B. and {Hahn}, C. and {Han}, J.~J. and {Hanif}, M.~M.~S. and {Herrera-Alcantar}, H.~K. and {Honscheid}, K. and {Hou}, J. and {Howlett}, C. and {Huterer}, D. and {Ir{\v{s}}i{\v{c}}}, V. and {Ishak}, M. and {Jacques}, A. and {Jana}, A. and {Jiang}, L. and {Jimenez}, J. and {Jing}, Y.~P. and {Joudaki}, S. and {Joyce}, R. and {Jullo}, E. and {Juneau}, S. and {Kara{\c{c}}ayl{\i}}, N.~G. and {Karim}, T. and {Kehoe}, R. and {Kent}, S. and {Khederlarian}, A. and {Kim}, S. and {Kirkby}, D. and {Kisner}, T. and {Kitaura}, F. and {Kizhuprakkat}, N. and {Kneib}, J. and {Koposov}, S.~E. and {Kov{\'a}cs}, A. and {Kremin}, A. and {Krolewski}, A. and {L'Huillier}, B. and {Lahav}, O. and {Lambert}, A. and {Lamman}, C. and {Lan}, T.-W. and {Landriau}, M. and {Lang}, D. and {Lange}, J.~U. and {Lasker}, J. and {Leauthaud}, A. and {Le Guillou}, L. and {Levi}, M.~E. and {Li}, T.~S. and {Linder}, E. and {Lyons}, A. and {Magneville}, C. and {Manera}, M. and {Manser}, C.~J. and {Margala}, D. and {Martini}, P. and {McDonald}, P. and {Medina}, G.~E. and {Medina-Varela}, L. and {Meisner}, A. and {Mena-Fern{\'a}ndez}, J. and {Meneses-Rizo}, J. and {Mezcua}, M. and {Miquel}, R. and {Montero-Camacho}, P. and {Moon}, J. and {Moore}, S. and {Moustakas}, J. and {Mueller}, E. and {Mundet}, J. and {Mu{\~n}oz-Guti{\'e}rrez}, A. and {Myers}, A.~D. and {Nadathur}, S. and {Napolitano}, L. and {Neveux}, R. and {Newman}, J.~A. and {Nie}, J. and {Nikutta}, R. and {Niz}, G. and {Norberg}, P. and {Noriega}, H.~E. and {Paillas}, E. and {Palanque-Delabrouille}, N. and {Palmese}, A. and {Pan}, Z. and {Parkinson}, D. and {Penmetsa}, S. and {Percival}, W.~J. and {P{\'e}rez-Fern{\'a}ndez}, A. and {P{\'e}rez-R{\`a}fols}, I. and {Pieri}, M. and {Poppett}, C. and {Porredon}, A. and {Pothier}, S.},
        title = "{The Early Data Release of the Dark Energy Spectroscopic Instrument}",
      journal = {\aj},
         year = 2024,
        month = aug,
       volume = {168},
       number = {2},
          eid = {58},
        pages = {58},
          doi = {10.3847/1538-3881/ad3217},
archivePrefix = {arXiv},
       eprint = {2306.06308},
 primaryClass = {astro-ph.CO},
       adsurl = {https://ui.adsabs.harvard.edu/abs/2024AJ....168...58D}
}

@ARTICLE{2024MNRAS.529.4840G,
       author = {{Garbutt}, J.~A. and {Parsons}, S.~G. and {Toloza}, O. and {G{\"a}nsicke}, B.~T. and {Hernandez}, M.~S. and {Koester}, D. and {Lagos}, F. and {Raddi}, R. and {Rebassa-Mansergas}, A. and {Ren}, J.~J. and {Schreiber}, M.~R. and {Zorotovic}, M.},
        title = "{The white dwarf binary pathways survey - X. Gaia orbits for known UV excess binaries}",
      journal = {\mnras},
         year = 2024,
        month = apr,
       volume = {529},
       number = {4},
        pages = {4840-4855},
          doi = {10.1093/mnras/stae807},
archivePrefix = {arXiv},
       eprint = {2403.07985},
 primaryClass = {astro-ph.SR},
       adsurl = {https://ui.adsabs.harvard.edu/abs/2024MNRAS.529.4840G}
}

@ARTICLE{2024ApJS..271...55K,
       author = {{Kirkpatrick}, J. Davy and {Marocco}, Federico and {Gelino}, Christopher R. and {Raghu}, Yadukrishna and {Faherty}, Jacqueline K. and {Bardalez Gagliuffi}, Daniella C. and {Schurr}, Steven D. and {Apps}, Kevin and {Schneider}, Adam C. and {Meisner}, Aaron M. and {Kuchner}, Marc J. and {Caselden}, Dan and {Smart}, R.~L. and {Casewell}, S.~L. and {Raddi}, Roberto and {Kesseli}, Aurora and {Stevnbak Andersen}, Nikolaj and {Antonini}, Edoardo and {Beaulieu}, Paul and {Bickle}, Thomas P. and {Bilsing}, Martin and {Chieng}, Raymond and {Colin}, Guillaume and {Deen}, Sam and {Dereveanco}, Alexandru and {Doll}, Katharina and {Durantini Luca}, Hugo A. and {Frazer}, Anya and {Gantier}, Jean Marc and {Gramaize}, L{\'e}opold and {Grant}, Kristin and {Hamlet}, Leslie K. and {Higashimura}, Hiro and {Hyogo}, Michiharu and {Ja{\l}owiczor}, Peter A. and {Jonkeren}, Alexander and {Kabatnik}, Martin and {Kiwy}, Frank and {Martin}, David W. and {Michaels}, Marianne N. and {Pendrill}, William and {Pessanha Machado}, Celso and {Pumphrey}, Benjamin and {Rothermich}, Austin and {Russwurm}, Rebekah and {Sainio}, Arttu and {Sanchez}, John and {Sapelkin-Tambling}, Fyodor Theo and {Sch{\"u}mann}, J{\"o}rg and {Selg-Mann}, Karl and {Singh}, Harshdeep and {Stenner}, Andres and {Sun}, Guoyou and {Tanner}, Christopher and {Th{\'e}venot}, Melina and {Ventura}, Maurizio and {Voloshin}, Nikita V. and {Walla}, Jim and {W{\k{e}}dracki}, Zbigniew and {Adorno}, Jose I. and {Aganze}, Christian and {Allers}, Katelyn N. and {Brooks}, Hunter and {Burgasser}, Adam J. and {Calamari}, Emily and {Connor}, Thomas and {Costa}, Edgardo and {Eisenhardt}, Peter R. and {Gagn{\'e}}, Jonathan and {Gerasimov}, Roman and {Gonzales}, Eileen C. and {Hsu}, Chih-Chun and {Kiman}, Rocio and {Li}, Guodong and {Low}, Ryan and {Mamajek}, Eric and {Pantoja}, Blake M. and {Popinchalk}, Mark and {Rees}, Jon M. and {Stern}, Daniel and {Su{\'a}rez}, Genaro and {Theissen}, Christopher and {Tsai}, Chao-Wei and {Vos}, Johanna M. and {Zurek}, David and {The Backyard Worlds: Planet 9 Collaboration}},
        title = "{The Initial Mass Function Based on the Full-sky 20 pc Census of {\ensuremath{\sim}}3600 Stars and Brown Dwarfs}",
      journal = {\apjs},
         year = 2024,
        month = apr,
       volume = {271},
       number = {2},
          eid = {55},
        pages = {55},
          doi = {10.3847/1538-4365/ad24e2},
archivePrefix = {arXiv},
       eprint = {2312.03639},
 primaryClass = {astro-ph.SR},
       adsurl = {https://ui.adsabs.harvard.edu/abs/2024ApJS..271...55K}
}

@ARTICLE{2024ApJ...965...96M,
       author = {{Mason}, Paul A. and {Hakala}, Pasi and {Wu}, Kinwah and {Barrett}, Paul E. and {I{\l}kiewicz}, Krystian and {Littlefield}, Colin and {Monroy}, Lorena C. and {Sezer}, Hasan C. and {Jablonski}, Francisco and {Garnavich}, Peter and {Szkody}, Paula and {Ramsay}, Gavin and {Duffy}, Christopher and {Scaringi}, Simone},
        title = "{TESS Photometry of AM Her and AR UMa: Binary Parameters, Cyclotron Emission Modeling, and Mass Transfer Duty Cycles}",
      journal = {\apj},
         year = 2024,
        month = apr,
       volume = {965},
       number = {1},
          eid = {96},
        pages = {96},
          doi = {10.3847/1538-4357/ad27d7},
       adsurl = {https://ui.adsabs.harvard.edu/abs/2024ApJ...965...96M}
}

@ARTICLE{2024AJ....167...62D,
       author = {{DESI Collaboration} and {Adame}, A.~G. and {Aguilar}, J. and {Ahlen}, S. and {Alam}, S. and {Aldering}, G. and {Alexander}, D.~M. and {Alfarsy}, R. and {Allende Prieto}, C. and {Alvarez}, M. and {Alves}, O. and {Anand}, A. and {Andrade-Oliveira}, F. and {Armengaud}, E. and {Asorey}, J. and {Avila}, S. and {Aviles}, A. and {Bailey}, S. and {Balaguera-Antol{\'\i}nez}, A. and {Ballester}, O. and {Baltay}, C. and {Bault}, A. and {Bautista}, J. and {Behera}, J. and {Beltran}, S.~F. and {BenZvi}, S. and {Beraldo e Silva}, L. and {Bermejo-Climent}, J.~R. and {Berti}, A. and {Besuner}, R. and {Beutler}, F. and {Bianchi}, D. and {Blake}, C. and {Blum}, R. and {Bolton}, A.~S. and {Brieden}, S. and {Brodzeller}, A. and {Brooks}, D. and {Brown}, Z. and {Buckley-Geer}, E. and {Burtin}, E. and {Cabayol-Garcia}, L. and {Cai}, Z. and {Canning}, R. and {Cardiel-Sas}, L. and {Carnero Rosell}, A. and {Castander}, F.~J. and {Cervantes-Cota}, J.~L. and {Chabanier}, S. and {Chaussidon}, E. and {Chaves-Montero}, J. and {Chen}, S. and {Chen}, X. and {Chuang}, C. and {Claybaugh}, T. and {Cole}, S. and {Cooper}, A.~P. and {Cuceu}, A. and {Davis}, T.~M. and {Dawson}, K. and {de Belsunce}, R. and {de la Cruz}, R. and {de la Macorra}, A. and {de Mattia}, A. and {Demina}, R. and {Demirbozan}, U. and {DeRose}, J. and {Dey}, A. and {Dey}, B. and {Dhungana}, G. and {Ding}, J. and {Ding}, Z. and {Doel}, P. and {Doshi}, R. and {Douglass}, K. and {Edge}, A. and {Eftekharzadeh}, S. and {Eisenstein}, D.~J. and {Elliott}, A. and {Escoffier}, S. and {Fagrelius}, P. and {Fan}, X. and {Fanning}, K. and {Fawcett}, V.~A. and {Ferraro}, S. and {Ereza}, J. and {Flaugher}, B. and {Font-Ribera}, A. and {Forero-S{\'a}nchez}, D. and {Forero-Romero}, J.~E. and {Frenk}, C.~S. and {G{\"a}nsicke}, B.~T. and {Garc{\'\i}a}, L. {\'A}. and {Garc{\'\i}a-Bellido}, J. and {Garcia-Quintero}, C. and {Garrison}, L.~H. and {Gil-Mar{\'\i}n}, H. and {Golden-Marx}, J. and {Gontcho A Gontcho}, S. and {Gonzalez-Morales}, A.~X. and {Gonzalez-Perez}, V. and {Gordon}, C. and {Graur}, O. and {Green}, D. and {Gruen}, D. and {Guy}, J. and {Hadzhiyska}, B. and {Hahn}, C. and {Han}, J.~J. and {Hanif}, M.~M.~S. and {Herrera-Alcantar}, H.~K. and {Honscheid}, K. and {Hou}, J. and {Howlett}, C. and {Huterer}, D. and {Ir{\v{s}}i{\v{c}}}, V. and {Ishak}, M. and {Jana}, A. and {Jiang}, L. and {Jimenez}, J. and {Jing}, Y.~P. and {Joudaki}, S. and {Jullo}, E. and {Joyce}, R. and {Juneau}, S. and {Kizhuprakkat}, N. and {Kara{\c{c}}ayl{\i}}, N.~G. and {Karim}, T. and {Kehoe}, R. and {Kent}, S. and {Khederlarian}, A. and {Kim}, S. and {Kirkby}, D. and {Kisner}, T. and {Kitaura}, F. and {Kneib}, J. and {Koposov}, S.~E. and {Kov{\'a}cs}, A. and {Kremin}, A. and {Krolewski}, A. and {L'Huillier}, B. and {Lahav}, O. and {Lambert}, A. and {Lamman}, C. and {Lan}, T.-W. and {Landriau}, M. and {Lang}, D. and {Lange}, J.~U. and {Lasker}, J. and {Le Guillou}, L. and {Leauthaud}, A. and {Levi}, M.~E. and {Li}, T.~S. and {Linder}, E. and {Lyons}, A. and {Magneville}, C. and {Manera}, M. and {Manser}, C.~J. and {Margala}, D. and {Martini}, P. and {McDonald}, P. and {Medina}, G.~E. and {Medina-Varela}, L. and {Meisner}, A. and {Mena-Fern{\'a}ndez}, J. and {Meneses-Rizo}, J. and {Mezcua}, M. and {Miquel}, R. and {Montero-Camacho}, P. and {Moon}, J. and {Moore}, S. and {Moustakas}, J. and {Mueller}, E. and {Mundet}, J. and {Mu{\~n}oz-Guti{\'e}rrez}, A. and {Myers}, A.~D. and {Nadathur}, S. and {Napolitano}, L. and {Neveux}, R. and {Newman}, J.~A. and {Nie}, J. and {Niz}, G. and {Norberg}, P. and {Noriega}, H.~E. and {Paillas}, E. and {Palanque-Delabrouille}, N. and {Palmese}, A. and {Zhiwei}, P. and {Parkinson}, D. and {Penmetsa}, S. and {Percival}, W.~J. and {P{\'e}rez-Fern{\'a}ndez}, A. and {P{\'e}rez-R{\`a}fols}, I. and {Pieri}, M. and {Poppett}, C. and {Porredon}, A. and {Prada}, F. and {Pucha}, R. and {Raichoor}, A. and {Ram{\'\i}rez-P{\'e}rez}, C.},
        title = "{Validation of the Scientific Program for the Dark Energy Spectroscopic Instrument}",
      journal = {\aj},
         year = 2024,
        month = feb,
       volume = {167},
       number = {2},
          eid = {62},
        pages = {62},
          doi = {10.3847/1538-3881/ad0b08},
archivePrefix = {arXiv},
       eprint = {2306.06307},
 primaryClass = {astro-ph.CO},
       adsurl = {https://ui.adsabs.harvard.edu/abs/2024AJ....167...62D}
}

@ARTICLE{2024A&A...682L...7S,
       author = {{Schreiber}, Matthias R. and {Belloni}, Diogo and {Schwope}, Axel D.},
        title = "{The cataclysmic variable orbital period gap: More evident than ever}",
      journal = {\aap},
         year = 2024,
        month = feb,
       volume = {682},
          eid = {L7},
        pages = {L7},
          doi = {10.1051/0004-6361/202348807},
archivePrefix = {arXiv},
       eprint = {2402.02076},
 primaryClass = {astro-ph.SR},
       adsurl = {https://ui.adsabs.harvard.edu/abs/2024A&A...682L...7S}
}

@ARTICLE{2023MNRAS.525.3597I,
       author = {{Inight}, K. and {G{\"a}nsicke}, Boris T. and {Schwope}, A. and {Anderson}, S.~F. and {Badenes}, C. and {Breedt}, E. and {Chandra}, V. and {Davies}, B.~D.~R. and {Gentile Fusillo}, N.~P. and {Green}, M.~J. and {Hermes}, J.~J. and {Huamani}, I. Achaica and {Hwang}, H. and {Knauff}, K. and {Kurpas}, J. and {Long}, K.~S. and {Malanushenko}, V. and {Morrison}, S. and {Quiroz C.}, I.~J. and {Ramos}, G.~N. Aichele and {Roman-Lopes}, A. and {Schreiber}, M.~R. and {Standke}, A. and {St{\"u}tz}, L. and {Thorstensen}, J.~R. and {Toloza}, O. and {Tovmassian}, G. and {Zakamska}, N.~L.},
        title = "{Cataclysmic Variables from Sloan Digital Sky Survey - V. The search for period bouncers continues}",
      journal = {\mnras},
         year = 2023,
        month = nov,
       volume = {525},
       number = {3},
        pages = {3597-3625},
          doi = {10.1093/mnras/stad2409},
archivePrefix = {arXiv},
       eprint = {2305.13371},
 primaryClass = {astro-ph.SR},
       adsurl = {https://ui.adsabs.harvard.edu/abs/2023MNRAS.525.3597I}
}

@ARTICLE{2023ApJS..269....3M,
       author = {{Moustakas}, John and {Lang}, Dustin and {Dey}, Arjun and {Juneau}, St{\'e}phanie and {Meisner}, Aaron and {Myers}, Adam D. and {Schlafly}, Edward F. and {Schlegel}, David J. and {Valdes}, Francisco and {Weaver}, Benjamin A. and {Zhou}, Rongpu},
        title = "{Siena Galaxy Atlas 2020}",
      journal = {\apjs},
         year = 2023,
        month = nov,
       volume = {269},
       number = {1},
          eid = {3},
        pages = {3},
          doi = {10.3847/1538-4365/acfaa2},
archivePrefix = {arXiv},
       eprint = {2307.04888},
 primaryClass = {astro-ph.GA},
       adsurl = {https://ui.adsabs.harvard.edu/abs/2023ApJS..269....3M}
}

@ARTICLE{2023MNRAS.525L..50S,
       author = {{Scaringi}, S. and {Breivik}, K. and {Littenberg}, T.~B. and {Knigge}, C. and {Groot}, P.~J. and {Veresvarska}, M.},
        title = "{Cataclysmic variables are a key population of gravitational wave sources for LISA}",
      journal = {\mnras},
         year = 2023,
        month = oct,
       volume = {525},
       number = {1},
        pages = {L50-L55},
          doi = {10.1093/mnrasl/slad093},
archivePrefix = {arXiv},
       eprint = {2307.02553},
 primaryClass = {astro-ph.HE},
       adsurl = {https://ui.adsabs.harvard.edu/abs/2023MNRAS.525L..50S}
}

@ARTICLE{2023MNRAS.524.4867I,
       author = {{Inight}, Keith and {G{\"a}nsicke}, Boris T. and {Breedt}, Elm{\'e} and {Israel}, Henry T. and {Littlefair}, Stuart P. and {Manser}, Christopher J. and {Marsh}, Tom R. and {Mulvany}, Tim and {Pala}, Anna Francesca and {Thorstensen}, John R.},
        title = "{A catalogue of cataclysmic variables from 20 yr of the Sloan Digital Sky Survey with new classifications, periods, trends, and oddities}",
      journal = {\mnras},
         year = 2023,
        month = oct,
       volume = {524},
       number = {4},
        pages = {4867-4898},
          doi = {10.1093/mnras/stad2018},
archivePrefix = {arXiv},
       eprint = {2304.06749},
 primaryClass = {astro-ph.SR},
       adsurl = {https://ui.adsabs.harvard.edu/abs/2023MNRAS.524.4867I}
}

@ARTICLE{2023MNRAS.523..305T,
       author = {{Toloza}, Odette and {G{\"a}nsicke}, Boris T. and {Guzm{\'a}n-Rinc{\'o}n}, Laura M. and {Marsh}, Tom R. and {Szkody}, Paula and {Schreiber}, Matthias R. and {de Martino}, Domitilla and {Zorotovic}, Monica and {El-Badry}, Kareem and {Koester}, Detlev and {Lagos}, Felipe},
        title = "{The C/N ratio from FUV spectroscopy as a constraint on evolution of the dwarf nova HS 0218 + 3229}",
      journal = {\mnras},
         year = 2023,
        month = jul,
       volume = {523},
       number = {1},
        pages = {305-326},
          doi = {10.1093/mnras/stad1306},
archivePrefix = {arXiv},
       eprint = {2209.06873},
 primaryClass = {astro-ph.SR},
       adsurl = {https://ui.adsabs.harvard.edu/abs/2023MNRAS.523..305T}
}

@ARTICLE{2023AJ....165..253H,
       author = {{Hahn}, ChangHoon and {Wilson}, Michael J. and {Ruiz-Macias}, Omar and {Cole}, Shaun and {Weinberg}, David H. and {Moustakas}, John and {Kremin}, Anthony and {Tinker}, Jeremy L. and {Smith}, Alex and {Wechsler}, Risa H. and {Ahlen}, Steven and {Alam}, Shadab and {Bailey}, Stephen and {Brooks}, David and {Cooper}, Andrew P. and {Davis}, Tamara M. and {Dawson}, Kyle and {Dey}, Arjun and {Dey}, Biprateep and {Eftekharzadeh}, Sarah and {Eisenstein}, Daniel J. and {Fanning}, Kevin and {Forero-Romero}, Jaime E. and {Frenk}, Carlos S. and {Gazta{\~n}aga}, Enrique and {A Gontcho}, Satya Gontcho and {Guy}, Julien and {Honscheid}, Klaus and {Ishak}, Mustapha and {Juneau}, St{\'e}phanie and {Kehoe}, Robert and {Kisner}, Theodore and {Lan}, Ting-Wen and {Landriau}, Martin and {Le Guillou}, Laurent and {Levi}, Michael E. and {Magneville}, Christophe and {Martini}, Paul and {Meisner}, Aaron and {Myers}, Adam D. and {Nie}, Jundan and {Norberg}, Peder and {Palanque-Delabrouille}, Nathalie and {Percival}, Will J. and {Poppett}, Claire and {Prada}, Francisco and {Raichoor}, Anand and {Ross}, Ashley J. and {Gaines}, Sasha and {Saulder}, Christoph and {Schlafly}, Eddie and {Schlegel}, David and {Sierra-Porta}, David and {Tarle}, Gregory and {Weaver}, Benjamin A. and {Y{\`e}che}, Christophe and {Zarrouk}, Pauline and {Zhou}, Rongpu and {Zhou}, Zhimin and {Zou}, Hu},
        title = "{The DESI Bright Galaxy Survey: Final Target Selection, Design, and Validation}",
      journal = {\aj},
         year = 2023,
        month = jun,
       volume = {165},
       number = {6},
          eid = {253},
        pages = {253},
          doi = {10.3847/1538-3881/accff8},
archivePrefix = {arXiv},
       eprint = {2208.08512},
 primaryClass = {astro-ph.CO},
       adsurl = {https://ui.adsabs.harvard.edu/abs/2023AJ....165..253H}
}

@ARTICLE{2023ApJ...947...37C,
       author = {{Cooper}, Andrew P. and {Koposov}, Sergey E. and {Allende Prieto}, Carlos and {Manser}, Christopher J. and {Kizhuprakkat}, Namitha and {Myers}, Adam D. and {Dey}, Arjun and {G{\"a}nsicke}, Boris T. and {Li}, Ting S. and {Rockosi}, Constance and {Valluri}, Monica and {Najita}, Joan and {Deason}, Alis and {Raichoor}, Anand and {Wang}, M.-Y. and {Ting}, Y.-S. and {Kim}, Bokyoung and {Carrillo}, Andreia and {Wang}, Wenting and {Beraldo e Silva}, Leandro and {Han}, Jiwon Jesse and {Ding}, Jiani and {S{\'a}nchez-Conde}, Miguel and {Aguilar}, Jessica N. and {Ahlen}, Steven and {Bailey}, Stephen and {Belokurov}, Vasily and {Brooks}, David and {Cunha}, Katia and {Dawson}, Kyle and {de la Macorra}, Axel and {Doel}, Peter and {Eisenstein}, Daniel J. and {Fagrelius}, Parker and {Fanning}, Kevin and {Font-Ribera}, Andreu and {Forero-Romero}, Jaime E. and {Gazta{\~n}aga}, Enrique and {Gontcho a Gontcho}, Satya and {Guy}, Julien and {Honscheid}, Klaus and {Kehoe}, Robert and {Kisner}, Theodore and {Kremin}, Anthony and {Landriau}, Martin and {Levi}, Michael E. and {Martini}, Paul and {Meisner}, Aaron M. and {Miquel}, Ramon and {Moustakas}, John and {Nie}, Jundan J.~D. and {Palanque-Delabrouille}, Nathalie and {Percival}, Will J. and {Poppett}, Claire and {Prada}, Francisco and {Rehemtulla}, Nabeel and {Schlafly}, Edward and {Schlegel}, David and {Schubnell}, Michael and {Sharples}, Ray M. and {Tarl{\'e}}, Gregory and {Wechsler}, Risa H. and {Weinberg}, David H. and {Zhou}, Zhimin and {Zou}, Hu},
        title = "{Overview of the DESI Milky Way Survey}",
      journal = {\apj},
         year = 2023,
        month = apr,
       volume = {947},
       number = {1},
          eid = {37},
        pages = {37},
          doi = {10.3847/1538-4357/acb3c0},
archivePrefix = {arXiv},
       eprint = {2208.08514},
 primaryClass = {astro-ph.GA},
       adsurl = {https://ui.adsabs.harvard.edu/abs/2023ApJ...947...37C}
}

@ARTICLE{2023AJ....165..126R,
       author = {{Raichoor}, A. and {Moustakas}, J. and {Newman}, Jeffrey A. and {Karim}, T. and {Ahlen}, S. and {Alam}, Shadab and {Bailey}, S. and {Brooks}, D. and {Dawson}, K. and {de la Macorra}, A. and {de Mattia}, A. and {Dey}, A. and {Dey}, Biprateep and {Dhungana}, G. and {Eftekharzadeh}, S. and {Eisenstein}, D.~J. and {Fanning}, K. and {Font-Ribera}, A. and {Garc{\'\i}a-Bellido}, J. and {Gazta{\~n}aga}, E. and {A Gontcho}, S. Gontcho and {Guy}, J. and {Honscheid}, K. and {Ishak}, M. and {Kehoe}, R. and {Kisner}, T. and {Kremin}, Anthony and {Lan}, Ting-Wen and {Landriau}, M. and {Le Guillou}, L. and {Levi}, Michael E. and {Magneville}, C. and {Manera}, M. and {Martini}, P. and {Meisner}, Aaron M. and {Myers}, Adam D. and {Nie}, Jundan and {Palanque-Delabrouille}, N. and {Percival}, W.~J. and {Poppett}, C. and {Prada}, F. and {Ross}, A.~J. and {Ruhlmann-Kleider}, V. and {Sabiu}, C.~G. and {Schlafly}, E.~F. and {Schlegel}, D. and {Tarl{\'e}}, Gregory and {Weaver}, B.~A. and {Y{\`e}che}, Christophe and {Zhou}, Rongpu and {Zhou}, Zhimin and {Zou}, H.},
        title = "{Target Selection and Validation of DESI Emission Line Galaxies}",
      journal = {\aj},
         year = 2023,
        month = mar,
       volume = {165},
       number = {3},
          eid = {126},
        pages = {126},
          doi = {10.3847/1538-3881/acb213},
archivePrefix = {arXiv},
       eprint = {2208.08513},
 primaryClass = {astro-ph.CO},
       adsurl = {https://ui.adsabs.harvard.edu/abs/2023AJ....165..126R}
}

@ARTICLE{2023A&A...671A.119C,
       author = {{Chabrier}, Gilles and {Baraffe}, Isabelle and {Phillips}, Mark and {Debras}, Florian},
        title = "{Impact of a new H/He equation of state on the evolution of massive brown dwarfs. New determination of the hydrogen burning limit}",
      journal = {\aap},
         year = 2023,
        month = mar,
       volume = {671},
          eid = {A119},
        pages = {A119},
          doi = {10.1051/0004-6361/202243832},
archivePrefix = {arXiv},
       eprint = {2212.07153},
 primaryClass = {astro-ph.SR},
       adsurl = {https://ui.adsabs.harvard.edu/abs/2023A&A...671A.119C}
}

@ARTICLE{2023ApJ...944..107C,
       author = {{Chaussidon}, Edmond and {Y{\`e}che}, Christophe and {Palanque-Delabrouille}, Nathalie and {Alexander}, David M. and {Yang}, Jinyi and {Ahlen}, Steven and {Bailey}, Stephen and {Brooks}, David and {Cai}, Zheng and {Chabanier}, Sol{\`e}ne and {Davis}, Tamara M. and {Dawson}, Kyle and {de laMacorra}, Axel and {Dey}, Arjun and {Dey}, Biprateep and {Eftekharzadeh}, Sarah and {Eisenstein}, Daniel J. and {Fanning}, Kevin and {Font-Ribera}, Andreu and {Gazta{\~n}aga}, Enrique and {A Gontcho}, Satya Gontcho and {Gonzalez-Morales}, Alma X. and {Guy}, Julien and {Herrera-Alcantar}, Hiram K. and {Honscheid}, Klaus and {Ishak}, Mustapha and {Jiang}, Linhua and {Juneau}, Stephanie and {Kehoe}, Robert and {Kisner}, Theodore and {Kov{\'a}cs}, Andras and {Kremin}, Anthony and {Lan}, Ting-Wen and {Landriau}, Martin and {Le Guillou}, Laurent and {Levi}, Michael E. and {Magneville}, Christophe and {Martini}, Paul and {Meisner}, Aaron M. and {Moustakas}, John and {Mu{\~n}oz-Guti{\'e}rrez}, Andrea and {Myers}, Adam D. and {Newman}, Jeffrey A. and {Nie}, Jundan and {Percival}, Will J. and {Poppett}, Claire and {Prada}, Francisco and {Raichoor}, Anand and {Ravoux}, Corentin and {Ross}, Ashley J. and {Schlafly}, Edward and {Schlegel}, David and {Tan}, Ting and {Tarl{\'e}}, Gregory and {Zhou}, Rongpu and {Zhou}, Zhimin and {Zou}, Hu},
        title = "{Target Selection and Validation of DESI Quasars}",
      journal = {\apj},
         year = 2023,
        month = feb,
       volume = {944},
       number = {1},
          eid = {107},
        pages = {107},
          doi = {10.3847/1538-4357/acb3c2},
archivePrefix = {arXiv},
       eprint = {2208.08511},
 primaryClass = {astro-ph.CO},
       adsurl = {https://ui.adsabs.harvard.edu/abs/2023ApJ...944..107C}
}

@ARTICLE{2023AJ....165...58Z,
       author = {{Zhou}, Rongpu and {Dey}, Biprateep and {Newman}, Jeffrey A. and {Eisenstein}, Daniel J. and {Dawson}, K. and {Bailey}, S. and {Berti}, A. and {Guy}, J. and {Lan}, Ting-Wen and {Zou}, H. and {Aguilar}, J. and {Ahlen}, S. and {Alam}, Shadab and {Brooks}, D. and {de la Macorra}, A. and {Dey}, A. and {Dhungana}, G. and {Fanning}, K. and {Font-Ribera}, A. and {Gontcho}, S. Gontcho A. and {Honscheid}, K. and {Ishak}, Mustapha and {Kisner}, T. and {Kov{\'a}cs}, A. and {Kremin}, A. and {Landriau}, M. and {Levi}, Michael E. and {Magneville}, C. and {Manera}, Marc and {Martini}, P. and {Meisner}, Aaron M. and {Miquel}, R. and {Moustakas}, J. and {Myers}, Adam D. and {Nie}, Jundan and {Palanque-Delabrouille}, N. and {Percival}, W.~J. and {Poppett}, C. and {Prada}, F. and {Raichoor}, A. and {Ross}, A.~J. and {Schlafly}, E. and {Schlegel}, D. and {Schubnell}, M. and {Tarl{\'e}}, Gregory and {Weaver}, B.~A. and {Wechsler}, R.~H. and {Y{\'e}che}, Christophe and {Zhou}, Zhimin},
        title = "{Target Selection and Validation of DESI Luminous Red Galaxies}",
      journal = {\aj},
         year = 2023,
        month = feb,
       volume = {165},
       number = {2},
          eid = {58},
        pages = {58},
          doi = {10.3847/1538-3881/aca5fb},
archivePrefix = {arXiv},
       eprint = {2208.08515},
 primaryClass = {astro-ph.CO},
       adsurl = {https://ui.adsabs.harvard.edu/abs/2023AJ....165...58Z}
}

@ARTICLE{2023AJ....165...50M,
       author = {{Myers}, Adam D. and {Moustakas}, John and {Bailey}, Stephen and {Weaver}, Benjamin A. and {Cooper}, Andrew P. and {Forero-Romero}, Jaime E. and {Abolfathi}, Bela and {Alexander}, David M. and {Brooks}, David and {Chaussidon}, Edmond and {Chuang}, Chia-Hsun and {Dawson}, Kyle and {Dey}, Arjun and {Dey}, Biprateep and {Dhungana}, Govinda and {Doel}, Peter and {Fanning}, Kevin and {Gazta{\~n}aga}, Enrique and {Gontcho A Gontcho}, Satya and {Gonzalez-Morales}, Alma X. and {Hahn}, ChangHoon and {Herrera-Alcantar}, Hiram K. and {Honscheid}, Klaus and {Ishak}, Mustapha and {Karim}, Tanveer and {Kirkby}, David and {Kisner}, Theodore and {Koposov}, Sergey E. and {Kremin}, Anthony and {Lan}, Ting-Wen and {Landriau}, Martin and {Lang}, Dustin and {Levi}, Michael E. and {Magneville}, Christophe and {Napolitano}, Lucas and {Martini}, Paul and {Meisner}, Aaron and {Newman}, Jeffrey A. and {Palanque-Delabrouille}, Nathalie and {Percival}, Will and {Poppett}, Claire and {Prada}, Francisco and {Raichoor}, Anand and {Ross}, Ashley J. and {Schlafly}, Edward F. and {Schlegel}, David and {Schubnell}, Michael and {Tan}, Ting and {Tarle}, Gregory and {Wilson}, Michael J. and {Y{\`e}che}, Christophe and {Zhou}, Rongpu and {Zhou}, Zhimin and {Zou}, Hu},
        title = "{The Target-selection Pipeline for the Dark Energy Spectroscopic Instrument}",
      journal = {\aj},
         year = 2023,
        month = feb,
       volume = {165},
       number = {2},
          eid = {50},
        pages = {50},
          doi = {10.3847/1538-3881/aca5f9},
archivePrefix = {arXiv},
       eprint = {2208.08518},
 primaryClass = {astro-ph.IM},
       adsurl = {https://ui.adsabs.harvard.edu/abs/2023AJ....165...50M}
}

@ARTICLE{2023MNRAS.518.4579P,
       author = {{Parsons}, S.~G. and {Hernandez}, M.~S. and {Toloza}, O. and {Zorotovic}, M. and {Schreiber}, M.~R. and {G{\"a}nsicke}, B.~T. and {Lagos}, F. and {Raddi}, R. and {Rebassa-Mansergas}, A. and {Ren}, J.~J. and {Koester}, D.},
        title = "{The white dwarf binary pathways survey - IX. Three long period white dwarf plus subgiant binaries}",
      journal = {\mnras},
         year = 2023,
        month = jan,
       volume = {518},
       number = {3},
        pages = {4579-4594},
          doi = {10.1093/mnras/stac3368},
archivePrefix = {arXiv},
       eprint = {2211.08440},
 primaryClass = {astro-ph.SR},
       adsurl = {https://ui.adsabs.harvard.edu/abs/2023MNRAS.518.4579P}
}

@ARTICLE{2023AJ....165....9S,
       author = {{Silber}, Joseph Harry and {Fagrelius}, Parker and {Fanning}, Kevin and {Schubnell}, Michael and {Aguilar}, Jessica Nicole and {Ahlen}, Steven and {Ameel}, Jon and {Ballester}, Otger and {Baltay}, Charles and {Bebek}, Chris and {Benton Beard}, Dominic and {Besuner}, Robert and {Cardiel-Sas}, Laia and {Casas}, Ricard and {Castander}, Francisco Javier and {Claybaugh}, Todd and {Dobson}, Carl and {Duan}, Yutong and {Dunlop}, Patrick and {Edelstein}, Jerry and {Emmet}, William T. and {Elliott}, Ann and {Evatt}, Matthew and {Gershkovich}, Irena and {Guy}, Julien and {Harris}, Stu and {Heetderks}, Henry and {Heetderks}, Ian and {Honscheid}, Klaus and {Illa}, Jose Maria and {Jelinsky}, Patrick and {Jelinsky}, Sharon R. and {Jimenez}, Jorge and {Karcher}, Armin and {Kent}, Stephen and {Kirkby}, David and {Kneib}, Jean-Paul and {Lambert}, Andrew and {Lampton}, Mike and {Leitner}, Daniela and {Levi}, Michael and {McCauley}, Jeremy and {Meisner}, Aaron and {Miller}, Timothy N. and {Miquel}, Ramon and {Mundet}, Juli{\'a} and {Poppett}, Claire and {Rabinowitz}, David and {Reil}, Kevin and {Roman}, David and {Schlegel}, David and {Serrano}, Santiago and {Van Shourt}, William and {Sprayberry}, David and {Tarl{\'e}}, Gregory and {Tie}, Suk Sien and {Weaverdyck}, Curtis and {Zhang}, Kai and {Azzaro}, Marco and {Bailey}, Stephen and {Becerril}, Santiago and {Blackwell}, Tami and {Bouri}, Mohamed and {Brooks}, David and {Buckley-Geer}, Elizabeth and {Castro}, Jose Pe{\~n}ate and {Derwent}, Mark and {Dey}, Arjun and {Dhungana}, Govinda and {Doel}, Peter and {Eisenstein}, Daniel J. and {Fahim}, Nasib and {Garcia-Bellido}, Juan and {Gazta{\~n}aga}, Enrique and {A Gontcho}, Satya Gontcho and {Gutierrez}, Gaston and {H{\"o}rler}, Philipp and {Kehoe}, Robert and {Kisner}, Theodore and {Kremin}, Anthony and {Kronig}, Luzius and {Landriau}, Martin and {Le Guillou}, Laurent and {Martini}, Paul and {Moustakas}, John and {Palanque-Delabrouille}, Nathalie and {Peng}, Xiyan and {Percival}, Will and {Prada}, Francisco and {Allende Prieto}, Carlos and {de Rivera}, Guillermo Gonzalez and {Sanchez}, Eusebio and {Sanchez}, Justo and {Sharples}, Ray and {Soares-Santos}, Marcelle and {Schlafly}, Edward and {Weaver}, Benjamin Alan and {Zhou}, Zhimin and {Zhu}, Yaling and {Zou}, Hu and {DESI Collaboration}},
        title = "{The Robotic Multiobject Focal Plane System of the Dark Energy Spectroscopic Instrument (DESI)}",
      journal = {\aj},
         year = 2023,
        month = jan,
       volume = {165},
       number = {1},
          eid = {9},
        pages = {9},
          doi = {10.3847/1538-3881/ac9ab1},
archivePrefix = {arXiv},
       eprint = {2205.09014},
 primaryClass = {astro-ph.IM},
       adsurl = {https://ui.adsabs.harvard.edu/abs/2023AJ....165....9S}
}

@ARTICLE{2022MNRAS.517.4916E,
       author = {{El-Badry}, Kareem and {Conroy}, Charlie and {Fuller}, Jim and {Kiman}, Rocio and {van Roestel}, Jan and {Rodriguez}, Antonio C. and {Burdge}, Kevin B.},
        title = "{Magnetic braking saturates: evidence from the orbital period distribution of low-mass detached eclipsing binaries from ZTF}",
      journal = {\mnras},
         year = 2022,
        month = dec,
       volume = {517},
       number = {4},
        pages = {4916-4939},
          doi = {10.1093/mnras/stac2945},
archivePrefix = {arXiv},
       eprint = {2208.05488},
 primaryClass = {astro-ph.SR},
       adsurl = {https://ui.adsabs.harvard.edu/abs/2022MNRAS.517.4916E}
}

@ARTICLE{2022MNRAS.517.2867H,
       author = {{Hernandez}, M.~S. and {Schreiber}, M.~R. and {Parsons}, S.~G. and {G{\"a}nsicke}, B.~T. and {Toloza}, O. and {Zorotovic}, M. and {Raddi}, R. and {Rebassa-Mansergas}, A. and {Ren}, J.~J.},
        title = "{The white dwarf binary pathways survey - VIII. A post-common envelope binary with a massive white dwarf and an active G-type secondary star}",
      journal = {\mnras},
         year = 2022,
        month = dec,
       volume = {517},
       number = {2},
        pages = {2867-2875},
          doi = {10.1093/mnras/stac2837},
archivePrefix = {arXiv},
       eprint = {2209.15591},
 primaryClass = {astro-ph.SR},
       adsurl = {https://ui.adsabs.harvard.edu/abs/2022MNRAS.517.2867H}
}

@ARTICLE{2022AJ....164..207D,
       author = {{DESI Collaboration} and {Abareshi}, B. and {Aguilar}, J. and {Ahlen}, S. and {Alam}, Shadab and {Alexander}, David M. and {Alfarsy}, R. and {Allen}, L. and {Allende Prieto}, C. and {Alves}, O. and {Ameel}, J. and {Armengaud}, E. and {Asorey}, J. and {Aviles}, Alejandro and {Bailey}, S. and {Balaguera-Antol{\'\i}nez}, A. and {Ballester}, O. and {Baltay}, C. and {Bault}, A. and {Beltran}, S.~F. and {Benavides}, B. and {BenZvi}, S. and {Berti}, A. and {Besuner}, R. and {Beutler}, Florian and {Bianchi}, D. and {Blake}, C. and {Blanc}, P. and {Blum}, R. and {Bolton}, A. and {Bose}, S. and {Bramall}, D. and {Brieden}, S. and {Brodzeller}, A. and {Brooks}, D. and {Brownewell}, C. and {Buckley-Geer}, E. and {Cahn}, R.~N. and {Cai}, Z. and {Canning}, R. and {Capasso}, R. and {Carnero Rosell}, A. and {Carton}, P. and {Casas}, R. and {Castander}, F.~J. and {Cervantes-Cota}, J.~L. and {Chabanier}, S. and {Chaussidon}, E. and {Chuang}, C. and {Circosta}, C. and {Cole}, S. and {Cooper}, A.~P. and {da Costa}, L. and {Cousinou}, M.-C. and {Cuceu}, A. and {Davis}, T.~M. and {Dawson}, K. and {de la Cruz-Noriega}, R. and {de la Macorra}, A. and {de Mattia}, A. and {Della Costa}, J. and {Demmer}, P. and {Derwent}, M. and {Dey}, A. and {Dey}, B. and {Dhungana}, G. and {Ding}, Z. and {Dobson}, C. and {Doel}, P. and {Donald-McCann}, J. and {Donaldson}, J. and {Douglass}, K. and {Duan}, Y. and {Dunlop}, P. and {Edelstein}, J. and {Eftekharzadeh}, S. and {Eisenstein}, D.~J. and {Enriquez-Vargas}, M. and {Escoffier}, S. and {Evatt}, M. and {Fagrelius}, P. and {Fan}, X. and {Fanning}, K. and {Fawcett}, V.~A. and {Ferraro}, S. and {Ereza}, J. and {Flaugher}, B. and {Font-Ribera}, A. and {Forero-Romero}, J.~E. and {Frenk}, C.~S. and {Fromenteau}, S. and {G{\"a}nsicke}, B.~T. and {Garcia-Quintero}, C. and {Garrison}, L. and {Gazta{\~n}aga}, E. and {Gerardi}, F. and {Gil-Mar{\'\i}n}, H. and {Gontcho A Gontcho}, S. and {Gonzalez-Morales}, Alma X. and {Gonzalez-de-Rivera}, G. and {Gonzalez-Perez}, V. and {Gordon}, C. and {Graur}, O. and {Green}, D. and {Grove}, C. and {Gruen}, D. and {Gutierrez}, G. and {Guy}, J. and {Hahn}, C. and {Harris}, S. and {Herrera}, D. and {Herrera-Alcantar}, Hiram K. and {Honscheid}, K. and {Howlett}, C. and {Huterer}, D. and {Ir{\v{s}}i{\v{c}}}, V. and {Ishak}, M. and {Jelinsky}, P. and {Jiang}, L. and {Jimenez}, J. and {Jing}, Y.~P. and {Joyce}, R. and {Jullo}, E. and {Juneau}, S. and {Kara{\c{c}}ayl{\i}}, N.~G. and {Karamanis}, M. and {Karcher}, A. and {Karim}, T. and {Kehoe}, R. and {Kent}, S. and {Kirkby}, D. and {Kisner}, T. and {Kitaura}, F. and {Koposov}, S.~E. and {Kov{\'a}cs}, A. and {Kremin}, A. and {Krolewski}, Alex and {L'Huillier}, B. and {Lahav}, O. and {Lambert}, A. and {Lamman}, C. and {Lan}, Ting-Wen and {Landriau}, M. and {Lane}, S. and {Lang}, D. and {Lange}, J.~U. and {Lasker}, J. and {Le Guillou}, L. and {Leauthaud}, A. and {Le Van Suu}, A. and {Levi}, Michael E. and {Li}, T.~S. and {Magneville}, C. and {Manera}, M. and {Manser}, Christopher J. and {Marshall}, B. and {Martini}, Paul and {McCollam}, W. and {McDonald}, P. and {Meisner}, Aaron M. and {Mena-Fern{\'a}ndez}, J. and {Meneses-Rizo}, J. and {Mezcua}, M. and {Miller}, T. and {Miquel}, R. and {Montero-Camacho}, P. and {Moon}, J. and {Moustakas}, J. and {Mueller}, E. and {Mu{\~n}oz-Guti{\'e}rrez}, Andrea and {Myers}, Adam D. and {Nadathur}, S. and {Najita}, J. and {Napolitano}, L. and {Neilsen}, E. and {Newman}, Jeffrey A. and {Nie}, J.~D. and {Ning}, Y. and {Niz}, G. and {Norberg}, P. and {Noriega}, Hern{\'a}n E. and {O'Brien}, T. and {Obuljen}, A. and {Palanque-Delabrouille}, N. and {Palmese}, A. and {Zhiwei}, P. and {Pappalardo}, D. and {PENG}, X. and {Percival}, W.~J. and {Perruchot}, S. and {Pogge}, R. and {Poppett}, C. and {Porredon}, A. and {Prada}, F. and {Prochaska}, J. and {Pucha}, R. and {P{\'e}rez-Fern{\'a}ndez}, A. and {P{\'e}rez-R{\`a}fols}, I. and {Rabinowitz}, D. and {Raichoor}, A.},
        title = "{Overview of the Instrumentation for the Dark Energy Spectroscopic Instrument}",
      journal = {\aj},
         year = 2022,
        month = nov,
       volume = {164},
       number = {5},
          eid = {207},
        pages = {207},
          doi = {10.3847/1538-3881/ac882b},
archivePrefix = {arXiv},
       eprint = {2205.10939},
 primaryClass = {astro-ph.IM},
       adsurl = {https://ui.adsabs.harvard.edu/abs/2022AJ....164..207D}
}

@ARTICLE{2022Natur.610..467B,
       author = {{Burdge}, Kevin B. and {El-Badry}, Kareem and {Marsh}, Thomas R. and {Rappaport}, Saul and {Brown}, Warren R. and {Caiazzo}, Ilaria and {Chakrabarty}, Deepto and {Dhillon}, V.~S. and {Fuller}, Jim and {G{\"a}nsicke}, Boris T. and {Graham}, Matthew J. and {Kara}, Erin and {Kulkarni}, S.~R. and {Littlefair}, S.~P. and {Mr{\'o}z}, Przemek and {Rodr{\'\i}guez-Gil}, Pablo and {Roestel}, Jan van and {Simcoe}, Robert A. and {Bellm}, Eric C. and {Drake}, Andrew J. and {Dekany}, Richard G. and {Groom}, Steven L. and {Laher}, Russ R. and {Masci}, Frank J. and {Riddle}, Reed and {Smith}, Roger M. and {Prince}, Thomas A.},
        title = "{A dense 0.1-solar-mass star in a 51-minute-orbital-period eclipsing binary}",
      journal = {\nat},
         year = 2022,
        month = oct,
       volume = {610},
       number = {7932},
        pages = {467-471},
          doi = {10.1038/s41586-022-05195-x},
archivePrefix = {arXiv},
       eprint = {2210.01809},
 primaryClass = {astro-ph.SR},
       adsurl = {https://ui.adsabs.harvard.edu/abs/2022Natur.610..467B}
}

@ARTICLE{2022ApJ...938...31S,
       author = {{Shen}, Ken J. and {Quataert}, Eliot},
        title = "{Binary Interaction Dominates Mass Ejection in Classical Novae}",
      journal = {\apj},
         year = 2022,
        month = oct,
       volume = {938},
       number = {1},
          eid = {31},
        pages = {31},
          doi = {10.3847/1538-4357/ac9136},
archivePrefix = {arXiv},
       eprint = {2205.06283},
 primaryClass = {astro-ph.SR},
       adsurl = {https://ui.adsabs.harvard.edu/abs/2022ApJ...938...31S}
}

@ARTICLE{2022MNRAS.512.5440V,
       author = {{van Roestel}, J. and {Kupfer}, T. and {Green}, M.~J. and {Wong}, T.~L.~S. and {Bildsten}, L. and {Burdge}, K. and {Prince}, T. and {Marsh}, T.~R. and {Szkody}, P. and {Fremling}, C. and {Graham}, M.~J. and {Dhillon}, V.~S. and {Littlefair}, S.~P. and {Bellm}, E.~C. and {Coughlin}, M. and {Duev}, D.~A. and {Goldstein}, D.~A. and {Laher}, R.~R. and {Rusholme}, B. and {Riddle}, R. and {Dekany}, R. and {Kulkarni}, S.~R.},
        title = "{Discovery and characterization of five new eclipsing AM CVn systems}",
      journal = {\mnras},
         year = 2022,
        month = jun,
       volume = {512},
       number = {4},
        pages = {5440-5461},
          doi = {10.1093/mnras/stab2421},
archivePrefix = {arXiv},
       eprint = {2107.07573},
 primaryClass = {astro-ph.SR},
       adsurl = {https://ui.adsabs.harvard.edu/abs/2022MNRAS.512.5440V}
}

@ARTICLE{2022MNRAS.512.1843H,
       author = {{Hernandez}, M.~S. and {Schreiber}, M.~R. and {Parsons}, S.~G. and {G{\"a}nsicke}, B.~T. and {Toloza}, O. and {Tovmassian}, G. and {Zorotovic}, M. and {Lagos}, F. and {Raddi}, R. and {Rebassa-Mansergas}, A. and {Ren}, J.~J. and {Tappert}, C.},
        title = "{The white dwarf binary pathways survey - VI. Two close post-common envelope binaries with TESS light curves}",
      journal = {\mnras},
         year = 2022,
        month = may,
       volume = {512},
       number = {2},
        pages = {1843-1856},
          doi = {10.1093/mnras/stac604},
archivePrefix = {arXiv},
       eprint = {2203.01745},
 primaryClass = {astro-ph.SR},
       adsurl = {https://ui.adsabs.harvard.edu/abs/2022MNRAS.512.1843H}
}

@ARTICLE{2022MNRAS.510.3605I,
       author = {{Inight}, K. and {G{\"a}nsicke}, B.~T. and {Blondel}, D. and {Boyd}, D. and {Ashley}, R.~P. and {Knigge}, C. and {Long}, K.~S. and {Marsh}, T.~R. and {McCleery}, J. and {Scaringi}, S. and {Steeghs}, D. and {Thorstensen}, J.~R. and {Vanmunster}, T. and {Wheatley}, P.~J.},
        title = "{ASAS J071404+7004.3 - a close, bright nova-like cataclysmic variable with gusty winds}",
      journal = {\mnras},
         year = 2022,
        month = mar,
       volume = {510},
       number = {3},
        pages = {3605-3621},
          doi = {10.1093/mnras/stab3662},
archivePrefix = {arXiv},
       eprint = {2109.14514},
 primaryClass = {astro-ph.SR},
       adsurl = {https://ui.adsabs.harvard.edu/abs/2022MNRAS.510.3605I}
}

@ARTICLE{2021MNRAS.508.4106E,
       author = {{El-Badry}, Kareem and {Rix}, Hans-Walter and {Quataert}, Eliot and {Kupfer}, Thomas and {Shen}, Ken J.},
        title = "{Birth of the ELMs: a ZTF survey for evolved cataclysmic variables turning into extremely low-mass white dwarfs}",
      journal = {\mnras},
         year = 2021,
        month = dec,
       volume = {508},
       number = {3},
        pages = {4106-4139},
          doi = {10.1093/mnras/stab2583},
archivePrefix = {arXiv},
       eprint = {2108.04255},
 primaryClass = {astro-ph.SR},
       adsurl = {https://ui.adsabs.harvard.edu/abs/2021MNRAS.508.4106E}
}

@ARTICLE{2021arXiv211115608K,
       author = {{Kulkarni}, S.~R. and {Harrison}, Fiona A. and {Grefenstette}, Brian W. and {Earnshaw}, Hannah P. and {Andreoni}, Igor and {Berg}, Danielle A. and {Bloom}, Joshua S. and {Cenko}, S. Bradley and {Chornock}, Ryan and {Christiansen}, Jessie L. and {Coughlin}, Michael W. and {Wuollet Criswell}, Alexander and {Darvish}, Behnam and {Das}, Kaustav K. and {De}, Kishalay and {Dessart}, Luc and {Dixon}, Don and {Dorsman}, Bas and {El-Badry}, Kareem and {Evans}, Christopher and {Ford}, K.~E. Saavik and {Fremling}, Christoffer and {Gansicke}, Boris T. and {Gezari}, Suvi and {Goetberg}, Y. and {Green}, Gregory M. and {Graham}, Matthew J. and {Heida}, Marianne and {Ho}, Anna Y.~Q. and {Jaodand}, Amruta D. and {Johns-Krull}, Christopher M. and {Kasliwal}, Mansi M. and {Lazzarini}, Margaret and {Lu}, Wenbin and {Margutti}, Raffaella and {Martin}, D. Christopher and {Masters}, Daniel Charles and {McKernan}, Barry and {Naze}, Yael and {Nissanke}, Samaya M. and {Parazin}, B. and {Perley}, Daniel A. and {Phinney}, E. Sterl and {Piro}, Anthony L. and {Raaijmakers}, G. and {Rauw}, Gregor and {Rodriguez}, Antonio C. and {Sana}, Hugues and {Senchyna}, Peter and {Singer}, Leo P. and {Spake}, Jessica J. and {Stassun}, Keivan G. and {Stern}, Daniel and {Teplitz}, Harry I. and {Weisz}, Daniel R. and {Yao}, Yuhan},
        title = "{Science with the Ultraviolet Explorer (UVEX)}",
      journal = {arXiv e-prints},
         year = 2021,
        month = nov,
          eid = {arXiv:2111.15608},
        pages = {arXiv:2111.15608},
          doi = {10.48550/arXiv.2111.15608},
archivePrefix = {arXiv},
       eprint = {2111.15608},
 primaryClass = {astro-ph.GA},
       adsurl = {https://ui.adsabs.harvard.edu/abs/2021arXiv211115608K}
}

@ARTICLE{2021AJ....162...94S,
       author = {{Szkody}, Paula and {Olde Loohuis}, Claire and {Koplitz}, Brad and {van Roestel}, Jan and {Dicenzo}, Brooke and {Ho}, Anna Y.~Q. and {Hillenbrand}, Lynne A. and {Bellm}, Eric C. and {Dekany}, Richard and {Drake}, Andrew J. and {Duev}, Dmitry A. and {Graham}, Matthew J. and {Kasliwal}, Mansi M. and {Mahabal}, Ashish A. and {Masci}, Frank J. and {Neill}, James D. and {Riddle}, Reed and {Rusholme}, Benjamin and {Sollerman}, Jesper and {Walters}, Richard},
        title = "{Cataclysmic Variables in the Second Year of the Zwicky Transient Facility}",
      journal = {\aj},
         year = 2021,
        month = sep,
       volume = {162},
       number = {3},
          eid = {94},
        pages = {94},
          doi = {10.3847/1538-3881/ac0efb},
archivePrefix = {arXiv},
       eprint = {2107.07051},
 primaryClass = {astro-ph.SR},
       adsurl = {https://ui.adsabs.harvard.edu/abs/2021AJ....162...94S}
}

@ARTICLE{2021MNRAS.505.2051E,
       author = {{El-Badry}, Kareem and {Quataert}, Eliot and {Rix}, Hans-Walter and {Weisz}, Daniel R. and {Kupfer}, Thomas and {Shen}, Ken J. and {Xiang}, Maosheng and {Yang}, Yong and {Liu}, Xiaowei},
        title = "{LAMOST J0140355 + 392651: an evolved cataclysmic variable donor transitioning to become an extremely low-mass white dwarf}",
      journal = {\mnras},
         year = 2021,
        month = aug,
       volume = {505},
       number = {2},
        pages = {2051-2073},
          doi = {10.1093/mnras/stab1318},
archivePrefix = {arXiv},
       eprint = {2104.07033},
 primaryClass = {astro-ph.SR},
       adsurl = {https://ui.adsabs.harvard.edu/abs/2021MNRAS.505.2051E}
}

@ARTICLE{2021MNRAS.504.2420I,
       author = {{Inight}, K. and {G{\"a}nsicke}, Boris T. and {Breedt}, E. and {Marsh}, T.~R. and {Pala}, A.~F. and {Raddi}, R.},
        title = "{Towards a volumetric census of close white dwarf binaries - I. Reference samples}",
      journal = {\mnras},
         year = 2021,
        month = jun,
       volume = {504},
       number = {2},
        pages = {2420-2442},
          doi = {10.1093/mnras/stab753},
archivePrefix = {arXiv},
       eprint = {2103.06892},
 primaryClass = {astro-ph.SR},
       adsurl = {https://ui.adsabs.harvard.edu/abs/2021MNRAS.504.2420I}
}

@ARTICLE{2021AJ....161..242F,
       author = {{F{\"o}rster}, F. and {Cabrera-Vives}, G. and {Castillo-Navarrete}, E. and {Est{\'e}vez}, P.~A. and {S{\'a}nchez-S{\'a}ez}, P. and {Arredondo}, J. and {Bauer}, F.~E. and {Carrasco-Davis}, R. and {Catelan}, M. and {Elorrieta}, F. and {Eyheramendy}, S. and {Huijse}, P. and {Pignata}, G. and {Reyes}, E. and {Reyes}, I. and {Rodr{\'\i}guez-Mancini}, D. and {Ruz-Mieres}, D. and {Valenzuela}, C. and {{\'A}lvarez-Maldonado}, I. and {Astorga}, N. and {Borissova}, J. and {Clocchiatti}, A. and {De Cicco}, D. and {Donoso-Oliva}, C. and {Hern{\'a}ndez-Garc{\'\i}a}, L. and {Graham}, M.~J. and {Jord{\'a}n}, A. and {Kurtev}, R. and {Mahabal}, A. and {Maureira}, J.~C. and {Mu{\~n}oz-Arancibia}, A. and {Molina-Ferreiro}, R. and {Moya}, A. and {Palma}, W. and {P{\'e}rez-Carrasco}, M. and {Protopapas}, P. and {Romero}, M. and {Sabatini-Gacitua}, L. and {S{\'a}nchez}, A. and {San Mart{\'\i}n}, J. and {Sep{\'u}lveda-Cobo}, C. and {Vera}, E. and {Vergara}, J.~R.},
        title = "{The Automatic Learning for the Rapid Classification of Events (ALeRCE) Alert Broker}",
      journal = {\aj},
         year = 2021,
        month = may,
       volume = {161},
       number = {5},
          eid = {242},
        pages = {242},
          doi = {10.3847/1538-3881/abe9bc},
archivePrefix = {arXiv},
       eprint = {2008.03303},
 primaryClass = {astro-ph.IM},
       adsurl = {https://ui.adsabs.harvard.edu/abs/2021AJ....161..242F}
}

@ARTICLE{2021NatAs...5..648S,
       author = {{Schreiber}, Matthias R. and {Belloni}, Diogo and {G{\"a}nsicke}, Boris T. and {Parsons}, Steven G. and {Zorotovic}, Monica},
        title = "{The origin and evolution of magnetic white dwarfs in close binary stars}",
      journal = {Nature Astronomy},
         year = 2021,
        month = apr,
       volume = {5},
        pages = {648-654},
          doi = {10.1038/s41550-021-01346-8},
archivePrefix = {arXiv},
       eprint = {2104.14607},
 primaryClass = {astro-ph.SR},
       adsurl = {https://ui.adsabs.harvard.edu/abs/2021NatAs...5..648S}
}

@ARTICLE{2021MNRAS.502..581C,
       author = {{Chote}, P. and {G{\"a}nsicke}, B.~T. and {McCormac}, J. and {Aungwerojwit}, A. and {Bayliss}, D. and {Burleigh}, M.~R. and {Casewell}, S.~L. and {Eigm{\"u}ller}, Ph and {Gill}, S. and {Goad}, M.~R. and {Hermes}, J.~J. and {Jenkins}, J.~S. and {Mukadam}, A.~S. and {Poshyachinda}, S. and {Raynard}, L. and {Reichart}, D.~E. and {Szkody}, P. and {Toloza}, O. and {West}, R.~G. and {Wheatley}, P.~J.},
        title = "{NGTS and HST insights into the long-period modulation in GW Librae}",
      journal = {\mnras},
         year = 2021,
        month = mar,
       volume = {502},
       number = {1},
        pages = {581-588},
          doi = {10.1093/mnras/staa4015},
archivePrefix = {arXiv},
       eprint = {2101.08786},
 primaryClass = {astro-ph.SR},
       adsurl = {https://ui.adsabs.harvard.edu/abs/2021MNRAS.502..581C}
}

@ARTICLE{2021AJ....161..147B,
       author = {{Bailer-Jones}, C.~A.~L. and {Rybizki}, J. and {Fouesneau}, M. and {Demleitner}, M. and {Andrae}, R.},
        title = "{Estimating Distances from Parallaxes. V. Geometric and Photogeometric Distances to 1.47 Billion Stars in Gaia Early Data Release 3}",
      journal = {\aj},
         year = 2021,
        month = mar,
       volume = {161},
       number = {3},
          eid = {147},
        pages = {147},
          doi = {10.3847/1538-3881/abd806},
archivePrefix = {arXiv},
       eprint = {2012.05220},
 primaryClass = {astro-ph.SR},
       adsurl = {https://ui.adsabs.harvard.edu/abs/2021AJ....161..147B}
}

@dataset{2021yCat.1352....0B,
       author = {{Bailer-Jones}, C.~A.~L. and {Rybizki}, J. and {Fouesneau}, M. and {Demleitner}, M. and {Andrae}, R.},
        title = "{VizieR Online Data Catalog: Distances to 1.47 billion stars in Gaia EDR3 (Bailer-Jones+, 2021)}",
 howpublished = {VizieR On-line Data Catalog: I/352.  Originally published in: 2021AJ....161..147B},
         year = 2021,
        month = feb,
          eid = {I/352},
       adsurl = {https://ui.adsabs.harvard.edu/abs/2021yCat.1352....0B}
}

@ARTICLE{2021MNRAS.501.1677H,
       author = {{Hernandez}, M.~S. and {Schreiber}, M.~R. and {Parsons}, S.~G. and {G{\"a}nsicke}, B.~T. and {Lagos}, F. and {Raddi}, R. and {Toloza}, O. and {Tovmassian}, G. and {Zorotovic}, M. and {Irawati}, P. and {Past{\'e}n}, E. and {Rebassa-Mansergas}, A. and {Ren}, J.~J. and {Rittipruk}, P. and {Tappert}, C.},
        title = "{The White Dwarf Binary Pathways Survey - IV. Three close white dwarf binaries with G-type secondary stars}",
      journal = {\mnras},
         year = 2021,
        month = feb,
       volume = {501},
       number = {2},
        pages = {1677-1689},
          doi = {10.1093/mnras/staa3815},
archivePrefix = {arXiv},
       eprint = {2012.04683},
 primaryClass = {astro-ph.SR},
       adsurl = {https://ui.adsabs.harvard.edu/abs/2021MNRAS.501.1677H}
}

@ARTICLE{2020RNAAS...4..188A,
       author = {{Allende Prieto}, Carlos and {Cooper}, Andrew P. and {Dey}, Arjun and {G{\"a}nsicke}, Boris T. and {Koposov}, Sergey E. and {Li}, Ting and {Manser}, Christopher and {Nidever}, David L. and {Rockosi}, Constance and {Wang}, Mei-Yu and {Aguado}, David S. and {Blum}, Robert and {Brooks}, David and {Eisenstein}, Daniel J. and {Duan}, Yutong and {Eftekharzadeh}, Sarah and {Gazta{\~n}aga}, Enrique and {Kehoe}, Robert and {Landriau}, Martin and {Lee}, Chien-Hsiu and {Levi}, Michael E. and {Meisner}, Aaron M. and {Myers}, Adam D. and {Najita}, Joan and {Olsen}, Knut and {Palanque-Delabrouille}, Nathalie and {Poppett}, Claire and {Prada}, Francisco and {Schlegel}, David J. and {Schubnell}, Michael and {Tarl{\'e}}, Gregory and {Valluri}, Monica and {Wechsler}, Risa H. and {Y{\`e}che}, Christophe},
        title = "{Preliminary Target Selection for the DESI Milky Way Survey (MWS)}",
      journal = {Research Notes of the American Astronomical Society},
         year = 2020,
        month = oct,
       volume = {4},
       number = {10},
          eid = {188},
        pages = {188},
          doi = {10.3847/2515-5172/abc1dc},
archivePrefix = {arXiv},
       eprint = {2010.11284},
 primaryClass = {astro-ph.GA},
       adsurl = {https://ui.adsabs.harvard.edu/abs/2020RNAAS...4..188A}
}

@ARTICLE{2020RNAAS...4..187R,
       author = {{Ruiz-Macias}, Omar and {Zarrouk}, Pauline and {Cole}, Shaun and {Norberg}, Peder and {Baugh}, Carlton and {Brooks}, David and {Dey}, Arjun and {Duan}, Yutong and {Eftekharzadeh}, Sarah and {Eisenstein}, Daniel J. and {Forero-Romero}, Jaime E. and {Gazta{\~n}aga}, Enrique and {Hahn}, ChangHoon and {Kehoe}, Robert and {Landriau}, Martin and {Lang}, Dustin and {Levi}, Michael E. and {Lucey}, John and {Meisner}, Aaron M. and {Moustakas}, John and {Myers}, Adam D. and {Palanque-Delabrouille}, Nathalie and {Poppett}, Claire and {Prada}, Francisco and {Raichoor}, Anand and {Schlegel}, David J. and {Schubnell}, Michael and {Tarl{\'e}}, Gregory and {Weinberg}, David H. and {Wilson}, M.~J. and {Y{\`e}che}, Christophe},
        title = "{Preliminary Target Selection for the DESI Bright Galaxy Survey (BGS)}",
      journal = {Research Notes of the American Astronomical Society},
         year = 2020,
        month = oct,
       volume = {4},
       number = {10},
          eid = {187},
        pages = {187},
          doi = {10.3847/2515-5172/abc25a},
archivePrefix = {arXiv},
       eprint = {2010.11283},
 primaryClass = {astro-ph.GA},
       adsurl = {https://ui.adsabs.harvard.edu/abs/2020RNAAS...4..187R}
}

@ARTICLE{2020ApJ...900L..37R,
       author = {{Rivera Sandoval}, L.~E. and {Maccarone}, T.~J. and {Pichardo Marcano}, M.},
        title = "{A Year-long Superoutburst from an Ultracompact White Dwarf Binary Reveals the Importance of Donor Star Irradiation}",
      journal = {\apjl},
         year = 2020,
        month = sep,
       volume = {900},
       number = {2},
          eid = {L37},
        pages = {L37},
          doi = {10.3847/2041-8213/abb130},
archivePrefix = {arXiv},
       eprint = {2008.09242},
 primaryClass = {astro-ph.HE},
       adsurl = {https://ui.adsabs.harvard.edu/abs/2020ApJ...900L..37R}
}

@ARTICLE{2020ApJS..249....3A,
       author = {{Ahumada}, Romina and {Allende Prieto}, Carlos and {Almeida}, Andr{\'e}s and {Anders}, Friedrich and {Anderson}, Scott F. and {Andrews}, Brett H. and {Anguiano}, Borja and {Arcodia}, Riccardo and {Armengaud}, Eric and {Aubert}, Marie and {Avila}, Santiago and {Avila-Reese}, Vladimir and {Badenes}, Carles and {Balland}, Christophe and {Barger}, Kat and {Barrera-Ballesteros}, Jorge K. and {Basu}, Sarbani and {Bautista}, Julian and {Beaton}, Rachael L. and {Beers}, Timothy C. and {Benavides}, B. Izamar T. and {Bender}, Chad F. and {Bernardi}, Mariangela and {Bershady}, Matthew and {Beutler}, Florian and {Bidin}, Christian Moni and {Bird}, Jonathan and {Bizyaev}, Dmitry and {Blanc}, Guillermo A. and {Blanton}, Michael R. and {Boquien}, M{\'e}d{\'e}ric and {Borissova}, Jura and {Bovy}, Jo and {Brandt}, W.~N. and {Brinkmann}, Jonathan and {Brownstein}, Joel R. and {Bundy}, Kevin and {Bureau}, Martin and {Burgasser}, Adam and {Burtin}, Etienne and {Cano-D{\'\i}az}, Mariana and {Capasso}, Raffaella and {Cappellari}, Michele and {Carrera}, Ricardo and {Chabanier}, Sol{\`e}ne and {Chaplin}, William and {Chapman}, Michael and {Cherinka}, Brian and {Chiappini}, Cristina and {Doohyun Choi}, Peter and {Chojnowski}, S. Drew and {Chung}, Haeun and {Clerc}, Nicolas and {Coffey}, Damien and {Comerford}, Julia M. and {Comparat}, Johan and {da Costa}, Luiz and {Cousinou}, Marie-Claude and {Covey}, Kevin and {Crane}, Jeffrey D. and {Cunha}, Katia and {Ilha}, Gabriele da Silva and {Dai}, Yu Sophia and {Damsted}, Sanna B. and {Darling}, Jeremy and {Davidson}, Jr., James W. and {Davies}, Roger and {Dawson}, Kyle and {De}, Nikhil and {de la Macorra}, Axel and {De Lee}, Nathan and {Queiroz}, Anna B{\'a}rbara de Andrade and {Deconto Machado}, Alice and {de la Torre}, Sylvain and {Dell'Agli}, Flavia and {du Mas des Bourboux}, H{\'e}lion and {Diamond-Stanic}, Aleksandar M. and {Dillon}, Sean and {Donor}, John and {Drory}, Niv and {Duckworth}, Chris and {Dwelly}, Tom and {Ebelke}, Garrett and {Eftekharzadeh}, Sarah and {Davis Eigenbrot}, Arthur and {Elsworth}, Yvonne P. and {Eracleous}, Mike and {Erfanianfar}, Ghazaleh and {Escoffier}, Stephanie and {Fan}, Xiaohui and {Farr}, Emily and {Fern{\'a}ndez-Trincado}, Jos{\'e} G. and {Feuillet}, Diane and {Finoguenov}, Alexis and {Fofie}, Patricia and {Fraser-McKelvie}, Amelia and {Frinchaboy}, Peter M. and {Fromenteau}, Sebastien and {Fu}, Hai and {Galbany}, Llu{\'\i}s and {Garcia}, Rafael A. and {Garc{\'\i}a-Hern{\'a}ndez}, D.~A. and {Garma Oehmichen}, Luis Alberto and {Ge}, Junqiang and {Geimba Maia}, Marcio Antonio and {Geisler}, Doug and {Gelfand}, Joseph and {Goddy}, Julian and {Gonzalez-Perez}, Violeta and {Grabowski}, Kathleen and {Green}, Paul and {Grier}, Catherine J. and {Guo}, Hong and {Guy}, Julien and {Harding}, Paul and {Hasselquist}, Sten and {Hawken}, Adam James and {Hayes}, Christian R. and {Hearty}, Fred and {Hekker}, S. and {Hogg}, David W. and {Holtzman}, Jon A. and {Horta}, Danny and {Hou}, Jiamin and {Hsieh}, Bau-Ching and {Huber}, Daniel and {Hunt}, Jason A.~S. and {Ider Chitham}, J. and {Imig}, Julie and {Jaber}, Mariana and {Jimenez Angel}, Camilo Eduardo and {Johnson}, Jennifer A. and {Jones}, Amy M. and {J{\"o}nsson}, Henrik and {Jullo}, Eric and {Kim}, Yerim and {Kinemuchi}, Karen and {Kirkpatrick}, IV, Charles C. and {Kite}, George W. and {Klaene}, Mark and {Kneib}, Jean-Paul and {Kollmeier}, Juna A. and {Kong}, Hui and {Kounkel}, Marina and {Krishnarao}, Dhanesh and {Lacerna}, Ivan and {Lan}, Ting-Wen and {Lane}, Richard R. and {Law}, David R. and {Le Goff}, Jean-Marc and {Leung}, Henry W. and {Lewis}, Hannah and {Li}, Cheng and {Lian}, Jianhui and {Lin}, Lihwai and {Long}, Dan and {Longa-Pe{\~n}a}, Pen{\'e}lope and {Lundgren}, Britt and {Lyke}, Brad W. and {Mackereth}, J. Ted and {MacLeod}, Chelsea L. and {Majewski}, Steven R. and {Manchado}, Arturo and {Maraston}, Claudia and {Martini}, Paul and {Masseron}, Thomas and {Masters}, Karen L. and {Mathur}, Savita and {McDermid}, Richard M. and {Merloni}, Andrea and {Merrifield}, Michael and {M{\'e}sz{\'a}ros}, Szabolcs and {Miglio}, Andrea and {Minniti}, Dante and {Minsley}, Rebecca and {Miyaji}, Takamitsu and {Mohammad}, Faizan Gohar and {Mosser}, Benoit and {Mueller}, Eva-Maria and {Muna}, Demitri and {Mu{\~n}oz-Guti{\'e}rrez}, Andrea and {Myers}, Adam D. and {Nadathur}, Seshadri and {Nair}, Preethi and {Nandra}, Kirpal and {Correa do Nascimento}, Janaina and {Nevin}, Rebecca Jean and {Newman}, Jeffrey A. and {Nidever}, David L. and {Nitschelm}, Christian and {Noterdaeme}, Pasquier and {O'Connell}, Julia E. and {Olmstead}, Matthew D. and {Oravetz}, Daniel and {Oravetz}, Audrey and {Osorio}, Yeisson and {Pace}, Zachary J. and {Padilla}, Nelson and {Palanque-Delabrouille}, Nathalie and {Palicio}, Pedro A.},
        title = "{The 16th Data Release of the Sloan Digital Sky Surveys: First Release from the APOGEE-2 Southern Survey and Full Release of eBOSS Spectra}",
      journal = {\apjs},
         year = 2020,
        month = jul,
       volume = {249},
       number = {1},
          eid = {3},
        pages = {3},
          doi = {10.3847/1538-4365/ab929e},
archivePrefix = {arXiv},
       eprint = {1912.02905},
 primaryClass = {astro-ph.GA},
       adsurl = {https://ui.adsabs.harvard.edu/abs/2020ApJS..249....3A}
}

@ARTICLE{2020AJ....160....6T,
       author = {{Thorstensen}, John R.},
        title = "{Spectroscopic Studies of 30 Short-period Cataclysmic Variable Stars and Remarks on the Evolution and Population of Similar Objects}",
      journal = {\aj},
         year = 2020,
        month = jul,
       volume = {160},
       number = {1},
          eid = {6},
        pages = {6},
          doi = {10.3847/1538-3881/ab911c},
archivePrefix = {arXiv},
       eprint = {2005.02150},
 primaryClass = {astro-ph.SR},
       adsurl = {https://ui.adsabs.harvard.edu/abs/2020AJ....160....6T}
}

@ARTICLE{2020MNRAS.494.3799P,
       author = {{Pala}, A.~F. and {G{\"a}nsicke}, B.~T. and {Breedt}, E. and {Knigge}, C. and {Hermes}, J.~J. and {Gentile Fusillo}, N.~P. and {Hollands}, M.~A. and {Naylor}, T. and {Pelisoli}, I. and {Schreiber}, M.~R. and {Toonen}, S. and {Aungwerojwit}, A. and {Cukanovaite}, E. and {Dennihy}, E. and {Manser}, C.~J. and {Pretorius}, M.~L. and {Scaringi}, S. and {Toloza}, O.},
        title = "{A Volume-limited Sample of Cataclysmic Variables from Gaia DR2: Space Density and Population Properties}",
      journal = {\mnras},
         year = 2020,
        month = may,
       volume = {494},
       number = {3},
        pages = {3799-3827},
          doi = {10.1093/mnras/staa764},
archivePrefix = {arXiv},
       eprint = {1907.13152},
 primaryClass = {astro-ph.SR},
       adsurl = {https://ui.adsabs.harvard.edu/abs/2020MNRAS.494.3799P}
}

@ARTICLE{2020MNRAS.494..915L,
       author = {{Lagos}, F. and {Schreiber}, M.~R. and {Parsons}, S.~G. and {Zurlo}, A. and {Mesa}, D. and {G{\"a}nsicke}, B.~T. and {Brahm}, R. and {Caceres}, C. and {Canovas}, H. and {Hernandez}, M.-S. and {Jordan}, A. and {Koester}, D. and {Schmidtobreick}, L. and {Tappert}, C. and {Zorotovic}, M.},
        title = "{The White Dwarf Binary Pathways Survey -III. Contamination from hierarchical triples containing a white dwarf}",
      journal = {\mnras},
         year = 2020,
        month = may,
       volume = {494},
       number = {1},
        pages = {915-922},
          doi = {10.1093/mnras/staa747},
archivePrefix = {arXiv},
       eprint = {2003.07290},
 primaryClass = {astro-ph.SR},
       adsurl = {https://ui.adsabs.harvard.edu/abs/2020MNRAS.494..915L}
}

@ARTICLE{2020MNRAS.492L..40A,
       author = {{Abril}, Javier and {Schmidtobreick}, Linda and {Ederoclite}, Alessandro and {L{\'o}pez-Sanjuan}, Carlos},
        title = "{Disentangling cataclysmic variables in Gaia's HR diagram}",
      journal = {\mnras},
         year = 2020,
        month = feb,
       volume = {492},
       number = {1},
        pages = {L40-L44},
          doi = {10.1093/mnrasl/slz181},
archivePrefix = {arXiv},
       eprint = {1912.01531},
 primaryClass = {astro-ph.SR},
       adsurl = {https://ui.adsabs.harvard.edu/abs/2020MNRAS.492L..40A}
}

@ARTICLE{2020AJ....159...43H,
       author = {{Hou}, Wen and {Luo}, A.-li and {Li}, Yin-Bi and {Qin}, Li},
        title = "{Spectroscopically Identified Cataclysmic Variables from the LAMOST Survey. I. The Sample}",
      journal = {\aj},
         year = 2020,
        month = feb,
       volume = {159},
       number = {2},
          eid = {43},
        pages = {43},
          doi = {10.3847/1538-3881/ab5962},
archivePrefix = {arXiv},
       eprint = {1911.08338},
 primaryClass = {astro-ph.SR},
       adsurl = {https://ui.adsabs.harvard.edu/abs/2020AJ....159...43H}
}

@ARTICLE{2020ApJ...889...49B,
       author = {{Brown}, Warren R. and {Kilic}, Mukremin and {Kosakowski}, Alekzander and {Andrews}, Jeff J. and {Heinke}, Craig O. and {Ag{\"u}eros}, Marcel A. and {Camilo}, Fernando and {Gianninas}, A. and {Hermes}, J.~J. and {Kenyon}, Scott J.},
        title = "{The ELM Survey. VIII. Ninety-eight Double White Dwarf Binaries}",
      journal = {\apj},
         year = 2020,
        month = jan,
       volume = {889},
       number = {1},
          eid = {49},
        pages = {49},
          doi = {10.3847/1538-4357/ab63cd},
archivePrefix = {arXiv},
       eprint = {2002.00064},
 primaryClass = {astro-ph.SR},
       adsurl = {https://ui.adsabs.harvard.edu/abs/2020ApJ...889...49B}
}

@ARTICLE{2019AJ....157..168D,
       author = {{Dey}, Arjun and {Schlegel}, David J. and {Lang}, Dustin and {Blum}, Robert and {Burleigh}, Kaylan and {Fan}, Xiaohui and {Findlay}, Joseph R. and {Finkbeiner}, Doug and {Herrera}, David and {Juneau}, St{\'e}phanie and {Landriau}, Martin and {Levi}, Michael and {McGreer}, Ian and {Meisner}, Aaron and {Myers}, Adam D. and {Moustakas}, John and {Nugent}, Peter and {Patej}, Anna and {Schlafly}, Edward F. and {Walker}, Alistair R. and {Valdes}, Francisco and {Weaver}, Benjamin A. and {Y{\`e}che}, Christophe and {Zou}, Hu and {Zhou}, Xu and {Abareshi}, Behzad and {Abbott}, T.~M.~C. and {Abolfathi}, Bela and {Aguilera}, C. and {Alam}, Shadab and {Allen}, Lori and {Alvarez}, A. and {Annis}, James and {Ansarinejad}, Behzad and {Aubert}, Marie and {Beechert}, Jacqueline and {Bell}, Eric F. and {BenZvi}, Segev Y. and {Beutler}, Florian and {Bielby}, Richard M. and {Bolton}, Adam S. and {Brice{\~n}o}, C{\'e}sar and {Buckley-Geer}, Elizabeth J. and {Butler}, Karen and {Calamida}, Annalisa and {Carlberg}, Raymond G. and {Carter}, Paul and {Casas}, Ricard and {Castander}, Francisco J. and {Choi}, Yumi and {Comparat}, Johan and {Cukanovaite}, Elena and {Delubac}, Timoth{\'e}e and {DeVries}, Kaitlin and {Dey}, Sharmila and {Dhungana}, Govinda and {Dickinson}, Mark and {Ding}, Zhejie and {Donaldson}, John B. and {Duan}, Yutong and {Duckworth}, Christopher J. and {Eftekharzadeh}, Sarah and {Eisenstein}, Daniel J. and {Etourneau}, Thomas and {Fagrelius}, Parker A. and {Farihi}, Jay and {Fitzpatrick}, Mike and {Font-Ribera}, Andreu and {Fulmer}, Leah and {G{\"a}nsicke}, Boris T. and {Gaztanaga}, Enrique and {George}, Koshy and {Gerdes}, David W. and {Gontcho}, Satya Gontcho A. and {Gorgoni}, Claudio and {Green}, Gregory and {Guy}, Julien and {Harmer}, Diane and {Hernandez}, M. and {Honscheid}, Klaus and {Huang}, Lijuan Wendy and {James}, David J. and {Jannuzi}, Buell T. and {Jiang}, Linhua and {Joyce}, Richard and {Karcher}, Armin and {Karkar}, Sonia and {Kehoe}, Robert and {Kneib}, Jean-Paul and {Kueter-Young}, Andrea and {Lan}, Ting-Wen and {Lauer}, Tod R. and {Le Guillou}, Laurent and {Le Van Suu}, Auguste and {Lee}, Jae Hyeon and {Lesser}, Michael and {Perreault Levasseur}, Laurence and {Li}, Ting S. and {Mann}, Justin L. and {Marshall}, Robert and {Mart{\'\i}nez-V{\'a}zquez}, C.~E. and {Martini}, Paul and {du Mas des Bourboux}, H{\'e}lion and {McManus}, Sean and {Meier}, Tobias Gabriel and {M{\'e}nard}, Brice and {Metcalfe}, Nigel and {Mu{\~n}oz-Guti{\'e}rrez}, Andrea and {Najita}, Joan and {Napier}, Kevin and {Narayan}, Gautham and {Newman}, Jeffrey A. and {Nie}, Jundan and {Nord}, Brian and {Norman}, Dara J. and {Olsen}, Knut A.~G. and {Paat}, Anthony and {Palanque-Delabrouille}, Nathalie and {Peng}, Xiyan and {Poppett}, Claire L. and {Poremba}, Megan R. and {Prakash}, Abhishek and {Rabinowitz}, David and {Raichoor}, Anand and {Rezaie}, Mehdi and {Robertson}, A.~N. and {Roe}, Natalie A. and {Ross}, Ashley J. and {Ross}, Nicholas P. and {Rudnick}, Gregory and {Safonova}, Sasha and {Saha}, Abhijit and {S{\'a}nchez}, F. Javier and {Savary}, Elodie and {Schweiker}, Heidi and {Scott}, Adam and {Seo}, Hee-Jong and {Shan}, Huanyuan and {Silva}, David R. and {Slepian}, Zachary and {Soto}, Christian and {Sprayberry}, David and {Staten}, Ryan and {Stillman}, Coley M. and {Stupak}, Robert J. and {Summers}, David L. and {Sien Tie}, Suk and {Tirado}, H. and {Vargas-Maga{\~n}a}, Mariana and {Vivas}, A. Katherina and {Wechsler}, Risa H. and {Williams}, Doug and {Yang}, Jinyi and {Yang}, Qian and {Yapici}, Tolga and {Zaritsky}, Dennis and {Zenteno}, A. and {Zhang}, Kai and {Zhang}, Tianmeng and {Zhou}, Rongpu and {Zhou}, Zhimin},
        title = "{Overview of the DESI Legacy Imaging Surveys}",
      journal = {\aj},
         year = 2019,
        month = may,
       volume = {157},
       number = {5},
          eid = {168},
        pages = {168},
          doi = {10.3847/1538-3881/ab089d},
archivePrefix = {arXiv},
       eprint = {1804.08657},
 primaryClass = {astro-ph.IM},
       adsurl = {https://ui.adsabs.harvard.edu/abs/2019AJ....157..168D}
}

@ARTICLE{2019A&A...625A..97R,
       author = {{Rimoldini}, L. and {Holl}, B. and {Audard}, M. and {Mowlavi}, N. and {Nienartowicz}, K. and {Evans}, D.~W. and {Guy}, L.~P. and {Lecoeur-Ta{\"\i}bi}, I. and {Jevardat de Fombelle}, G. and {Marchal}, O. and {Roelens}, M. and {De Ridder}, J. and {Sarro}, L.~M. and {Regibo}, S. and {Lopez}, M. and {Clementini}, G. and {Ripepi}, V. and {Molinaro}, R. and {Garofalo}, A. and {Moln{\'a}r}, L. and {Plachy}, E. and {Juh{\'a}sz}, {\'A}. and {Szabados}, L. and {Lebzelter}, T. and {Teyssier}, D. and {Eyer}, L.},
        title = "{Gaia Data Release 2. All-sky classification of high-amplitude pulsating stars}",
      journal = {\aap},
         year = 2019,
        month = may,
       volume = {625},
          eid = {A97},
        pages = {A97},
          doi = {10.1051/0004-6361/201834616},
archivePrefix = {arXiv},
       eprint = {1811.03919},
 primaryClass = {astro-ph.SR},
       adsurl = {https://ui.adsabs.harvard.edu/abs/2019A&A...625A..97R}
}

@ARTICLE{2019MNRAS.482.4570G,
       author = {{Gentile Fusillo}, Nicola Pietro and {Tremblay}, Pier-Emmanuel and {G{\"a}nsicke}, Boris T. and {Manser}, Christopher J. and {Cunningham}, Tim and {Cukanovaite}, Elena and {Hollands}, Mark and {Marsh}, Thomas and {Raddi}, Roberto and {Jordan}, Stefan and {Toonen}, Silvia and {Geier}, Stephan and {Barstow}, Martin and {Cummings}, Jeffrey D.},
        title = "{A Gaia Data Release 2 catalogue of white dwarfs and a comparison with SDSS}",
      journal = {\mnras},
         year = 2019,
        month = feb,
       volume = {482},
       number = {4},
        pages = {4570-4591},
          doi = {10.1093/mnras/sty3016},
archivePrefix = {arXiv},
       eprint = {1807.03315},
 primaryClass = {astro-ph.SR},
       adsurl = {https://ui.adsabs.harvard.edu/abs/2019MNRAS.482.4570G}
}

@ARTICLE{2019PASP..131a8002B,
       author = {{Bellm}, Eric C. and {Kulkarni}, Shrinivas R. and {Graham}, Matthew J. and {Dekany}, Richard and {Smith}, Roger M. and {Riddle}, Reed and {Masci}, Frank J. and {Helou}, George and {Prince}, Thomas A. and {Adams}, Scott M. and {Barbarino}, C. and {Barlow}, Tom and {Bauer}, James and {Beck}, Ron and {Belicki}, Justin and {Biswas}, Rahul and {Blagorodnova}, Nadejda and {Bodewits}, Dennis and {Bolin}, Bryce and {Brinnel}, Valery and {Brooke}, Tim and {Bue}, Brian and {Bulla}, Mattia and {Burruss}, Rick and {Cenko}, S. Bradley and {Chang}, Chan-Kao and {Connolly}, Andrew and {Coughlin}, Michael and {Cromer}, John and {Cunningham}, Virginia and {De}, Kishalay and {Delacroix}, Alex and {Desai}, Vandana and {Duev}, Dmitry A. and {Eadie}, Gwendolyn and {Farnham}, Tony L. and {Feeney}, Michael and {Feindt}, Ulrich and {Flynn}, David and {Franckowiak}, Anna and {Frederick}, S. and {Fremling}, C. and {Gal-Yam}, Avishay and {Gezari}, Suvi and {Giomi}, Matteo and {Goldstein}, Daniel A. and {Golkhou}, V. Zach and {Goobar}, Ariel and {Groom}, Steven and {Hacopians}, Eugean and {Hale}, David and {Henning}, John and {Ho}, Anna Y.~Q. and {Hover}, David and {Howell}, Justin and {Hung}, Tiara and {Huppenkothen}, Daniela and {Imel}, David and {Ip}, Wing-Huen and {Ivezi{\'c}}, {\v{Z}}eljko and {Jackson}, Edward and {Jones}, Lynne and {Juric}, Mario and {Kasliwal}, Mansi M. and {Kaspi}, S. and {Kaye}, Stephen and {Kelley}, Michael S.~P. and {Kowalski}, Marek and {Kramer}, Emily and {Kupfer}, Thomas and {Landry}, Walter and {Laher}, Russ R. and {Lee}, Chien-De and {Lin}, Hsing Wen and {Lin}, Zhong-Yi and {Lunnan}, Ragnhild and {Giomi}, Matteo and {Mahabal}, Ashish and {Mao}, Peter and {Miller}, Adam A. and {Monkewitz}, Serge and {Murphy}, Patrick and {Ngeow}, Chow-Choong and {Nordin}, Jakob and {Nugent}, Peter and {Ofek}, Eran and {Patterson}, Maria T. and {Penprase}, Bryan and {Porter}, Michael and {Rauch}, Ludwig and {Rebbapragada}, Umaa and {Reiley}, Dan and {Rigault}, Mickael and {Rodriguez}, Hector and {van Roestel}, Jan and {Rusholme}, Ben and {van Santen}, Jakob and {Schulze}, S. and {Shupe}, David L. and {Singer}, Leo P. and {Soumagnac}, Maayane T. and {Stein}, Robert and {Surace}, Jason and {Sollerman}, Jesper and {Szkody}, Paula and {Taddia}, F. and {Terek}, Scott and {Van Sistine}, Angela and {van Velzen}, Sjoert and {Vestrand}, W. Thomas and {Walters}, Richard and {Ward}, Charlotte and {Ye}, Quan-Zhi and {Yu}, Po-Chieh and {Yan}, Lin and {Zolkower}, Jeffry},
        title = "{The Zwicky Transient Facility: System Overview, Performance, and First Results}",
      journal = {\pasp},
         year = 2019,
        month = jan,
       volume = {131},
       number = {995},
        pages = {018002},
          doi = {10.1088/1538-3873/aaecbe},
archivePrefix = {arXiv},
       eprint = {1902.01932},
 primaryClass = {astro-ph.IM},
       adsurl = {https://ui.adsabs.harvard.edu/abs/2019PASP..131a8002B}
}

@ARTICLE{2018MNRAS.480..302K,
       author = {{Kupfer}, T. and {Korol}, V. and {Shah}, S. and {Nelemans}, G. and {Marsh}, T.~R. and {Ramsay}, G. and {Groot}, P.~J. and {Steeghs}, D.~T.~H. and {Rossi}, E.~M.},
        title = "{LISA verification binaries with updated distances from Gaia Data Release 2}",
      journal = {\mnras},
         year = 2018,
        month = oct,
       volume = {480},
       number = {1},
        pages = {302-309},
          doi = {10.1093/mnras/sty1545},
archivePrefix = {arXiv},
       eprint = {1805.00482},
 primaryClass = {astro-ph.SR},
       adsurl = {https://ui.adsabs.harvard.edu/abs/2018MNRAS.480..302K}
}

@ARTICLE{2018A&A...616A...2L,
       author = {{Lindegren}, L. and {Hern{\'a}ndez}, J. and {Bombrun}, A. and {Klioner}, S. and {Bastian}, U. and {Ramos-Lerate}, M. and {de Torres}, A. and {Steidelm{\"u}ller}, H. and {Stephenson}, C. and {Hobbs}, D. and {Lammers}, U. and {Biermann}, M. and {Geyer}, R. and {Hilger}, T. and {Michalik}, D. and {Stampa}, U. and {McMillan}, P.~J. and {Casta{\~n}eda}, J. and {Clotet}, M. and {Comoretto}, G. and {Davidson}, M. and {Fabricius}, C. and {Gracia}, G. and {Hambly}, N.~C. and {Hutton}, A. and {Mora}, A. and {Portell}, J. and {van Leeuwen}, F. and {Abbas}, U. and {Abreu}, A. and {Altmann}, M. and {Andrei}, A. and {Anglada}, E. and {Balaguer-N{\'u}{\~n}ez}, L. and {Barache}, C. and {Becciani}, U. and {Bertone}, S. and {Bianchi}, L. and {Bouquillon}, S. and {Bourda}, G. and {Br{\"u}semeister}, T. and {Bucciarelli}, B. and {Busonero}, D. and {Buzzi}, R. and {Cancelliere}, R. and {Carlucci}, T. and {Charlot}, P. and {Cheek}, N. and {Crosta}, M. and {Crowley}, C. and {de Bruijne}, J. and {de Felice}, F. and {Drimmel}, R. and {Esquej}, P. and {Fienga}, A. and {Fraile}, E. and {Gai}, M. and {Garralda}, N. and {Gonz{\'a}lez-Vidal}, J.~J. and {Guerra}, R. and {Hauser}, M. and {Hofmann}, W. and {Holl}, B. and {Jordan}, S. and {Lattanzi}, M.~G. and {Lenhardt}, H. and {Liao}, S. and {Licata}, E. and {Lister}, T. and {L{\"o}ffler}, W. and {Marchant}, J. and {Martin-Fleitas}, J.-M. and {Messineo}, R. and {Mignard}, F. and {Morbidelli}, R. and {Poggio}, E. and {Riva}, A. and {Rowell}, N. and {Salguero}, E. and {Sarasso}, M. and {Sciacca}, E. and {Siddiqui}, H. and {Smart}, R.~L. and {Spagna}, A. and {Steele}, I. and {Taris}, F. and {Torra}, J. and {van Elteren}, A. and {van Reeven}, W. and {Vecchiato}, A.},
        title = "{Gaia Data Release 2. The astrometric solution}",
      journal = {\aap},
         year = 2018,
        month = aug,
       volume = {616},
          eid = {A2},
        pages = {A2},
          doi = {10.1051/0004-6361/201832727},
archivePrefix = {arXiv},
       eprint = {1804.09366},
 primaryClass = {astro-ph.IM},
       adsurl = {https://ui.adsabs.harvard.edu/abs/2018A&A...616A...2L}
}

@ARTICLE{2018A&A...616A...1G,
       author = {{Gaia Collaboration} and {Brown}, A.~G.~A. and {Vallenari}, A. and {Prusti}, T. and {de Bruijne}, J.~H.~J. and {Babusiaux}, C. and {Bailer-Jones}, C.~A.~L. and {Biermann}, M. and {Evans}, D.~W. and {Eyer}, L. and {Jansen}, F. and {Jordi}, C. and {Klioner}, S.~A. and {Lammers}, U. and {Lindegren}, L. and {Luri}, X. and {Mignard}, F. and {Panem}, C. and {Pourbaix}, D. and {Randich}, S. and {Sartoretti}, P. and {Siddiqui}, H.~I. and {Soubiran}, C. and {van Leeuwen}, F. and {Walton}, N.~A. and {Arenou}, F. and {Bastian}, U. and {Cropper}, M. and {Drimmel}, R. and {Katz}, D. and {Lattanzi}, M.~G. and {Bakker}, J. and {Cacciari}, C. and {Casta{\~n}eda}, J. and {Chaoul}, L. and {Cheek}, N. and {De Angeli}, F. and {Fabricius}, C. and {Guerra}, R. and {Holl}, B. and {Masana}, E. and {Messineo}, R. and {Mowlavi}, N. and {Nienartowicz}, K. and {Panuzzo}, P. and {Portell}, J. and {Riello}, M. and {Seabroke}, G.~M. and {Tanga}, P. and {Th{\'e}venin}, F. and {Gracia-Abril}, G. and {Comoretto}, G. and {Garcia-Reinaldos}, M. and {Teyssier}, D. and {Altmann}, M. and {Andrae}, R. and {Audard}, M. and {Bellas-Velidis}, I. and {Benson}, K. and {Berthier}, J. and {Blomme}, R. and {Burgess}, P. and {Busso}, G. and {Carry}, B. and {Cellino}, A. and {Clementini}, G. and {Clotet}, M. and {Creevey}, O. and {Davidson}, M. and {De Ridder}, J. and {Delchambre}, L. and {Dell'Oro}, A. and {Ducourant}, C. and {Fern{\'a}ndez-Hern{\'a}ndez}, J. and {Fouesneau}, M. and {Fr{\'e}mat}, Y. and {Galluccio}, L. and {Garc{\'\i}a-Torres}, M. and {Gonz{\'a}lez-N{\'u}{\~n}ez}, J. and {Gonz{\'a}lez-Vidal}, J.~J. and {Gosset}, E. and {Guy}, L.~P. and {Halbwachs}, J.-L. and {Hambly}, N.~C. and {Harrison}, D.~L. and {Hern{\'a}ndez}, J. and {Hestroffer}, D. and {Hodgkin}, S.~T. and {Hutton}, A. and {Jasniewicz}, G. and {Jean-Antoine-Piccolo}, A. and {Jordan}, S. and {Korn}, A.~J. and {Krone-Martins}, A. and {Lanzafame}, A.~C. and {Lebzelter}, T. and {L{\"o}ffler}, W. and {Manteiga}, M. and {Marrese}, P.~M. and {Mart{\'\i}n-Fleitas}, J.~M. and {Moitinho}, A. and {Mora}, A. and {Muinonen}, K. and {Osinde}, J. and {Pancino}, E. and {Pauwels}, T. and {Petit}, J.-M. and {Recio-Blanco}, A. and {Richards}, P.~J. and {Rimoldini}, L. and {Robin}, A.~C. and {Sarro}, L.~M. and {Siopis}, C. and {Smith}, M. and {Sozzetti}, A. and {S{\"u}veges}, M. and {Torra}, J. and {van Reeven}, W. and {Abbas}, U. and {Abreu Aramburu}, A. and {Accart}, S. and {Aerts}, C. and {Altavilla}, G. and {{\'A}lvarez}, M.~A. and {Alvarez}, R. and {Alves}, J. and {Anderson}, R.~I. and {Andrei}, A.~H. and {Anglada Varela}, E. and {Antiche}, E. and {Antoja}, T. and {Arcay}, B. and {Astraatmadja}, T.~L. and {Bach}, N. and {Baker}, S.~G. and {Balaguer-N{\'u}{\~n}ez}, L. and {Balm}, P. and {Barache}, C. and {Barata}, C. and {Barbato}, D. and {Barblan}, F. and {Barklem}, P.~S. and {Barrado}, D. and {Barros}, M. and {Barstow}, M.~A. and {Bartholom{\'e} Mu{\~n}oz}, S. and {Bassilana}, J.-L. and {Becciani}, U. and {Bellazzini}, M. and {Berihuete}, A. and {Bertone}, S. and {Bianchi}, L. and {Bienaym{\'e}}, O. and {Blanco-Cuaresma}, S. and {Boch}, T. and {Boeche}, C. and {Bombrun}, A. and {Borrachero}, R. and {Bossini}, D. and {Bouquillon}, S. and {Bourda}, G. and {Bragaglia}, A. and {Bramante}, L. and {Breddels}, M.~A. and {Bressan}, A. and {Brouillet}, N. and {Br{\"u}semeister}, T. and {Brugaletta}, E. and {Bucciarelli}, B. and {Burlacu}, A. and {Busonero}, D. and {Butkevich}, A.~G. and {Buzzi}, R. and {Caffau}, E. and {Cancelliere}, R. and {Cannizzaro}, G. and {Cantat-Gaudin}, T. and {Carballo}, R. and {Carlucci}, T. and {Carrasco}, J.~M. and {Casamiquela}, L. and {Castellani}, M. and {Castro-Ginard}, A. and {Charlot}, P. and {Chemin}, L. and {Chiavassa}, A. and {Cocozza}, G. and {Costigan}, G. and {Cowell}, S. and {Crifo}, F. and {Crosta}, M. and {Crowley}, C. and {Cuypers}, J. and {Dafonte}, C. and {Damerdji}, Y. and {Dapergolas}, A. and {David}, P. and {David}, M. and {de Laverny}, P. and {De Luise}, F.},
        title = "{Gaia Data Release 2. Summary of the contents and survey properties}",
      journal = {\aap},
         year = 2018,
        month = aug,
       volume = {616},
          eid = {A1},
        pages = {A1},
          doi = {10.1051/0004-6361/201833051},
archivePrefix = {arXiv},
       eprint = {1804.09365},
 primaryClass = {astro-ph.GA},
       adsurl = {https://ui.adsabs.harvard.edu/abs/2018A&A...616A...1G}
}

@ARTICLE{2018MNRAS.473.3241H,
       author = {{Hern{\'a}ndez Santisteban}, J.~V. and {Knigge}, C. and {Pretorius}, M.~L. and {Sullivan}, M. and {Warner}, B.},
        title = "{The space density of post-period minimum Cataclysmic Variables}",
      journal = {\mnras},
         year = 2018,
        month = jan,
       volume = {473},
       number = {3},
        pages = {3241-3250},
          doi = {10.1093/mnras/stx2296},
archivePrefix = {arXiv},
       eprint = {1708.09601},
 primaryClass = {astro-ph.SR},
       adsurl = {https://ui.adsabs.harvard.edu/abs/2018MNRAS.473.3241H}
}

@ARTICLE{2017RNAAS...1...29T,
       author = {{Thorstensen}, John R. and {Ringwald}, Frederick A. and {Taylor}, Cynthia J. and {Sheets}, Holly A. and {Peters}, Christopher S. and {Skinner}, Julie N. and {Alper}, Erek H. and {Weil}, Kathryn E.},
        title = "{New or Improved Orbital Periods of Cataclysmic Binaries}",
      journal = {Research Notes of the American Astronomical Society},
         year = 2017,
        month = dec,
       volume = {1},
       number = {1},
          eid = {29},
        pages = {29},
          doi = {10.3847/2515-5172/aa9d2a},
archivePrefix = {arXiv},
       eprint = {1711.09094},
 primaryClass = {astro-ph.SR},
       adsurl = {https://ui.adsabs.harvard.edu/abs/2017RNAAS...1...29T}
}

@ARTICLE{2017MNRAS.472.4193R,
       author = {{Rebassa-Mansergas}, A. and {Ren}, J.~J. and {Irawati}, P. and {Garc{\'\i}a-Berro}, E. and {Parsons}, S.~G. and {Schreiber}, M.~R. and {G{\"a}nsicke}, B.~T. and {Rodr{\'\i}guez-Gil}, P. and {Liu}, X. and {Manser}, C. and {Nevado}, S.~P. and {Jim{\'e}nez-Ibarra}, F. and {Costero}, R. and {Echevarr{\'\i}a}, J. and {Michel}, R. and {Zorotovic}, M. and {Hollands}, M. and {Han}, Z. and {Luo}, A. and {Villaver}, E. and {Kong}, X.},
        title = "{The white dwarf binary pathways survey - II. Radial velocities of 1453 FGK stars with white dwarf companions from LAMOST DR 4}",
      journal = {\mnras},
         year = 2017,
        month = dec,
       volume = {472},
       number = {4},
        pages = {4193-4203},
          doi = {10.1093/mnras/stx2259},
archivePrefix = {arXiv},
       eprint = {1708.09480},
 primaryClass = {astro-ph.SR},
       adsurl = {https://ui.adsabs.harvard.edu/abs/2017MNRAS.472.4193R}
}

@ARTICLE{2017ApJS..230...24B,
       author = {{Bianchi}, Luciana and {Shiao}, Bernie and {Thilker}, David},
        title = "{Revised Catalog of GALEX Ultraviolet Sources. I. The All-Sky Survey: GUVcat\_AIS}",
      journal = {\apjs},
         year = 2017,
        month = jun,
       volume = {230},
       number = {2},
          eid = {24},
        pages = {24},
          doi = {10.3847/1538-4365/aa7053},
archivePrefix = {arXiv},
       eprint = {1704.05903},
 primaryClass = {astro-ph.GA},
       adsurl = {https://ui.adsabs.harvard.edu/abs/2017ApJS..230...24B}
}

@dataset{2017yCat....102027W,
       author = {{Watson}, C. and {Henden}, A.~A. and {Price}, A.},
        title = "{VizieR Online Data Catalog: AAVSO International Variable Star Index VSX (Watson+, 2006-2014)}",
 howpublished = {VizieR On-line Data Catalog: B/vsx.  Originally published in: 2006SASS...25...47W},
         year = 2017,
        month = may,
          eid = {B/vsx},
       adsurl = {https://ui.adsabs.harvard.edu/abs/2017yCat....102027W}
}

@ARTICLE{2016MNRAS.463.2125P,
       author = {{Parsons}, S.~G. and {Rebassa-Mansergas}, A. and {Schreiber}, M.~R. and {G{\"a}nsicke}, B.~T. and {Zorotovic}, M. and {Ren}, J.~J.},
        title = "{The white dwarf binary pathways survey - I. A sample of FGK stars with white dwarf companions}",
      journal = {\mnras},
         year = 2016,
        month = dec,
       volume = {463},
       number = {2},
        pages = {2125-2136},
          doi = {10.1093/mnras/stw2143},
archivePrefix = {arXiv},
       eprint = {1604.01613},
 primaryClass = {astro-ph.SR},
       adsurl = {https://ui.adsabs.harvard.edu/abs/2016MNRAS.463.2125P}
}

@ARTICLE{2016AJ....152..226T,
       author = {{Thorstensen}, John R. and {Alper}, Erek H. and {Weil}, Kathryn E.},
        title = "{A Trip to the Cataclysmic Binary Zoo: Detailed Follow-up of 35 Recently Discovered Systems}",
      journal = {\aj},
         year = 2016,
        month = dec,
       volume = {152},
       number = {6},
          eid = {226},
        pages = {226},
          doi = {10.3847/1538-3881/152/6/226},
archivePrefix = {arXiv},
       eprint = {1609.02215},
 primaryClass = {astro-ph.SR},
       adsurl = {https://ui.adsabs.harvard.edu/abs/2016AJ....152..226T}
}

@ARTICLE{2016MNRAS.458.3808R,
       author = {{Rebassa-Mansergas}, A. and {Ren}, J.~J. and {Parsons}, S.~G. and {G{\"a}nsicke}, B.~T. and {Schreiber}, M.~R. and {Garc{\'\i}a-Berro}, E. and {Liu}, X.-W. and {Koester}, D.},
        title = "{The SDSS spectroscopic catalogue of white dwarf-main-sequence binaries: new identifications from DR 9-12}",
      journal = {\mnras},
         year = 2016,
        month = jun,
       volume = {458},
       number = {4},
        pages = {3808-3819},
          doi = {10.1093/mnras/stw554},
archivePrefix = {arXiv},
       eprint = {1603.01017},
 primaryClass = {astro-ph.SR},
       adsurl = {https://ui.adsabs.harvard.edu/abs/2016MNRAS.458.3808R}
}

@ARTICLE{2016MNRAS.455L..16S,
       author = {{Schreiber}, M.~R. and {Zorotovic}, M. and {Wijnen}, T.~P.~G.},
        title = "{Three in one go: consequential angular momentum loss can solve major problems of CV evolution}",
      journal = {\mnras},
         year = 2016,
        month = jan,
       volume = {455},
       number = {1},
        pages = {L16-L20},
          doi = {10.1093/mnrasl/slv144},
archivePrefix = {arXiv},
       eprint = {1510.04294},
 primaryClass = {astro-ph.SR},
       adsurl = {https://ui.adsabs.harvard.edu/abs/2016MNRAS.455L..16S}
}



\newpage

\appendix 

\section*{Affiliations}
\label{sec:affiliations}

\vspace{0.4cm}

$^{1}$Department of Physics, University of Warwick, Gibbet Hill Road, Coventry, CV4 7AL, UK\\
$^{2}$Department of Physics, Faculty of Science, Naresuan University, Phitsanulok 65000, Thailand\\
$^{3}$Department of Physics \& Astronomy, University  of Wyoming, 1000 E. University, Dept.~3905, Laramie, WY 82071, USA\\
$^{4}$Department of Physics and Astronomy, Dartmouth College, Hanover NH 03755, USA \\
$^{5}$Lawrence Berkeley National Laboratory, 1 Cyclotron Road, Berkeley, CA 94720, USA\\
$^{6}$Department of Physics, Boston University, 590 Commonwealth Avenue, Boston, MA 02215 USA\\
$^{7}$Dipartimento di Fisica ``Aldo Pontremoli'', Universit\`a degli Studi di Milano, Via Celoria 16, I-20133 Milano, Italy\\
$^{8}$INAF-Osservatorio Astronomico di Brera, Via Brera 28, 20122 Milano, Italy\\
$^{9}$Department of Physics \& Astronomy, University College London, Gower Street, London, WC1E 6BT, UK\\
$^{10}$Instituto de F\'{\i}sica, Universidad Nacional Aut\'{o}noma de M\'{e}xico,  Circuito de la Investigaci\'{o}n Cient\'{\i}fica, Ciudad Universitaria, Cd. de M\'{e}xico  C.~P.~04510,  M\'{e}xico\\
$^{11}$Department of Astronomy \& Astrophysics, University of Toronto, Toronto, ON M5S 3H4, Canada\\
$^{12}$Department of Physics \& Astronomy and Pittsburgh Particle Physics, Astrophysics, and Cosmology Center (PITT PACC), University of Pittsburgh, 3941 O'Hara Street, Pittsburgh, PA 15260, USA\\
$^{13}$Instituci\'{o} Catalana de Recerca i Estudis Avan\c{c}ats, Passeig de Llu\'{\i}s Companys, 23, 08010 Barcelona, Spain\\
$^{14}$Institut de F\'{i}sica dâ€™Altes Energies (IFAE), The Barcelona Institute of Science and Technology, Edifici Cn, Campus UAB, 08193, Bellaterra (Barcelona), Spain\\
$^{15}$Departamento de F\'isica, Universidad de los Andes, Cra. 1 No. 18A-10, Edificio Ip, CP 111711, Bogot\'a, Colombia\\
$^{16}$Observatorio Astron\'omico, Universidad de los Andes, Cra. 1 No. 18A-10, Edificio H, CP 111711 Bogot\'a, Colombia\\
$^{17}$University of Virginia, Department of Astronomy, Charlottesville, VA 22904, USA\\
$^{18}$Fermi National Accelerator Laboratory, PO Box 500, Batavia, IL 60510, USA\\
$^{19}$NSF NOIRLab, 950 N. Cherry Ave., Tucson, AZ 85719, USA\\
$^{20}$Institute for Astronomy, University of Edinburgh, Royal Observatory, Blackford Hill, Edinburgh EH9 3HJ, UK\\
$^{21}$Institute of Astronomy, University of Cambridge, Madingley Road, Cambridge CB3 0HA, UK\\
$^{22}$Sorbonne Universit\'{e}, CNRS/IN2P3, Laboratoire de Physique Nucl\'{e}aire et de Hautes Energies (LPNHE), FR-75005 Paris, France\\
$^{23}$Department of Physics and Astronomy, Siena University, 515 Loudon Road, Loudonville, NY 12211, USA\\
$^{24}$Departament de F\'isica, EEBE, Universitat Polit\`ecnica de Catalunya, c/Eduard Maristany 10, 08930 Barcelona, Spain\\
$^{25}$Department of Physics and Astronomy, University of Waterloo, 200 University Ave W, Waterloo, ON N2L 3G1, Canada\\
$^{26}$Perimeter Institute for Theoretical Physics, 31 Caroline St. North, Waterloo, ON N2L 2Y5, Canada\\
$^{27}$Waterloo Centre for Astrophysics, University of Waterloo, 200 University Ave W, Waterloo, ON N2L 3G1, Canada\\
$^{28}$Instituto de Astrof\'{i}sica de Andaluc\'{i}a (CSIC), Glorieta de la Astronom\'{i}a, s/n, E-18008 Granada, Spain\\
$^{29}$Department of Physics and Astronomy, Sejong University, 209 Neungdong-ro, Gwangjin-gu, Seoul 05006, Republic of Korea\\
$^{30}$CIEMAT, Avenida Complutense 40, E-28040 Madrid, Spain\\
$^{31}$Department of Physics, University of Michigan, 450 Church Street, Ann Arbor, MI 48109, USA\\
$^{32}$University of Michigan, 500 S. State Street, Ann Arbor, MI 48109, USA\\
$^{33}$Department of Physics \& Astronomy, Ohio University, 139 University Terrace, Athens, OH 45701, USA\\
$^{34}$National Astronomical Observatories, Chinese Academy of Sciences, A20 Datun Road, Chaoyang District, Beijing, 100101, P.~R.~China\\

\clearpage

\section{The Close White Dwarf Binary survey}\label{sec:CWDBC}

\begin{figure*} 
\includegraphics[width=1\columnwidth]{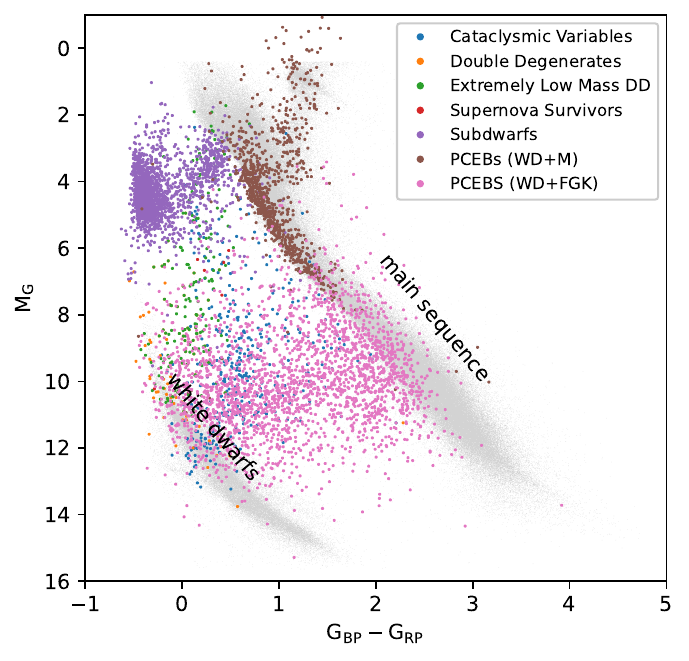} 
\hfill
\includegraphics[width=1\columnwidth]{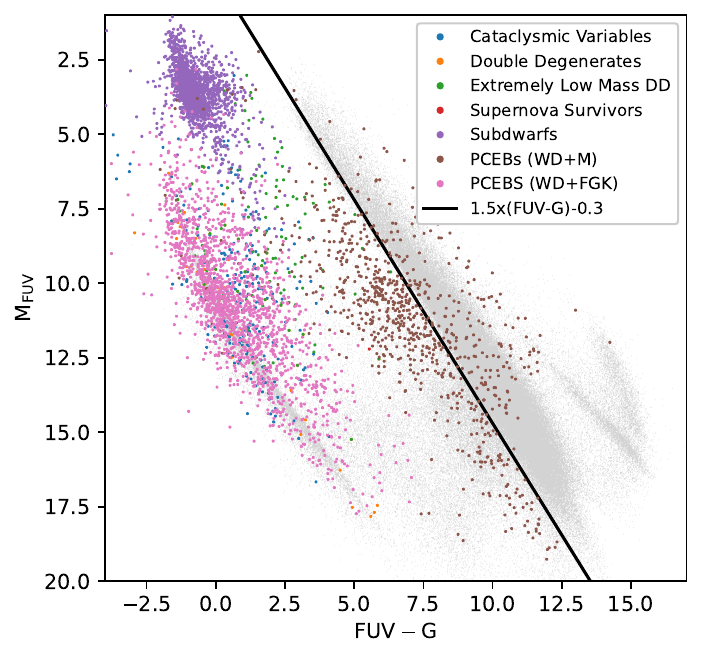}
\caption{\label{fig:desi_wdbinaries_hrd} \textit{Left:} The majority of known close white dwarf binaries in the \textit{Gaia} HR diagram fall between the main sequence and the white dwarf cooling track. The sub-types include interacting binaries and detached systems such as post common envelope binaries (PCEBs). \textit{Right:} The population of known close white dwarf binaries in a UV-optical colour-magnitude diagram, where a simple linear cut (Eq.\,\ref{eq:fuv-cut}), recovers the overwhelming majority of known white dwarf binaries while minimizing the number of contaminants.}
\end{figure*} 

\begin{figure*} 
\includegraphics[width=1\columnwidth]{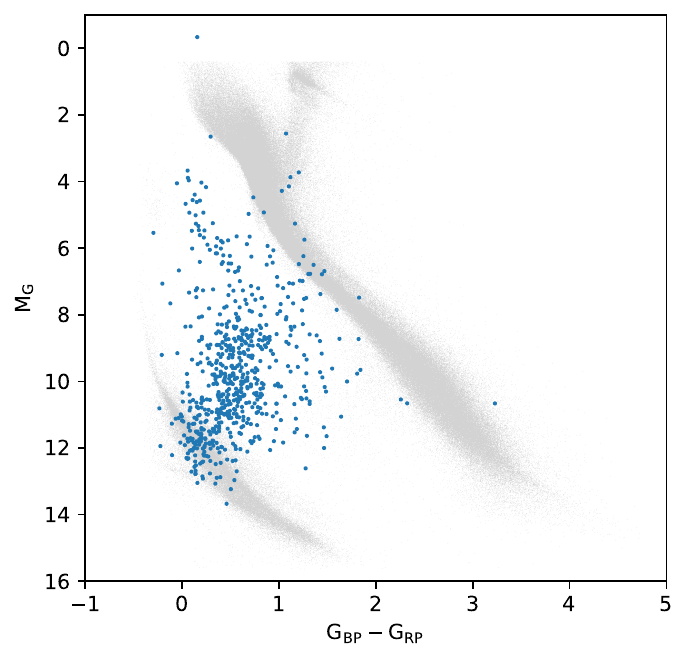} 
\hfill
\includegraphics[width=1\columnwidth]{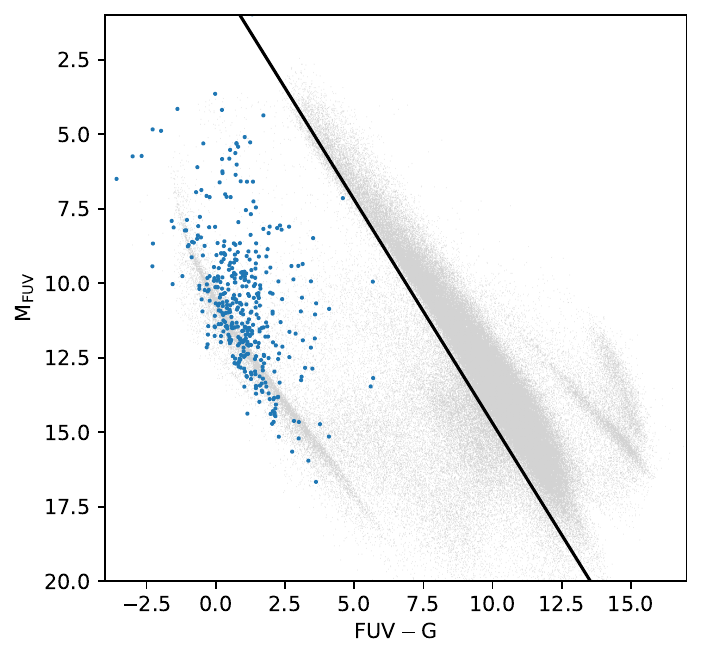} 
\caption{\label{fig:desi_ourCVs_hrd}
Same as Fig.\,\ref{fig:desi_wdbinaries_hrd}, but showing the DESI CVs from this work, the majority of which fall between the main sequence and the white dwarf cooling track (left). However, only a small fraction of the DESI CVs were captured by the CWDB target class (right), which is due to a combination of a lack of depth and coverage of the \textit{GALEX} observations and the fact that a large fraction of the CVs that have a UV detection are already contained within the single white dwarf target class.}
\end{figure*} 

It is tempting to select white dwarf binary candidates from the \textit{Gaia} HR diagram as ``systems that lie between the white dwarf cooling sequence and the main sequence'', and that approach has been used in a number of investigations \citep[e.g.][]{2025A&A...699A.153R, 2021MNRAS.508.4106E}. Upon closer inspection, however, it becomes clear that the parameter space between the two stellar sequences is contaminated by quasars, flaring stars, young stellar objects, and more generally astronomical objects that have poor \textit{Gaia} astrometry and colours. In addition, a fraction of genuine white dwarf binaries have poor astrometry and photometry because of variability and/or unusual colours \citep{2018A&A...616A...2L}, hence a cut on the quality of \textit{Gaia} data would remove genuine CVs. 

We therefore designed a target selection that can drastically reduce the number of false positives in a simple \textit{Gaia}-only selection of white dwarf binary candidates by making use of the fact that (almost) all white dwarf binaries are moderately bright UV sources due to the white dwarf and/or the accretion flow. In other words, our selection aims to (1) strengthen the likelihood that an object between the two stellar sequence is a white dwarf binary and (2) detect white dwarf binaries that are ``hidden'' within the main sequence, or so close to it that they cannot be reliably selected from optical data alone (see \citealt{2016MNRAS.463.2125P, 2017MNRAS.472.4193R, 2026ApJS..284...38Y} for a similar approach to find detached white dwarf plus FGK-type binaries). 

The first step in designing this UV-\textit{Gaia} based target selection was to carefully map the parameter space spanned by white dwarf binaries. For this purpose, we collected clean samples of all white dwarf binary sub-types from the literature.  Spectroscopically confirmed systems include cataclysmic variables and detached white dwarf plus M-dwarf binaries with SDSS spectroscopy from \citet{2023MNRAS.524.4867I} and \citet{2010MNRAS.402..620R, 2012MNRAS.419..806R, 2013MNRAS.433.3398R, 2016MNRAS.463.2125P}, extremely low-mass white dwarfs from \citet{2020ApJ...889...49B}, double-degenerates from Marsh (unpublished). We added the white dwarf plus FGK binary candidates from \citet{2016MNRAS.463.2125P} and \citet{2016MNRAS.463.2125P}, where the main-sequence component is spectroscopically confirmed, but the existence of a white dwarf companion is based on the detection of a UV excess. Follow-up with \textit{HST} demonstrated that the majority of these systems are indeed white dwarf binaries \citep{2015MNRAS.452.1754P, 2021MNRAS.501.1677H, 2022MNRAS.512.1843H, 2022MNRAS.517.2867H, 2023MNRAS.518.4579P, 2024MNRAS.529.4840G} with a small number of contaminants in the form of wide white dwarf tertiaries \citep{2020MNRAS.494..915L} and active stars \citep{2024MNRAS.529.4840G}. 

Next, we cross-matched these samples with \textit{Gaia} DR2 (this DESI target class was defined before \textit{Gaia} DR3 was available). Whereas the vast majority of white dwarf binaries are located between the main sequence and the white dwarf cooling sequence (Fig.\,\ref{fig:desi_wdbinaries_hrd}, left), a significant number are scattered across, and above, the main sequence and giant branch; these are systems with a luminous companion, and/or a high accretion rate. Capturing these latter systems requires the target selection to go beyond \textit{Gaia}. 

Finally, we cross-matched the \textit{GALEX}~GR7 source catalogue \citep{2017ApJS..230...24B} with \textit{Gaia}~DR2, resulting in $\simeq 1.4 \times 10^6$ \textit{Gaia} sources with far-ultraviolet (FUV) detections. Transposing the known white dwarf binaries into a $FUV$ vs $FUV-G$ colour-magnitude diagram (Fig.\,\ref{fig:desi_wdbinaries_hrd}, right), we found that the vast majority of all known white dwarf binaries are recovered by a simple linear cut:
\begin{equation}
\label{eq:fuv-cut}
M_{FUV}>1.5 \times (FUV-G)-0.3
\end{equation}
Systems excluded by this cut are exclusively giant stars with white dwarf companions, that are extremely difficult to identify by any means other than ultraviolet spectroscopy. In addition, we applied a basic quality cut on the \textit{Gaia} parallax,
($\varpi / \sigma_{\varpi} >5$) to remove contaminants scattered into the selected region, resulting in $\simeq 126\,000$ candidates. Finally, we removed single white dwarf candidates (which would have duplicated targets in the DESI white dwarf targeting category) leaving a list of $34\,056$ targets in total. These were split into 7057 faint ($G>18$) and 26\,999 bright ($16<G\le18$) targets to be observed in the dark and bright-time surveys.

\bsp	
\label{lastpage}
\end{document}